\documentclass[letterpaper,11pt]{article}
\pdfoutput=1 

\usepackage{jheppub} 

\usepackage{subfig}
\usepackage{xspace}

\newcommand{\EC}[2]{{\textrm{ECF}(#1,#2)}}
\newcommand{\rN}[2]{r_{#1}^{(#2)}}
\newcommand{\EEC}{\textrm{EEC}}
\newcommand{\GeV}{\textrm{GeV}}
\newcommand{\as}{\alpha_s}

\newcommand{\Nsub}[2]{\tau_{#1}^{(#2)}}
\newcommand{\Nsubnobeta}[1]{\tau_{#1}}

\newcommand{\C}[2]{C^{(#2)}_{#1}}
\newcommand{\Cnobeta}[1]{C_{#1}}

\usepackage{color}
\definecolor{darkgreen}{rgb}{0,0.5,0}
\definecolor{darkblue}{rgb}{0,0,0.5}
\definecolor{darkred}{rgb}{0.5,0,0.0}

\DeclareRobustCommand{\Sec}[1]{Sec.~\ref{#1}}

\DeclareRobustCommand{\App}[1]{App.~\ref{#1}}

\DeclareRobustCommand{\Fig}[1]{Fig.~\ref{#1}}
\DeclareRobustCommand{\Figs}[2]{Figs.~\ref{#1} and \ref{#2}}
\DeclareRobustCommand{\Eq}[1]{Eq.~(\ref{#1})}

\DeclareRobustCommand{\Ref}[1]{Ref.~\cite{#1}}
\DeclareRobustCommand{\Refs}[1]{Refs.~\cite{#1}}

\newcommand{\pythia}{\textsc{Pythia}\xspace}
\newcommand{\madgraph}{\textsc{MadGraph5}\xspace}
\newcommand{\fastjet}{\textsc{FastJet}\xspace}
\newcommand{\herwig}{\textsc{Herwig}\xspace}
\newcommand{\herwigpp}{\textsc{Herwig++}\xspace}

\newcommand{\be}{\begin{equation}}
\newcommand{\ee}{\end{equation}}

\newcommand{\order}[1]{{\cal O}\left(#1\right)}

\title{Energy Correlation Functions for Jet Substructure}

 \author[a]{Andrew J. Larkoski,}
 \author[b,c,1]{Gavin P. Salam\note{On leave from Department of Physics, Princeton University, Princeton, NJ 08544, USA.}}
 \author[a]{and Jesse Thaler}
  \affiliation[a]{Center for Theoretical Physics, Massachusetts Institute of Technology,\\ Cambridge, MA 02139, USA}
\affiliation[b]{CERN, PH-TH, CH-1211 Geneva 23, Switzerland}
\affiliation[c]{LPTHE; CNRS UMR 7589; UPMC Univ. Paris 6; Paris 75252, France}

\emailAdd{larkoski@mit.edu}
\emailAdd{gavin.salam@cern.ch}
\emailAdd{jthaler@mit.edu}

\preprint{
  MIT--CTP 4446\\
  \mbox{ } \hfill CERN--PH--TH/2013--066\\
  \mbox{ } \hfill LPN13--026
}

\abstract{
We show how generalized energy correlation functions can be used as a powerful probe of jet substructure.  These correlation functions are based on the energies and pair-wise angles of particles within a jet, with $(N+1)$-point correlators sensitive to $N$-prong substructure.  Unlike many previous jet substructure methods, these correlation functions do not require the explicit identification of subjet regions.  In addition, the correlation functions are better probes of certain soft and collinear features that are masked by other methods.  We present three Monte Carlo case studies to illustrate the utility of these observables:  2-point correlators for quark/gluon discrimination, 3-point correlators for boosted $W$/$Z$/Higgs boson identification, and 4-point correlators for boosted top quark identification.  For quark/gluon discrimination, the 2-point correlator is particularly powerful, as can be understood via a next-to-leading logarithmic calculation. For boosted 2-prong resonances the benefit depends on the mass of the resonance.
}

\begin{document} 
\maketitle
\flushbottom

\section{Introduction}

The field of jet substructure has evolved significantly over the last
few years \cite{Abdesselam:2010pt,Altheimer:2012mn}.  Many procedures
have been developed not only for identifying and classifying jets
\cite{Seymour:1993mx,Butterworth:2002tt,Butterworth:2008iy,Brooijmans:1077731,Thaler:2008ju,Kaplan:2008ie,Almeida:2008yp}
but also for removing jet contamination due to underlying event or pile-up \cite{Cacciari:2007fd,Butterworth:2008iy,Krohn:2009th,Ellis:2009me,Alon:2011xb,Soyez:2012hv}.   On the theoretical side, there has been substantial progress in computing and understanding these observables and procedures in perturbative QCD \cite{Ellis:2010rwa,Banfi:2010pa,Walsh:2011fz,Li:2012bw,Dasgupta:2012hg,Feige:2012vc,Larkoski:2012eh,Chien:2012ur,Jouttenus:2013hs}.  On the experimental side, the ATLAS and CMS experiments at the Large Hadron Collider (LHC) have begun measuring and testing jet substructure ideas \cite{ATLAS-CONF-2011-073,Miller:2011qg,ATLAS-CONF-2011-053,ATLAS:2012am,ATLAS:2012xna,Aad:2012meb,Aad:2012raa,ATLAS:2012dp,ATLAS-CONF-2012-066,ATLAS-CONF-2012-065,CMS-PAS-QCD-10-041,CMS-PAS-JME-10-013,CMS:2011bqa,Chatrchyan:2012mec,Chatrchyan:2012tt,Chatrchyan:2012sn}, with pile-up suppression becoming increasingly important at higher luminosities.  With the recent discovery of a Higgs-like particle \cite{:2012gk,:2012gu}, jet substructure methods for identifying the $H\to b\bar{b}$ decay mode \cite{Butterworth:2008iy} (and potentially the $H \to gg$ decay mode) could be vital for testing Higgs properties.

A common strategy for jet substructure studies is to first identity
subjets, namely, localized subclusters of energy within a jet.  Jet
discrimination then involves studying the properties of and
relationship between the subjets.  For example, BDRS
\cite{Butterworth:2008iy} and related methods
\cite{Kaplan:2008ie,Plehn:2009rk,Plehn:2010st} involve first
reclustering the jet with the Cambridge/Aachen
\cite{Dokshitzer:1997in,Wobisch:1998wt,Wobisch:2000dk} or $k_T$
\cite{Catani:1993hr,Ellis:1993tq}  jet algorithm and then stepping
through the  clustering history to identify a hard splitting in the
jet; pruning~\cite{Ellis:2009me} is similar. 
$N$-subjettiness \cite{Thaler:2010tr,Thaler:2011gf} relies on a (quasi-)minimization procedure to identify $N$ subjet directions in the jet.  Of course, there are jet shapes such as jet angularities \cite{Berger:2003iw,Almeida:2008yp}, planar flow \cite{Thaler:2008ju,Almeida:2008yp}, Zernike coefficients \cite{GurAri:2011vx}, and Fox-Wolfram moments \cite{Bernaciak:2012nh} that can be used for classifying jets without subjet finding.  Considered individually, however,  these jet shapes tend not to yield the same discrimination power as subjet methods, since they are sensitive mainly to exotic kinematic configurations and not directly to prong-like substructure.

In this paper, we introduce generalized energy correlation functions
that can identify $N$-prong jet substructure without requiring a
subjet finding procedure.  These correlators only use information
about the energies and pair-wise angles of particles within a jet, but
yield discrimination power comparable to methods based on subjets.  As
we will see, $(N+1)$-point correlation functions are sensitive to
$N$-prong substructure, with an angular exponent $\beta$ that can be
adjusted to optimize the discrimination power.  To our knowledge, the
2-point correlators---schematically $\sum_{i,j} E_i E_j
\theta_{ij}^\beta$ where the sum runs over all particles $i$ and $j$ in a
jet or event---first appeared in \Ref{Banfi:2004yd} and independently
in \Ref{Jankowiak:2011qa}, with no previous studies of higher-point
correlators.\footnote{Our definition of the energy correlation
  function should not be confused with
  \Refs{Basham:1977iq,Basham:1978bw,Basham:1978zq,Hofman:2008ar},
  which refer to an ensemble average of products of energies measured
  at fixed angles.  Here, energy correlation functions are measured on
  an event-by-event basis.}

Besides the novelty of not requiring subjet finding,  a key feature of the generalized energy correlation functions is that
the angular exponent $\beta$ can be set to any value consistent with infrared and collinear safety, namely $\beta > 0$.  In contrast, observables like angularities \cite{Berger:2003iw,Almeida:2008yp} are required to have $\beta > 1$ to avoid being dominated by recoil effects.\footnote{As will be discussed in \Sec{subsec:benefits} and a forthcoming publication \cite{broadening}, $N$-subjettiness may or may not have recoil sensitivity depending on how the axes are chosen.}  By choosing values of $\beta \simeq 0.2$, the correlators are able to more effectively probe small-scale collinear splittings, which will turn out to be useful for quark/gluon discrimination.

To put our work in perspective, it is worth remembering that the basic idea for using energy correlation functions to determine the number of jets in an event is actually quite old.  As we will review, the $C$-parameter for $e^+e^-$ collisions \cite{Parisi:1978eg,Donoghue:1979vi} is essentially a 3-point energy correlation function that can be used to identify events that have two jets.  However, the $C$-parameter is defined as a function of the eigenvalues of the sphericity tensor and therefore only gives sensible values for systems that have zero total momentum and for events that are nearly dijet-like.  In contrast, our generalized energy correlation functions give sensible results in any Lorentz frame and can be used to identify any number of jets in an event (or subjets within a jet).  In addition, they can be defined in any number of spacetime dimensions.

The remainder of the paper is organized as follows.  In \Sec{sec:eec}, we introduce arbitrary-point energy correlation functions and define appropriate energy correlation double ratios $\C{N}{\beta}$ (built from the $(N+1)$-point correlator), which can be used to identify a system with $N$ (sub)jets.  We also contrast the behavior of $\C{N}{\beta}$ with $N$-subjettiness ratios. We then present three case studies to show how these generalized energy correlation functions work for different types of jet discrimination.  
\begin{itemize}
\item \textit{Quark/gluon discrimination.}  Using $\C{1}{\beta}$ (built from the 2-point correlator) in \Sec{sec:oneprong}, we perform both an analytic study and a Monte Carlo study of quark/gluon separation.  Through a next-to-leading logarithmic study, we explain why quark/gluon discrimination greatly improves as the angular exponent approaches zero (at least down to $\beta \simeq 0.2$),
highlighting the importance of working with recoil-free observables.
\item \textit{Boosted $W$/$Z$/Higgs identification.}  Using
  $\C{2}{\beta}$ (built from the 3-point correlator) in
  \Sec{sec:twoprong}, we will see that the discrimination power
  between QCD jets and jets with two intrinsic
  subjets from a colour-singlet decay depends strongly on the ratio of the jet mass to its
  transverse momentum.
  This occurs because a QCD jet obtains mass in different ways depending on this ratio.  In particular, we will see that the energy correlation function performs better than $N$-subjettiness in situations where the jet mass is dominated by soft wide-angle emissions.
\item \textit{Boosted top quark identification.}  Using $\C{3}{\beta}$ (built from the 4-point correlator) in \Sec{sec:threeprong}, we find comparable discrimination power to other top-tagging methods. While one might worry that the 4-point correlators would face a high computational cost, we find that a boosted top event can be analyzed for a single value of $\beta$ in a few milliseconds.
\end{itemize}
We conclude in \Sec{sec:conc} with an experimental and theoretical outlook.  The energy correlation functions are available as an add-on to \fastjet~3 \cite{Cacciari:2011ma} as part of the \fastjet contrib project (\url{http://fastjet.hepforge.org/contrib/}).

\section{Generalized Energy Correlation Functions}
\label{sec:eec}

The basis for our analysis is the $N$-point energy correlation function (ECF) 
\begin{equation}
\label{eq:preECF}
\EC{N}{\beta} = \sum_{i_1 < i_2< \ldots < i_N \in J}  \left( \prod_{a = 1}^N E_{i_a} \right)  \left(  \prod_{b=1}^{N-1} \prod_{c = b+1}^N  \theta_{i_b i_c} \right)^\beta \ .
\end{equation}
Here, the sum runs over all particles within the system $J$ (either a jet or the whole event).   Each term consists of $N$ energies multiplied together with $\binom{N}{2}$ pairwise angles raised to the angular exponent $\beta$.  This function is well-defined in any number of space-time dimensions as well as for systems that do not have zero total momentum.  Note that it is infrared and collinear (IRC) safe for all $\beta> 0$.  Moreover, $\EC{N}{\beta}$ goes to zero in all possible soft and collinear limits of $N$ partons.

As written, \Eq{eq:preECF} is most appropriate for $e^+e^-$ colliders where energies and angles are the usual experimental observables.  For hadron colliders, it is more natural to define $\EC{N}{\beta}$ as a transverse momentum correlation function:\footnote{We will continue to use the notation ECF, though we will mainly use the transverse momentum version in this paper.} 
\begin{equation}
\EC{N}{\beta} = \sum_{i_1 < i_2 < \ldots <  i_N \in J}  \left( \prod_{a = 1}^N {p_T}_{i_a} \right)  \left(  \prod_{b=1}^{N-1} \prod_{c = b+1}^N  R_{i_b i_c} \right)^\beta \ ,
\end{equation}
where $R_{ij}$ is the Euclidean distance between $i$ and $j$ in the
rapidity-azimuth angle plane, $R^2_{ij} = (y_i - y_j)^2 + (\phi_i -
\phi_j)^2$, with $y_i = \frac12 \ln\frac{E_i+p_{zi}}{E_i-p_{zi}}$.
In this paper, we will only consider up to 4-point correlation functions:
\begin{align}
\EC{0}{\beta} &= 1, \\
\EC{1}{\beta} &= \sum_{i \in J} {p_T}_{i}, \\
\EC{2}{\beta} &= \sum_{i<j \in J} {p_T}_{i} \, {p_T}_{j} (R_{ij})^\beta, \\
\EC{3}{\beta}&= \sum_{i<j<k \in J} {p_T}_{i} \,{p_T}_{j} \,{p_T}_{k} \left( R_{ij}  R_{ik}  R_{jk} \right)^\beta, \\
\EC{4}{\beta}&= \sum_{i<j<k<\ell \in J} {p_T}_{i}\, {p_T}_{j} \,{p_T}_{k} \,{p_T}_{\ell} \left( R_{ij}  R_{ik}  R_{i\ell}  R_{jk}  R_{j\ell}  R_{k\ell} \right)^\beta.
\end{align}
If a jet has fewer than $N$ constituents then $\EC{N}{\beta} = 0$.
Note that the computational cost for $\EC{N}{\beta}$
with $k$ particles scales like $k^N/N!$. 

From the $\EC{N}{\beta}$, we would like to define a dimensionless observable that can be used to determine if a system has $N$ subjets.  The key observation is that the $(N+1)$-point correlators go to zero if there are only $N$ particles.  More generally, if a system has $N$ subjets, then $\EC{N+1}{\beta}$ should be significantly smaller than $\EC{N}{\beta}$.  One potentially interesting ratio is
\be
\label{eq:ECsingleratio}
\rN{N}{\beta} \equiv \frac{\EC{N+1}{\beta}}{\EC{N}{\beta}},
\ee
which behaves much like $N$-subjettiness $\tau_N$ in that for a system
of $N$ partons plus soft radiation, the observable is linear in the
energy of the soft radiation.\footnote{Unlike $N$-subjettiness, this
  ratio scales like $\gamma^{1-N \beta}$ under transverse Lorentz
  boosts $\gamma$, which is somewhat undesirable when considering
  systems with several subjets. }  Of course, this is but one choice for an interesting combination of the energy correlation functions, and one can imagine using the whole set of energy correlation functions in a multivariate analysis.

In this paper, we will work exclusively with the energy correlation double ratio
\begin{equation}
\C{N}{\beta}  \equiv \frac{\rN{N}{\beta}}{\rN{N-1}{\beta}}= \frac{\EC{N+1}{\beta} \,  \EC{N-1}{\beta}}{ \EC{N}{\beta}^2} \ ,
\end{equation}
which is dimensionless.\footnote{This double ratio scales as $\gamma^{-\beta}$ under transverse Lorentz boosts. }  
One way to motivate this observable is that we already know that $N$-subjettiness ratios $\tau_N/\tau_{N-1}$ are good probes of $N$-prong substructure \cite{Thaler:2010tr,Thaler:2011gf}.  As we will see, the notation ``$C$'' is motivated by the fact that this variable generalizes the $C$-parameter \cite{Parisi:1978eg,Donoghue:1979vi}.  One should keep in mind that $\C{N}{\beta}$ involves $(N+1)$-point correlators, and when clear from context, we will drop the $^{(\beta)}$ superscript.  

The energy correlation double ratio $\Cnobeta{N}$ effectively measures
higher-order radiation from leading order (LO) substructure.  For a
system with $N$ subjets, the LO substructure consists of $N$ hard
prongs, so if $\Cnobeta{N}$ is small, then the higher-order radiation
must be soft or collinear with respect to the LO structure.  If
$\Cnobeta{N}$ is large, then the higher-order radiation is not
strongly-ordered with respect to the LO structure, so the system has
more than $N$ subjets.
Thus, if $\Cnobeta{N}$ is small and $\Cnobeta{N-1}$ is large, then we can say that a system has $N$ subjets.  In this way, the energy correlation double ratio $\Cnobeta{N}$ behaves like $N$-subjettiness ratios $\tau_N/\tau_{N-1}$, with key advantages to be discussed in \Sec{subsec:benefits}.

\subsection{Relationship to Previous Observables}
\label{subsec:previous}

While the definition of the energy correlation double ratio $\Cnobeta{N}$ is new, it is related to previous observables for $e^+e^-$ and hadron colliders that
have been studied in great detail.

An energy-energy correlation ($\EEC$) function for $e^+ e^-$ events was introduced in \Ref{Banfi:2004yd} for its particularly nice factorization and resummation properties.  It is defined as
\begin{equation}
\EEC_a = \frac{1}{E_{\text{tot}}^2} \sum_{i\neq j} E_i E_j|\sin \theta_{ij}|^a (1-|\cos\theta_{ij}|)^{1-a}  \ \Theta[(\vec{q_i}\cdot\vec{n}_T)(\vec{q_j}\cdot\vec{n}_T)] \ ,
\end{equation}
where the sum runs over all particles in the event and $\vec{n}_T$ is the direction of the thrust axis.  This variable is IRC safe for all $a<2$.  The $\Theta$-function is only non-zero if the pair of particles is in the same hemisphere.  This removes the large correlation of the two initial hard partons which would otherwise dominate the sum, and means that $\EEC_a$ behaves much like the jet angularities \cite{Berger:2003iw,Almeida:2008yp} with the same angular exponent $a$.  The $\EEC$ was introduced because it is insensitive to recoil effects and has smooth behavior for all allowed values of $a$.  In particular, $\EEC_a$ has a smooth transition through $a=1$, whereas angularities exhibit non-smooth behavior and also are increasingly sensitive to recoil effects as the angular power $a$ increases.  If one considers only one hemisphere of a dijet event, then $\EEC_a$ is approximately the same as $\C{1}{\beta}$ in our notation with $\beta = 2-a$.  Both observables are sensitive to 1-prong (sub)structure, and we will discuss the issue of recoil further in \Sec{subsec:benefits}.

A related two-particle angular correlation function was introduced in \Refs{Jankowiak:2011qa,Jankowiak:2012na,Larkoski:2012eh} for discrimination of jets initiated by QCD from jets from boosted heavy particle decays. The angular correlation function is defined as
\begin{equation}
{\cal G}_\beta(R) = \sum_{i,j} p_{Ti} \ p_{Tj} R_{ij}^\beta \Theta[R- R_{ij}] \ ,
\end{equation}
where the $\Theta$-function only allows pairs of particles separated by an angular scale of $R$ or less to contribute to the observable.  The behavior of the observable can be studied as a function of $R$, and jets that are approximately scale invariant should have an angular correlation function that scales as a power of $R$. For a fixed value of $R$, the properties of the angular correlation function are very similar to that of $\EEC_a$ and $\C{1}{\beta}$.

As mentioned above, the notation $\C{N}{\beta}$ was chosen because of its relation to the $C$-parameter from $e^+e^-$ collisions \cite{Parisi:1978eg,Donoghue:1979vi}.  The $C$-parameter is used to identify two-jet configurations without recourse to a
jet algorithm or explicit jet axes choice.  It is defined as
\begin{equation}
C = \frac{3}{2}\frac{\sum_{i,j}|{\bf p}_i||{\bf p}_j|\sin^2\theta_{ij}}{\left( \sum_i |{\bf p}_i| \right)^2} \ ,
\end{equation}
which can also be expressed in terms of the eigenvalues of the sphericity tensor.  At first glance, this looks very much like $C^{(2)}_1$ in the sense that the numerator looks like a 2-point correlation function with $\beta = 2$.  There is a crucial difference between the behavior of $\sin^2\theta_{ij}$ and $\theta_{ij}^2$, however, such that the $C$-parameter vanishes for dijet configurations when the jets are back-to-back (i.e.~$\theta_{ij} = \pi$).  If we expand around the dijet limit, then the $C$-parameter really behaves like a 3-point correlation function (i.e.~like $C_2$).  To see this, note that for $e^+e^- \to q\bar{q}g$, the $C$ parameter has the simple form
\begin{equation}
C = 6\frac{(1-x_1)(1-x_2)(1-x_3)}{x_1x_2x_3} \ ,
\end{equation}
where $x_i = 2 p_i \cdot Q / Q^2$ and $Q$ is the total four-vector of the system.\footnote{The $C$-parameter only properly makes sense if 
the total momentum of the system is zero, and so is not immediately applicable for hadron collisions.}  The angle between final-state particles $i$ and $j$ in the $e^+e^- \to q\bar{q}g$ system is
\begin{equation}
1-\cos\theta_{ij} = 2\frac{1-x_k}{x_i x_j}  \ .
\end{equation}
Thus, if we change $\theta\to 2\sin\frac{\theta}{2}$ in the definition of $\EC{N}{\beta}$, then $\C{2}{2}$ can be expressed as
\begin{equation}
\C{2}{2} \propto \frac{(1-x_1)(1-x_2)(1-x_3)}{x_1x_2x_3} \ ,
\end{equation}
which, up to normalization, is the traditional $C$-parameter.  Of course, at higher orders in perturbation theory the definitions of the $C$-parameter and $\C{2}{2}$ diverge.  Both observables are sensitive to 2-prong (sub)structure, though $\C{2}{\beta}$ gives sensible answers even for systems with non-zero total momentum and has an adjustable angular exponent $\beta$.

Higher-point energy correlation functions have been studied very little in the literature.  Two early studies for $e^+ e^-$ collisions are in \Refs{Donoghue:1979vi,Fox:1979id}.  However, both define observables that only make sense for systems with total momentum equal to zero and explicitly use operations only defined in three-dimensional space, such as cross-products and the properties of momentum tensors with rank greater than 2.  Thus, these observables cannot be easily generalized to determine if a (boosted) system has $N$ (sub)jets.  Historically, observables like the $D$-parameter \cite{Parisi:1978eg,Donoghue:1979vi,Ellis:1980wv} have been used to identify peculiar phase space configurations such as a planar configuration of particles.  However, this is not directly related to the number of jets in the event.  Recent substructure variables like planar flow \cite{Thaler:2008ju,Almeida:2008yp}, Zernike coefficients \cite{GurAri:2011vx}, and Fox-Wolfram moments \cite{Bernaciak:2012nh} are similarly sensitive to peculiar phase space configurations rather than prong-like substructure.  Planar flow, for example, vanishes if the constituents of the jet lie along a line, which is a good probe for some (but not all) 3-prong configurations.  The energy correlation double ratio $\Cnobeta{N}$ is designed to directly probe $N$-prong configurations, though the high computational cost of $\EC{N+1}{\beta}$ likely limits the practical range to $N \le 3$ (i.e.~up to three-prongs).

\subsection{Advantages Compared to $N$-subjettiness}
\label{subsec:benefits}

The variable $N$-subjettiness \cite{Thaler:2010tr,Thaler:2011gf} (based on $N$-jettiness \cite{Stewart:2010tn}) is a jet observable that can be used to test whether a jet has $N$ subjets, and it has been used in a number of theoretical \cite{Englert:2011iz,Bai:2011mr,Feige:2012vc,Curtin:2012rm,Cohen:2012yc,Soyez:2012hv,Ellis:2012zp} and experimental \cite{ATLAS:2012dp,ATLAS:2012am} substructure studies.  Since both $N$-subjettiness and the energy correlation double ratio $\Cnobeta{N}$ share the same motivation, it is worth highlighting some of the advantages of the energy correlation double ratio.

First, a quick review of $N$-subjettiness.  It is defined in terms of $N$ subjet axes $\hat{n}_A$ as\footnote{In \Refs{Thaler:2010tr,Thaler:2011gf}, $N$-subjettiness was defined with an overall normalization factor to make it dimensionless.  Here, we remove the normalization factor so it has the same dimensions as \Eq{eq:ECsingleratio}.}
\begin{equation}
\label{eq:Nsubdef}
\Nsub{N}{\beta} =\sum_{i} p_{T i}\min\left\{  R_{1,i}^\beta, R_{2,i}^\beta,\dotsc, R_{N,i}^\beta   \right\} \ ,
\end{equation}
where the sum runs over all particles in the jet and $R_{A,i}$ is the distance from axis $A$ to particle $i$.  There are a variety of methods to determine the subjet directions, with arguably the most elegant way being to minimize $\Nsubnobeta{N}$ over all possible subjet directions $\hat{n}_A$ \cite{Thaler:2011gf}. If a jet has $N$ subjets, then $\Nsubnobeta{N-1}$ should be much larger than $\Nsubnobeta{N}$, so the observable that is typically used for jet discrimination studies is the ratio
\begin{equation}
\label{eq:NsubRatio}
\Nsub{N,N-1}{\beta} \equiv \frac{\Nsub{N}{\beta}}{\Nsub{N-1}{\beta}} \ .
\end{equation}
As discussed above, this ratio is directly analogous to the energy correlation double ratio $\C{N}{\beta} \equiv \rN{N}{\beta}/\rN{N-1}{\beta}$.  

One immediate point of contrast between $N$-subjettiness and the energy correlation double ratio is that $\Cnobeta{N}$ does not require a separate procedure (such as minimization) to determine the subjet directions.  While novel, this by itself does not necessarily imply that $\Cnobeta{N}$ will have better discrimination power than $\Nsubnobeta{N,N-1}$, though it does mean that $\Cnobeta{N}$ is a simpler variable to study.\footnote{In particular, $\beta$ serves two different roles for $N$-subjettiness.  As in $\C{N}{\beta}$, $\beta$ controls the weight given to collinear or wide-angle emissions.  In addition, when the minimization procedure is used, $\beta$ controls the location of the axes which minimize $N$-subjettiness.  When trying to determine the optimal value for $\beta$ for subjet discrimination, it is difficult to disentangle these two effects.}  We now explain two test cases where $\Cnobeta{N}$ can perform better than $\Nsubnobeta{N,N-1}$:  insensitivity to recoil for $C_1$ and sensitivity to soft wide-angle emissions for $C_2$.

\begin{figure}
\begin{center}
\subfloat[]{\label{fig:recoil2} 
\includegraphics[width=7.0cm]{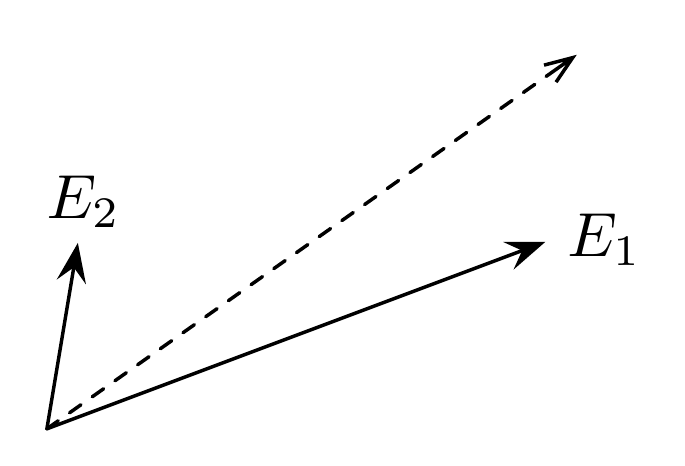}
}
$\qquad$
\subfloat[]{\label{fig:recoil1}
\includegraphics[width=7.0cm]{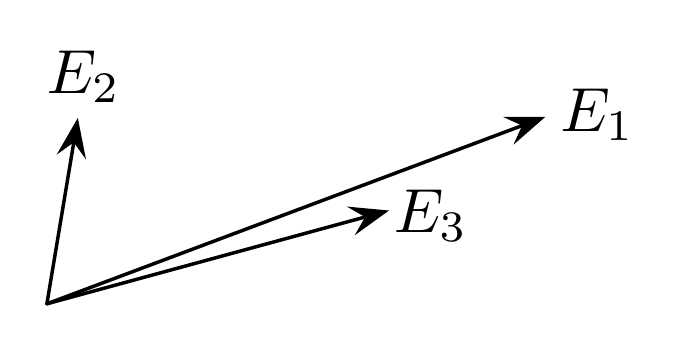}
}
\end{center}
\caption{Example kinematics with soft wide-angle radiation.  Left:  recoil of the jet axis (dashed)
  away from the hard jet core ($E_1$) due to soft wide-angle radiation
  ($E_2$), which is relevant for small values of $\beta$.  Right:  a
  three-particle configuration that highlights the difference between
 $\Cnobeta{2}$ and $\Nsubnobeta{2,1}$.  
}
\label{fig:examplekinematics}
\end{figure}

\subsubsection{Insensitivity to Recoil}
\label{subsubsec:recoil}

A recoil-sensitive observable is one for which soft emissions have an indirect effect on the observable.  In addition to the direct contribution to the observable, soft radiation in a recoil-sensitive observable changes the collinear contribution by an ${\cal O}(1)$ amount.  An example of a recoil-sensitive observable is angularities for the angular exponent $a \ge 1$ ($\beta \le 1$), which was studied in \Ref{Banfi:2004yd}.  Because $\Cnobeta{N}$ is insensitive to recoils, it is better able to resolve the collinear singularity of QCD.

For 1-prong jets, the effect of recoil on an observable is illustrated in \Fig{fig:recoil2}.  Because of conservation of momentum, soft wide-angle radiation displaces the hard jet core from the jet axis.  Angularities (i.e.~1-subjettiness) are sensitive to this displacement since they are measured with respect to the jet center.   For a jet with two constituents separated by an angle $\theta_{12}$ (using the notation in \Eq{eq:preECF} for simplicity),
\begin{equation}
\tau_1^{(\beta)} = \frac{E_2 E_1^\beta}{(E_1 + E_2)^\beta} (\theta_{12})^\beta + \frac{E_1 E_2^\beta}{(E_1 + E_2)^\beta} (\theta_{12})^\beta \ .
\end{equation}
Taking the $E_2 \ll E_1$ limit one can view the first term as the
contribution directly from the emission $E_2$, while the second term
comes about because particle $1$ recoils when it emits
particle $2$.
The dependence of $\tau_1^{(\beta)}$ on the energies and emission
angle is different according to the value of $\beta$.  For $\beta > 1$, the second term is negligible, and the angularities become
\begin{equation}
\tau_1^{(\beta > 1)} \simeq E_2 (\theta_{12})^\beta,
\end{equation}
such that $\tau_1$ is linear in the soft radiation $E_2$.  However, for smaller values of $\beta$, the expression for angularities changes because recoil effects become important.  For $\beta = 1$, both terms are identical  in the $E_2 \ll E_1$ limit and angularities become
\begin{equation}
\tau_1^{(1)} \simeq 2 E_2 (\theta_{12})^\beta .
\end{equation}
For $\beta < 1$, the first term is negligible in the $E_2 \ll E_1$ limit, and the angularities are dominated by the effect of recoil of the hard radiation
\begin{equation}
\tau_1^{(\beta < 1)} \simeq E_1^{1-\beta} E_2^\beta (\theta_{12})^\beta.
\end{equation}
By contrast, the observable $\C{1}{\beta}$ has the same behavior for all values of $\beta$:
\begin{equation}
\EC{1}{\beta} = E_1,
\qquad
\EC{2}{\beta} = E_1 E_2 \left(\theta_{12}\right)^\beta,
\qquad \Rightarrow
\qquad
\C{1}{\beta} = \frac{E_2 \left(\theta_{12}\right)^\beta}{E_1},
\end{equation}
which is dominated by the splitting angle and energy of the softer particle in the jet for all values of $\beta>0$.

Because $N$-subjettiness is essentially a sum over $N$ subjet angularities, $\Nsubnobeta{N}$ can suffer from the same recoil-sensitivity as angularities for $\beta\leq1$, depending on how the subjet axes are determined.  For example, if $N$-subjettiness is defined using $k_T$ subjet axes, then $\Nsubnobeta{N}$ is recoil sensitive.  $N$-subjettiness is also recoil sensitive if the subjet axes are aligned with the subjet momenta.  A related issue is that if the subjet axes are determined using the minimization procedure, then the minimization is only well-behaved for $\beta \ge 1$.\footnote{That said, the minimization procedure does eliminate the recoil effect.}  In all of these cases, the useful range of $\beta$ is limited to $\beta \ge 1$.  In contrast, the energy correlation double ratio is recoil-free and well-behaved for the whole IRC-safe range $\beta > 0$.  As we will see in \Sec{sec:oneprong} (and demonstrated recently in \Ref{Gallicchio:2012ez}), this is relevant for quark/gluon discrimination, where $\beta \simeq 0.2$ for $C_1$ is the optimal choice.

It should be noted that one can construct a recoil-free version of
$N$-subjettiness where the subjet axes are always chosen to minimize
the $\beta = 1$ measure (see forthcoming work in \Ref{broadening}),
regardless of which $\beta$ is used in \Eq{eq:Nsubdef}.  We refer to
axes defined in this way as ``broadening axes'', since $\beta = 1$ is
closely related to the jet broadening measure \cite{Catani:1992jc}.
We will make use of this fact later when comparing $C_1$ to 1-subjettiness in \Sec{subsec:quarkgluonMCstudy}.

\subsubsection{Sensitivity to Soft Wide-Angle Emissions}
\label{subsubsec:softwide}

Another point of contrast between $\Cnobeta{N}$ and
$\Nsubnobeta{N,N-1}$ is in how the two variables behave in the
presence of emissions at multiple angular scales.
The way $N$-subjettiness is defined, every jet is partitioned into $N$ subjets, even if there are fewer than $N$ ``real'' subjets.  For example, when a jet has a soft subjet separated at large angle (as one might expect from the radiation off a quark or gluon), $N$-subjettiness will still identify that soft subjet region, yielding a relatively low value of $\Nsubnobeta{N,N-1}$ (and therefore making the jet look more $N$-prong-like than it really is).  In contrast, because the energy correlation function is sensitive to all possible soft and collinear singularities, $\Cnobeta{N}$ takes on a relatively high value in the presence of a soft wide-angle subjet, making the jet look less $N$-prong like (as desired).  

We can show this concretely for $C_2$ using the configuration in \Fig{fig:recoil1} where there is the following hierarchy of the energies and angles:\footnote{ Roughly the same conclusions about $C_2$ versus $\Nsubnobeta{2,1}$ hold for the limit $E_1 \simeq E_2 \gg E_3$ as well, which is relevant for the $Z$ boson discussion below.}
\be
E_1 \gg E_2, E_3, \qquad \theta_{13} \ll \theta_{12} \simeq \theta_{23}.
\ee
Again using the notation in \Eq{eq:preECF}, the energy correlation functions are
\begin{align}
&\EC{1}{\beta} \simeq E_1,
\qquad
\EC{2}{\beta} \simeq E_1 \max\left[E_2 \left(\theta_{12}\right)^\beta,E_3 \left(\theta_{13}\right)^\beta  \right],
\nonumber \\
&\EC{3}{\beta} = E_1 E_2 E_3 \left(\theta_{12} \theta_{23} \theta_{13} \right)^\beta, 
\end{align}
yielding
\be\label{eq:C2_config}
\C{2}{\beta} = \frac{\EC{3}{\beta}\EC{1}{\beta}}{\EC{2}{\beta}^2}\simeq \frac{E_2 E_3 \left(\theta_{12}\right)^{2\beta} \left(\theta_{13}\right)^\beta}{\max\left[E_2 \left(\theta_{12}\right)^\beta,E_3 \left(\theta_{13}\right)^\beta  \right]^2} \ .
\ee
For $N$-subjettiness with three jet constituents, it is consistent to choose axes that lie along the hardest particle in a subjet.  For 1-subjettiness, the axis lies along particle 1.  For $2$-subjettiness, one axis lies along particle 1 and the other axis lies along particle 2 or particle 3, depending on the relationship between $E_3 \theta_{13}$ and $E_2 \theta_{12}$.  This gives
\begin{align}
&\Nsub{1}{\beta} \simeq  \max\left[E_2 (\theta_{12})^\beta,E_3 (\theta_{13})^\beta\right], \qquad \Nsub{2}{\beta} \simeq \min\left[E_2 (\theta_{12})^\beta,E_3 (\theta_{13})^\beta\right] 
\nonumber \\
& \Rightarrow \qquad \Nsub{2,1}{\beta}= \frac{\min\left[E_2 (\theta_{12})^\beta,E_3 (\theta_{13})^\beta\right]}{\max\left[E_2 (\theta_{12})^\beta,E_3 (\theta_{13})^\beta\right]}\ .  \label{eq:tau21limit}
\end{align}
Regardless of the ordering of $E_3 \theta_{13}$ and $E_2 \theta_{12}$ we see that:
\be
\label{eq:c2tau2relation}
\C{2}{\beta} \simeq \Nsub{2,1}{\beta} \times
(\theta_{12})^\beta, 
\ee
so in the presence of a soft subjet at large
angle $\theta_{12}$, $\Cnobeta{2}$ yields a larger value than
$\Nsubnobeta{2,1}$ (i.e.~more background-like as desired).  As we will
see in \Sec{sec:twoprong}, this allows $\Cnobeta{2}$ to perform
better than $\Nsubnobeta{2,1}$ for background rejection in
regions of phase space where soft wide-angle radiation plays an
important role.

One way to understand the improved performance of $\Cnobeta{2}$ with respect to $\tau_{2,1}$ is to consider the concrete example of $\beta=2$ at fixed jet mass $m$.\footnote{We thank Gregory Soyez for helpful discussions on these points.}  Using the kinematic limit above, the jet mass-squared is given approximately by
\be
m^2 \simeq  E_1 \max\left[E_2 \left(\theta_{12}\right)^2,E_3 \left(\theta_{13}\right)^2  \right],
\ee
and it is convenient to define $z$ as the energy fraction of the emission that dominates the mass (e.g. $z = E_2 / E_1$ if $E_2 \left(\theta_{12}\right)^2 > E_3 \left(\theta_{13}\right)^2$).  For fixed jet mass, QCD backgrounds tend to peak at small values of $z$, but we see from \Eq{eq:tau21limit} that $\tau_{2,1}$ does not have any $z$-dependence for fixed jet mass.  For $\Cnobeta{2}$, if particle 2 dominates the mass (i.e.~if a soft wide-angle emission dominates the mass), then
\be
\C{2}{2} \simeq \Nsub{2,1}{2} \times \frac{m^2}{ (E_1)^2} \frac{1}{z},
\ee
so $\Cnobeta{2}$ penalizes small values of $z$.   In this way,
$\Cnobeta{2}$ acts similarly to taggers that reject jets if the
kinematics of the dominant splitting of the jet is consistent with
background
\cite{Seymour:1993mx,Butterworth:2002tt,Brooijmans:1077731,Butterworth:2008iy,Ellis:2009me,Krohn:2009th}.   
In contrast, $\tau_{2,1}$ only exploits the degree to which radiation is collimated with respect to the two subjet directions, and does not take into account the $z$-dependence at fixed jet mass.

If particle 3 dominates the mass (i.e.~if the mass is dominated by a
hard core of energy), then $\Cnobeta{2}$ is constant in the energy
fraction $z$, and so is no longer affected by the kinematics of the
emission that generated the mass.
However, there is still the potential for improved performance in
identifying boosted color singlet resonances like $Z$ bosons.  For a
boosted $Z$ boson, emissions at wide angle with respect to the angle
between decay products are suppressed by color coherence.  As one goes
to higher boosts where the ratio of jet mass to jet $p_T$ decreases
for fixed jet radius, the volume of phase space for allowed emissions
decreases, which can also be seen as a consequence of angular ordering.
It is therefore less likely for a $Z$ boson signal to generate final
state radiation at large $\theta_{12}$, while background QCD jets will
emit at large angle independently of the $p_T$.
Because radiation at large angles has an enhanced effect on
$\Cnobeta{2}$ as compared to $\tau_{2,1}$, cf.\
\Eq{eq:c2tau2relation}, we expect $\Cnobeta{2}$ to be more effective
at discriminating color-singlet signals from background QCD jets.

\section{Quark vs.~Gluon Discrimination with ${\Cnobeta{1}}$}
\label{sec:oneprong}

Our first case study is to use the energy correlation functions to
discriminate between quark jets and gluon jets. 
The observable $\Cnobeta{1}$ contains the 2-point energy correlation
function $\EC{2}{\beta}$ and so is sensitive to radiation in a jet
about a single hard core.\footnote{The CMS experiment uses an
  observable they call $p_T D = \sum_{i} p_{ti}^2 / (\sum_{i}
  p_{ti})^2$ for quark versus gluon
  discrimination~\cite{Chatrchyan:2012sn,Pandolfi:1480598}. It is  
  related to the $\beta=0$ limit of $\C{1}{\beta}$ as $p_T D = 1 -
  2\C{1}{0}$.}  This case study is 
simple enough that we can predict the quark/gluon discrimination power
through an analytic calculation, which we will subsequently validate
with Monte Carlo simulations.  In our later case studies involving
higher-point correlators, we will rely on Monte Carlo alone.

In any discussion of quark--gluon discrimination, one should
start with a reminder that defining what is meant by a quark or a
gluon jet is a subtle task, since the one existing infrared-safe way
of defining quark and gluon jets~\cite{Banfi:2006hf} works only at
parton level.
Existing work on practical aspects of quark--gluon discrimination
in \Refs{Gallicchio:2011xq,Gallicchio:2012ez,Krohn:2012fg,Chatrchyan:2012sn,Pandolfi:1480598} 
  has not entered into these issues.
Instead the discussion has relied on Monte Carlo simulations, defining a
quark (gluon) jet to be whatever results from the showering of a quark
(gluon) parton.
We will adopt a variant of this methodology in our Monte Carlo
studies.
Our analytic approach will instead define a quark or gluon jet in
terms of the sum of the flavors of the partons contained inside it. 
It is based on resummation and therefore contains similar physics to
the Monte Carlo parton shower.

\subsection{Leading Logarithmic Analysis}

We begin our analysis by considering the leading logarithmic (LL) structure of the cross section for the observable $\Cnobeta{1}$.  With $L$ equal to the logarithm of $\Cnobeta{1}$, we define LL order as including all terms in the cross section that scale like $\as^n L^{2n}$, for $n\geq1$.
At LL order,
quark versus gluon jet discrimination can be understood as a
consequence of quarks and gluons having different color charges.  To
LL order, the strong coupling constant $\alpha_s$ can be taken fixed
and only the most singular term in the splitting function need be
retained.  With only one soft-collinear gluon emission, the
normalized differential cross section for any infrared and collinear safe
observable $e$ has the same form for both quark and gluon jets: 
\begin{equation}
\label{eq:masterLL1prong}
\frac1{\sigma}\frac{d\sigma}{ d e}  = 2 \frac{\alpha_s}{\pi} C \int_0^{R_0} \frac{d\theta}{\theta}\int_0^1 \frac{dz}{z} \delta(e-\hat{e}) \ ,
\end{equation}
where $C$ is the color factor, $R_0$ is the jet radius,\footnote{We
  use this somewhat non-standard notation because $R$ will later be
  used with a different meaning.} $z$ is the energy fraction of the emitted gluon, $\theta$ is its splitting angle, and $\hat{e}$ is a function of $z$ and $\theta$.  Recall that $C_F = 4/3$ for quarks and $C_A = 3$ for gluons.

At this order, the observable $\C{1}{\beta}$ is
\begin{equation}
\hat{C}_{1}^{(\beta)} = z(1-z)\theta^\beta ,
\end{equation}
which takes a maximum value of $\frac{1}{4}R_0^\beta$. So integrating
\Eq{eq:masterLL1prong} yields, for small $\C{1}{\beta}$, 
the cross section 
\begin{equation}
\frac1{\sigma}\frac{d\sigma}{d \C{1}{\beta}} = \frac{2\alpha_s}{\pi}\frac{C}{\beta}\frac{1}{\C{1}{\beta}}\ln\frac{R_0^\beta}{\C{1}{\beta}} \ .
\end{equation} 
We identify the logarithm $L$ as
\begin{equation}
L \equiv \ln\frac{R_0^\beta}{\C{1}{\beta}} \ ,
\end{equation}
which we use in the following expressions for compactness.
This distribution can be resummed to LL order by exponentiating the cumulative $\C{1}{\beta}$ distribution.
The resummed distribution that follows is then
\begin{equation}
\frac1{\sigma}\frac{d\sigma^{\rm LL}}{d\C{1}{\beta}} = \frac{2\alpha_s}{\pi}\frac{C}{\beta}\frac{L}{\C{1}{\beta}} e^{-\frac{\alpha_s}{\pi}\frac{C}{\beta}L^2} .
\end{equation}
Because the quark color factor is smaller than the gluon color factor, the Sudakov suppression is less for quarks.  Thus, the $\C{1}{\beta}$ distribution for quark jets is peaked at smaller values than for gluon jets.

To figure out the quark/gluon discrimination power from this
$\C{1}{\beta}$ resummed distribution, we will make a sliding cut on
$\C{1}{\beta}$ and count the number of events that lie to the left of
the cut.  Adjusting this cut then defines a ROC curve relating the
signal (quark) jet efficiency to the background (gluon) jet rejection.
To LL accuracy, the (normalized) cumulative distributions for quarks and gluons are:
\begin{equation}\label{eq:LLcumulant}
\Sigma_q(\C{1}{\beta})  =  e^{-\frac{\alpha_s}{\pi}\frac{C_F}{\beta}L^2}, \quad \Sigma_g(\C{1}{\beta})  =  e^{-\frac{\alpha_s}{\pi}\frac{C_A}{\beta}L^2}.
\end{equation}  
Note
that at LL order, there is a simple relationship between these
cumulative distributions:
\be
\label{eq:LLcumulantrelation}
\Sigma_g(\C{1}{\beta}) = \left( \Sigma_q(\C{1}{\beta}) \right)^{C_A/C_F}.
\ee
Thus, if a sliding cut on $\C{1}{\beta}$ retains a fraction $x$ of the quarks, it will retain a fraction $x^{C_A/C_F}$ of the gluons. The quark/gluon discrimination curve is then
\begin{equation}\label{eq:rocqvg}
\text{disc}(x) = x^{C_A/C_F}=x^{9/4} ,
\end{equation}
which (perhaps surprisingly) is independent of $\beta$.  This LL discrimination result holds for a wide class of IRC safe observables sensitive to the overall jet color factor, including the jet mass.  Only beyond LL order does the discrimination curve depend on $\beta$.

\subsection{Next-to-Leading Logarithmic Analysis}
\label{subsec:NLLanalysis}

We continue our analysis to next-to-leading logarithmic (NLL) order, which we define as including all terms that scale as $\as^n L^{n+1}$ and $\as^n L^n$
in $\ln \Sigma$.  In addition, we will also include the non-logarithmically enhanced term arising at ${\cal O}(\as)$.
At NLL order,
there are several new effects that must be included,
which together turn out to improve the quark/gluon discrimination
power of $\C{1}{\beta}$ compared to the LL estimate.  The dominant
effects are subleading terms in the splitting functions and phase
space restrictions due to multiple emissions.  In addition, one must
account for the running of $\alpha_s$, fixed-order corrections, and
non-global logarithms \cite{Dasgupta:2001sh} arising from the phase
space cut of the jet algorithm.  We will consider how these
affect the discrimination power of $\C{1}{\beta}$, ultimately showing
that small values of $\beta$ improve quark/gluon discrimination. 
We will work in an approximation of small jet radius, $R_0 \ll 1$,
which will allow us to consider only the effects of radiation from the
jet, while neglecting modifications associated with the full antenna
structure of initial and final-state partons. 

The resummation to NLL for generic (global) observables was carried
out in \Ref{Banfi:2004yd}.  The central result of that analysis was an
expression for the NLL cumulative distribution for an arbitrary
observable (satisfying certain basic conditions, e.g.\ recursive infrared
safety).  From \Ref{Banfi:2004yd}, the probability that the value 
of an observable is less than $e^{-L}$ takes a form such as
\begin{equation}
  \label{eq:master}
  \Sigma(e^{-L}) = N \frac{e^{-\gamma_E R'}}{\Gamma(1+R')} e^{-R} \,,  \qquad R' \equiv \frac{dR}{dL} \ ,
\end{equation}
where $N$ is a matching factor to fixed order, $N = 1 +
\order{\alpha_s}$, and $\gamma_E \simeq
0.5772$ is the Euler-Mascheroni constant.  In a fixed-coupling
approximation, the ``radiator'' function $R$ for the observable
$\C{1}{\beta}$ is 
\begin{equation}\label{eq:Rdef}
  R = \frac{\alpha_s}{\pi}\, \frac{C}{\beta} (L + B)^2\,,
\end{equation}
where $C$ is
the color factor of the jet and $B$ encodes subleading terms in the
splitting functions.\footnote{To obtain \Eq{eq:Rdef}, we used the fact that, for a general jet observable that takes the form
$$
{\cal O} = \sum_{i\in J}\left( \frac{p_{Ti}}{p_{T J}} \right)^A \left( \frac{R_i}{R_0} \right)^B \ ,
$$
where $R_i$ is the angle of the emission, Eq.~(2.19) in \Ref{Banfi:2004yd} applies for $a = A$, $b= B-A$, and $d= 1$,
where we identify the scales $Q = Q_{12}=2E_\ell=p_{TJ} R_0$.  The sum over $\ell=1,2$ in Eq.~(2.19) is replaced by the individual contribution $\ell=1$.
}
 For quark jets $B_q = -\frac{3}{4}$ and for
gluon jets $B_g= -\frac{11}{12} + \frac{n_f}{6 C_A}$, where $n_f$ is
the number of light quark flavors.  The specific NLL resummed formula in
\Eq{eq:master} holds for observables that are global, recursively
infrared and collinear safe (rIRC), and additive.
The last two conditions are satisfied by
$\C{1}{\beta}$.
The general expression for $R$ with running $\alpha_s$ appears in
\Ref{Banfi:2004yd}.  The scale at which $\alpha_s$ is evaluated is
$p_T R_0$, and we will use the shorthand 
\begin{equation}
\alpha_s \equiv \alpha_s (p_T R_0) \ ,
\end{equation}
unless an explicit scale is used as the argument of $\alpha_s$.
Because $\C{1}{\beta}$ for a jet is non-global, it is necessary to
include an extra factor in the resummation, discussed in
detail in \Sec{sec:non-global}.
We will also include information obtained by matching to the ${\cal
  O}(\alpha_s)$ fixed-order cross section, where the
matching procedure is described in \App{app:match}.

\begin{figure}
\begin{center}
\subfloat[]{\label{fig:NLLdiscriminationROC}
\includegraphics[width=7.0cm]{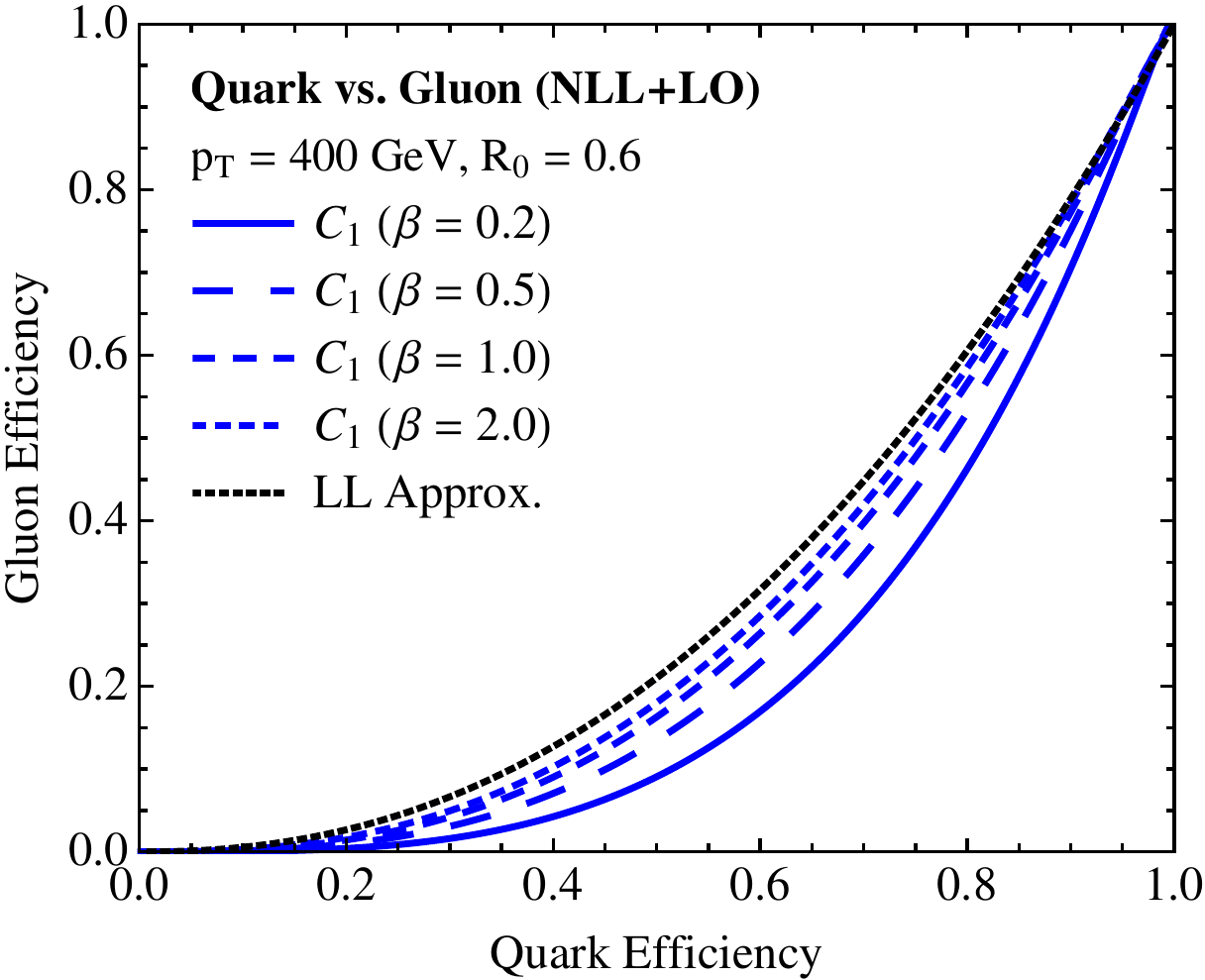}
}
$\qquad$
\subfloat[]{\label{fig:NLLdiscriminationSB} 
\includegraphics[width=7.0cm]{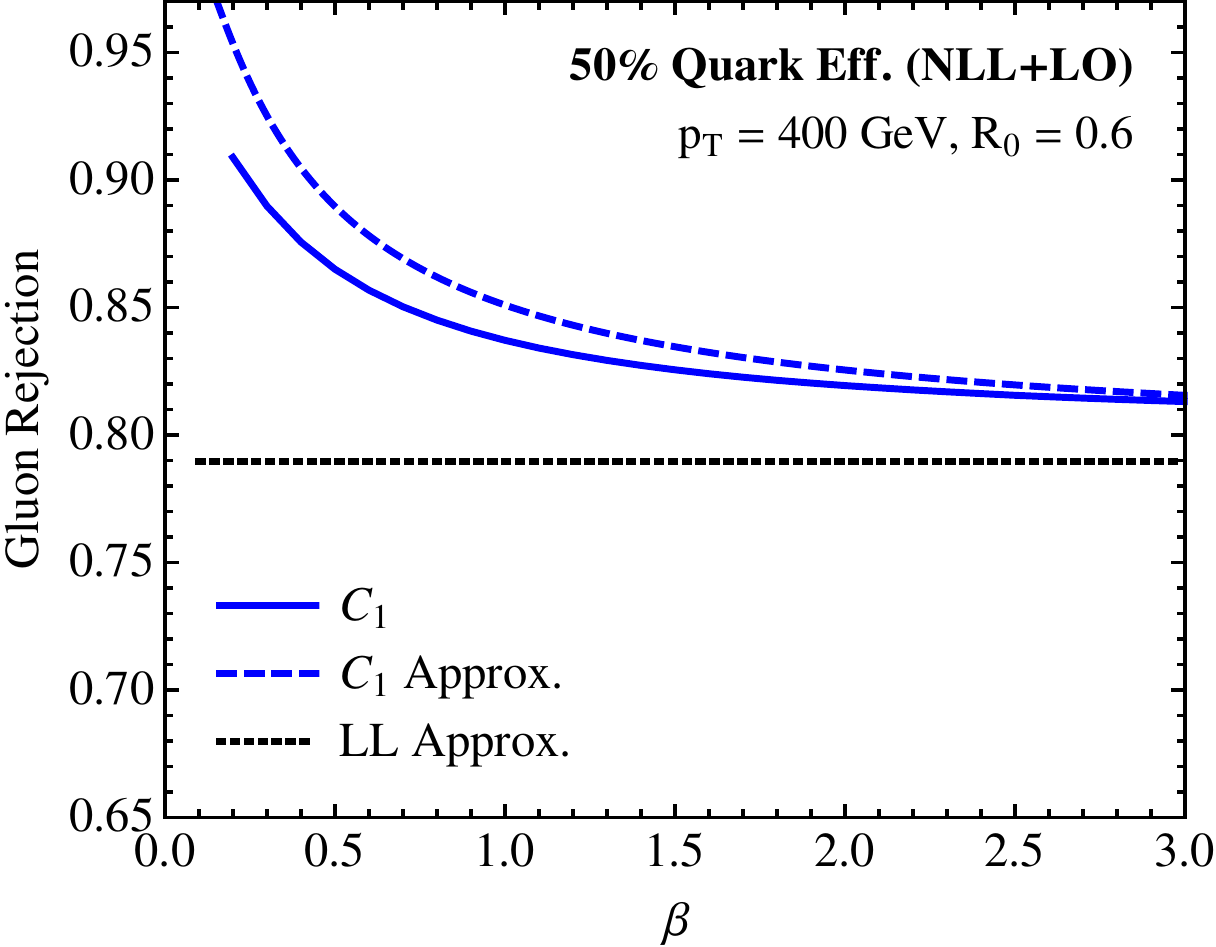}
}
\end{center}
\caption{ Left:  Quark/gluon discrimination curves using $\C{1}{\beta}$, calculated at NLL order matched to fixed order for various values of $\beta$.  The $\beta$-independent LL prediction is shown for comparison.  Right:  Gluon rejection rates at 50\% quark efficiency, as a function of $\beta$, demonstrating that $\beta \simeq 0.2$ is optimal at NLL order (for smaller values of $\beta$, non-perturbative effects become important).  Also shown is an analytic approximation from \Eq{eq:summaryequation} ($\Cnobeta{1}$ Approx.)\ that includes the most important physics that enters at NLL.
}
\label{fig:NLLdiscrimination}
\end{figure}

Armed with the matched NLL cumulative distribution, including the
non-global and $\order{\as}$ corrections, we can now determine the quark versus gluon discrimination curve by numerically inverting $\Sigma_q$ and plugging it into the expression for $\Sigma_g$.  This is shown for various values of $\beta$ in \Fig{fig:NLLdiscriminationROC}.  
In \Fig{fig:NLLdiscriminationSB}, we fix 50\% quark efficiency and
show the gluon rejection rate (i.e.~one minus the gluon efficiency) as
a function of $\beta$ for $R_0 = 0.6$.   Also on
this plot is an approximate analytic expression for the rejection rate
as a function of $\beta$ that we derive below in
\Eq{eq:summaryequation}. 
We see that the discrimination power improves as $\beta$ decreases. 
It is, however, not sensible to take $\beta$ too small: for $\beta=0$
our observable is collinear unsafe, and large non-perturbative effects
can be expected as $\beta$ approaches zero.
Furthermore for $\beta \lesssim \alpha_s$ the convergence of our
calculation breaks down (cf.~\App{app:nonpert}).

To understand the behavior of \Fig{fig:NLLdiscriminationSB}
semi-analytically, we will study the impact of different physical effects on
the discrimination.  To do so, we will again express $\Sigma_g$ in terms of
$\Sigma_q$ so as to determine the discrimination power of a cut on
$\Cnobeta{1}$.  
In fact, we are most interested in the exponent
relating $\Sigma_g$ to $\Sigma_q$ (as in \Eq{eq:LLcumulantrelation}),
so we will actually relate the logarithms of the two cumulative distributions to one another.  We are interested in the regime where $\ln1/ \Sigma \sim 1$, which, from \Eq{eq:LLcumulant}, implies that $\alpha_s L^2\sim 1$.  The logarithm of the cumulative distribution has the schematic expansion
\begin{equation}
  \label{eq:lnSigma-expansion}
\ln \Sigma \sim \alpha_s L^2 + \alpha_s L +\alpha_s + \alpha_s^2 L^3 + \alpha_s^2 L^2 + \alpha_s^2 L + \alpha_s^2 + {\cal O}(\alpha_s^3)  \ .
\end{equation}
With the power counting of $\alpha_s L^2\sim 1$, we will consider all
terms from \Eq{eq:lnSigma-expansion} that scale as $\alpha_s^0$,
$\alpha_s^{1/2}$, or $\alpha_s^1$.  This corresponds to all terms at
order $\alpha_s$ from \Eq{eq:lnSigma-expansion}, as well as the terms
at $\alpha_s^2 L^3$, $\alpha_s^2 L^2$, and $\alpha_s^3 L^4$.  To
illustrate this power counting, consider, for example, the term
$\alpha_s L$, which scales as $\alpha_s^{1/2}$ as one varies
$\alpha_s$ while keeping $\alpha_s L^2$ fixed and of order
$1$.

In what follows we will pay special attention to the terms at order $\alpha_s
L$ and $\alpha_s^2 L^2$, which turn out to be the most relevant ones
when establishing deviations from our LL analysis and whose dominant
contributions have clearly identifiable physical origins.
The terms at order $\alpha_s^2 L^3$ and $\alpha_s^3 L^4$ are
simply proportional to the LL color factor, multiplied by powers of
the $\beta$-function, and so do not significantly modify the LL
analysis. 

\subsubsection{Subleading Terms in Splitting Functions}
\label{sec:subleading}

We first consider the effect on the discrimination from the subleading terms in the splitting functions.
In the observable $\C{1}{\beta}$, $\beta$ controls the weight given to
collinear and wide-angle emissions in the jet.  At large values of
$\beta$, wide-angle emissions are given greater weight, and at small
values of $\beta$, collinear emissions are given greater
weight. Wide-angle soft radiation is controlled by the term in the
splitting function that diverges as the energy fraction goes to zero;
i.e., the term $dz / z$.  Both quarks and gluons have the same
functional form for the soft limit of the splitting function, with the
only difference being the overall color factor.  By contrast,
collinear emissions are controlled by the subleading terms in the
splitting function, which differ for quarks and gluons (i.e.~different
values of the $B$ coefficient).  Therefore, as $\beta$ goes to zero
and the collinear emissions become more important in $\C{1}{\beta}$,
the differences between the quark and gluon splitting functions are accentuated.

To see this behavior directly from \Eq{eq:master}, we can ignore the
$R'$-dependent prefactor and focus on the $e^{-R}$ factor.  We can
write $B_g = B_q + \delta B$, where 
\begin{equation}
\delta B=\frac{n_f-C_A}{6 C_A} \ , 
\end{equation}
which is $\frac{1}{9}$ for $n_f=5$.  We then have
\begin{eqnarray}
  \label{eq:3}
  R_g &\simeq& \frac{C_A}{C_F} R_q \left(1 + \frac{2\delta B}{L+B_q} \right) \nonumber \\
      &=&  \frac{C_A}{C_F} R_q \left(1 + 2\delta B\sqrt{\frac{\alpha_s
            C_F}{\pi \beta R_q}} \right) \ .
\end{eqnarray}
This last form allows us to relate the cumulative distribution for gluons to that of quarks, in the same spirit as \Eq{eq:LLcumulantrelation}:  
\begin{equation}
  \label{eq:4}
  \ln \Sigma_g \simeq {\frac{C_A}{C_F}
    \left(1 + 2\delta B\sqrt{\frac{\alpha_s
            C_F}{\pi \beta \ln 1/\Sigma_q}} \right)} \ln \Sigma_q,
\end{equation}
This implies that the separation between the quark and gluon
distributions increases as $\beta$ decreases and so smaller values of
$\beta$ result in better discrimination.  Because this effect first
arises at $\order{\sqrt{\as}}$, there will be corrections at
$\order{\as}$ due to the running coupling.  Note also that the
coefficient $\delta B$ is quite small in QCD, and so the total effect from the subleading terms in the splitting functions on the discrimination power is minimal.

\subsubsection{Multiple Emissions}

Next, consider the effect of multiple emissions.  The Sudakov
logarithm corresponds to the integral of the area (in $\ln k_t$, $\ln
\theta$ space) over which emissions are forbidden.  At LL, any number
of emissions can lie arbitrarily close to the lower boundary of the
phase space region without changing the value of the observable.  At
NLL, one must consider the cumulative effect of the emissions that lie
near the phase space boundary.  Multiple emissions tend to increase
the value of the observable $C_1$, and so, for a fixed value of $C_1$,
they must be suppressed.  This introduces an extra degree of
discrimination between quarks and gluons; there are likely to be more
such emissions for gluons than quarks and so it costs more to
``accept'' a gluon jet.  For a given LL Sudakov factor, the extent of
the boundary region is effectively increased as $\beta$ is decreased,
leading to better quark versus gluon discrimination at small
$\beta$.

In \Eq{eq:master}, the effect of muliple emissions is seen in the $R'$-dependent prefactor.  For small values of $R'$, the prefactor has the expansion
\begin{eqnarray}
  \label{eq:expansion}
   \frac{e^{-\gamma_E R'}}{\Gamma(1+R')} 
   &=& 1 - \frac{\pi^2}{12} R'^2 + {\cal O}\left(R'^3\right) \nonumber \\
   &=& 1 - \frac{\pi^2}{12} \frac{4 \alpha_s}{\pi} \frac{C}{\beta} R +
   {\cal O}\left(\frac{\alpha_s^2 L R}{\beta^2}\right) \ .
\end{eqnarray}
We will drop terms at ${\cal O}\left(\alpha_s^2 L R/\beta^2\right)$ and higher, which constrains us to consider $\beta  \gtrsim \alpha_s L$.  The cumulative distribution can then be written approximately as
\begin{equation}
  \label{eq:prefactor-rewritten}
  \ln \Sigma \simeq - R \left(1 + \frac{4\pi }{12}\frac{C}{\beta} \alpha_s\right) ,
\end{equation}
which allows us to relate $\Sigma_g$ in terms of $\Sigma_q$ as
\begin{equation}
  \label{eq:prefactor-impact}
   \ln \Sigma_g = {\frac{C_A}{C_F}
    \frac{1 + \frac{4\pi }{12}\frac{C_A}{\beta} \alpha_s}{1 + \frac{4\pi
      }{12}\frac{C_F}{\beta} \alpha_s}} \ln \Sigma_q
  \simeq
  \frac{C_A}{C_F}
    \left(1 + \frac{4\pi }{12}\frac{C_A-C_F}{\beta} \alpha_s\right) \ln \Sigma_q.
\end{equation}
This again suggests an increase in discrimination power for relatively small $\beta$.
While this effect appears at order $\alpha_s$ rather than
$\sqrt{\alpha_s}$, it has a substantially larger coefficient.

\subsubsection{Non-Global Logarithms}
\label{sec:non-global}

Because jets are defined in a restricted phase space, non-global logarithms may contribute to the quark versus gluon discrimination power.  The effect of non-global logarithms
on the cumulative distribution can, for our purposes, be approximated in the large-$N_C$
limit as~\cite{Dasgupta:2001sh,Dasgupta:2002bw,Banfi:2002hw}
\begin{equation}
  \label{eq:NG}
  \Sigma_\text{with NG} 
  = e^{ - C C_A \frac{\pi^2}{12} \frac{\alpha_s^2}{\pi^2} L^2} \Sigma 
  = e^{ - C_A \frac{\pi^2}{12} \frac{\alpha_s}{\pi} \beta R} \Sigma \,.
\end{equation}
This neglects some contributions starting at order $\alpha_s^3 L^3$
in the exponent, but these would not affect the quark-gluon
discrimination at our accuracy.
Recently a first numerical calculation has been performed including
the full-$N_C$ structure~\cite{Hatta:2013iba} and it suggests that
finite-$N_C$ corrections are small.

If we temporarily ignore the $R'$-dependent prefactor in
\Eq{eq:master}, the inclusion of non-global logarithms leads to
\begin{equation}
  \label{eq:NF-again}
  \ln \Sigma_\text{with NG} \simeq -R \left( 1 - C_A \frac{\pi^2}{12} \frac{\alpha_s}{\pi} \beta\right) \ .
\end{equation}
All quark/gluon dependence resides in the color factor inside $R$, so we still have the property from the LL calculation (again, ignoring the prefactor and setting $\delta B = 0$)
\begin{equation}
  \label{eq:ratios-with-NG}
  \Sigma_{g,\text{with NG}}(L) = \left[\Sigma_{q,\text{with NG}}(L) \right]^{C_A/C_F}\ .
\end{equation}
Hence non-global logarithms do not modify the above arguments in any significant way.  

This analysis holds for the anti-$k_T$ jet algorithm, whose boundary
is unaffected by soft radiation at angles $\sim R_0$.
For other algorithms of the generalized-$k_T$ family, which have
irregular, soft-emission-dependent boundaries, there are additional
terms, clustering logarithms~\cite{Appleby:2002ke,Delenda:2006nf}, which
also appear starting from order $\as^2 L^2$.
Some of the $\order{\as^2}$ clustering logarithms involve color
factor combinations such as $C_F^2$ and $C_A^2$ for quarks and gluons
respectively, and so presumably would have an impact on quark-gluon
discrimination at our accuracy.
We leave the study of these terms for future work.

\subsubsection{Summary of NLL Result}\label{sec:NLLsumm}

Using the results of Ref.~\cite{Banfi:2004yd} and \App{app:match} to
include all effects up through ${\cal O}(\alpha_s)$ in the logarithm
of the cumulative distribution we find
\begin{align}
\label{eq:summaryequation}
\ln \Sigma_g \simeq& \ \frac{C_A}{C_F}\left( 1 + \frac{n_F-C_A}{3C_A}\sqrt{\frac{\alpha_s
            C_F}{\pi \beta \ln 1/\Sigma_q}}  + \frac{n_F-C_A}{C_A} \frac{\alpha_s}{36\pi} \frac{b_0}{\beta}(2-\beta) \right.\nonumber \\
            & \qquad\quad \left.+ \  \frac{\alpha_s\pi }{3}\frac{C_A-C_F}{\beta} - \frac{17}{36}\frac{\alpha_s}{\pi}\frac{C_F}{C_A}\frac{n_f - C_A}{\beta\ln 1/\Sigma_q}  + \ldots \right)\ln \Sigma_q \ .
\end{align}
This expression includes two terms beyond those discussed in the subsections above.
The one proportional to $b_0$, where $b_0 = \frac{11}{3}C_A -
\frac{2}{3}n_f$ is the one-loop $\beta$-function coefficient, 
has two origins: it comes from the running coupling corrections to the
contribution from the subleading terms in the splitting functions and
from the running-coupling corrections to the relation between the
logarithm $L$ and $\ln \Sigma_q$.
The last term in parentheses comes from $\order{\as}$ matrix element
corrections, discussed in detail in \App{app:match}.
It depends on the choice of the jet definition, including the
procedure by which one defines quark versus gluon jets at parton level. 
Specifically, we assume any algorithm is equivalent to the
generalized $k_T$ family of jet algorithms at order $\alpha_s$, and at
this order define a quark jet to be one that contains a quark and a
gluon, while jets containing $gg$ or $q\bar q$ are considered to be
gluon jets.\footnote{In contrast to the situation with LO studies, at
  $\order{\alpha_s}$ it does not makes sense to discuss jet flavor
  based on the flavor of the parton that ``initiates'' the jet, since
  interference effects between diagrams mean that the initiating
  parton cannot be uniquely identified. The question of quark-gluon
  jet definition at fixed order is discussed further in
  \App{app:match}. }
Beyond $\order{\alpha_s}$, the calculation assumes that the algorithm
maintains a rigid circular boundary in the presence of multiple soft
particles at angles of order $R_0$, i.e.\ that it behaves like the
anti-$k_T$ algorithm. 

Note that every subleading term in \Eq{eq:summaryequation} is
proportional to a difference of color or quark number factors and so
the discrimination power depends sensitively on these differences. 
The overall quark versus gluon discrimination power increases as $\beta$ is decreased (even though the last term favors larger values of $\beta$ for $n_f > C_A$).   Numerically, this behavior is dominated by the subleading terms in the splitting functions and the multiple emissions effect.  The effect of the subleading terms in the splitting functions goes like $\sqrt{\alpha_s}$ and so is formally more important than the multiple emissions effect which is ${\cal O}(\alpha_s)$.  However, the effect of the subleading terms is multiplied by the small number $\delta B$ and so is numerically smaller than the contribution from multiple emissions.  Running-coupling and fixed-order effects are significantly smaller.

Robustly, then, smaller values of $\beta$ lead to better
discrimination between quark and gluon jets.
One explicitly sees that we have an expansion in powers of
$\sqrt{\alpha_s/\beta}$, and so it can only be trusted for $\beta$
substantially larger than $\alpha_s$; in practice, perhaps $\beta
\gtrsim (2\sim4)\times \alpha_s$ (see \App{app:nonpert}).
It is interesting to comment also on traditional angularities: for
$\beta > 1$ most of \Eq{eq:summaryequation} still holds, and
only the last term in parentheses would be modified.
However, for $\beta \le 1$ angularities are dominated by recoil
effects, with a structure that is independent of $\beta$, and so we
expect that the discrimination should saturate.
Because the energy correlation double ratio $\C{1}{\beta}$ is
recoil-free for all values of $\beta$, it is better able to probe the
collinear singularity and multiple emission effects that
distinguish quarks from gluons.

\subsection{Monte Carlo Study}
\label{subsec:quarkgluonMCstudy}

We now use a showering Monte Carlo simulation to validate the above
NLL analysis of $\C{1}{\beta}$.  A similar study of the EEC function
appears in \Ref{Gallicchio:2012ez}, where it was called the two-point
moment.\footnote{\Ref{Gallicchio:2012ez} examined the $\Cnobeta{1}$
  quark-gluon discrimination for a range of $\beta$ values and reached
  a conclusion that is consistent with ours.  While their initial
  analysis na\"ively suggests that $\Cnobeta{1}$, Fig.~18, performs worse than
  jet broadening (``girth'', or equivalently $\tau_1$ with $\beta = 1$),
  Fig.~13, that comparison involves
  different Monte Carlo event samples. 
  Table 1 of \Ref{Gallicchio:2012ez} compares the observables on equal
  footing, which shows that $\Cnobeta{1}$ indeed has better
  discrimination power than jet broadening, consistent with our
  discussion here.}
Through this paper, jets are identified with the anti-$k_T$ algorithm \cite{Cacciari:2008gp} using \fastjet~3.0.3 \cite{Cacciari:2011ma}.  No detector simulations are used other than to remove muons and neutrinos from the event samples before jet finding, as was done in analyses for the BOOST 2010 report \cite{Abdesselam:2010pt}.  

We generate pure quark and gluon dijet samples from the processes
$qq\to qq$ and $gg\to gg$ in \pythia~8.165
\cite{Sjostrand:2006za,Sjostrand:2007gs} at the 8 TeV LHC using tune
4C \cite{Corke:2010yf}.  While \pythia{} is not fully accurate to NLL, it does include subleading terms in the splitting functions and multiple emissions, so not surprisingly we
find improved discrimination at smaller values of $\beta$, in agreement with
\Sec{subsec:NLLanalysis}. We scan over various jet radii and $p_T$
cuts to study the dependence of the quark/gluon discrimination on
these parameters.  For this study, we only use the hardest
reconstructed hadron-level jet in the event with a transverse momentum
in the ranges of $ p_T \in [200,300]~\GeV$, $[400,500]~\GeV$, or
$[800,900]~\GeV$.\footnote{The reason for focusing only on the leading
  jet is that we want to minimize ambiguities related to defining
  quark and gluon jets.  The subleading jet is the one more likely to
  have undergone radiation, and with radiation, quark jets may change
  into gluon jets, and vice-versa. 
  Additionally, the local emission environment is changed (e.g.\
  non-global logs may become more important).
  The probability that an event has radiation in the vicinity of the
  subleading jet is $\order{\alpha_s}$, while it is
  $\order{\alpha_s^2}$ near the leading jet.
  As a cross-check on the flavour composition of our events, we have
  clustered the parton-level showered events with the flavor-$k_t$
  algorithm~\cite{Banfi:2006hf}. 
  We find that the flavor of the leading jet is consistent with
  expectations except in a small fraction of events, between a few
  percent and ten percent depending on the generator.
}  If the hardest jet in the event lies outside the $p_T$ range of interest, the event is ignored. In addition, we scan over jet radii values of $R_0=0.4$, $0.6$, and $0.8$.   Because our broad conclusions hold for all samples generated, we only show representative plots to illustrate the quark/gluon performance of $\Cnobeta{1}$. 

\begin{figure}
\centering
\subfloat[]{\label{fig:qvgC} \includegraphics[width=7.0cm]{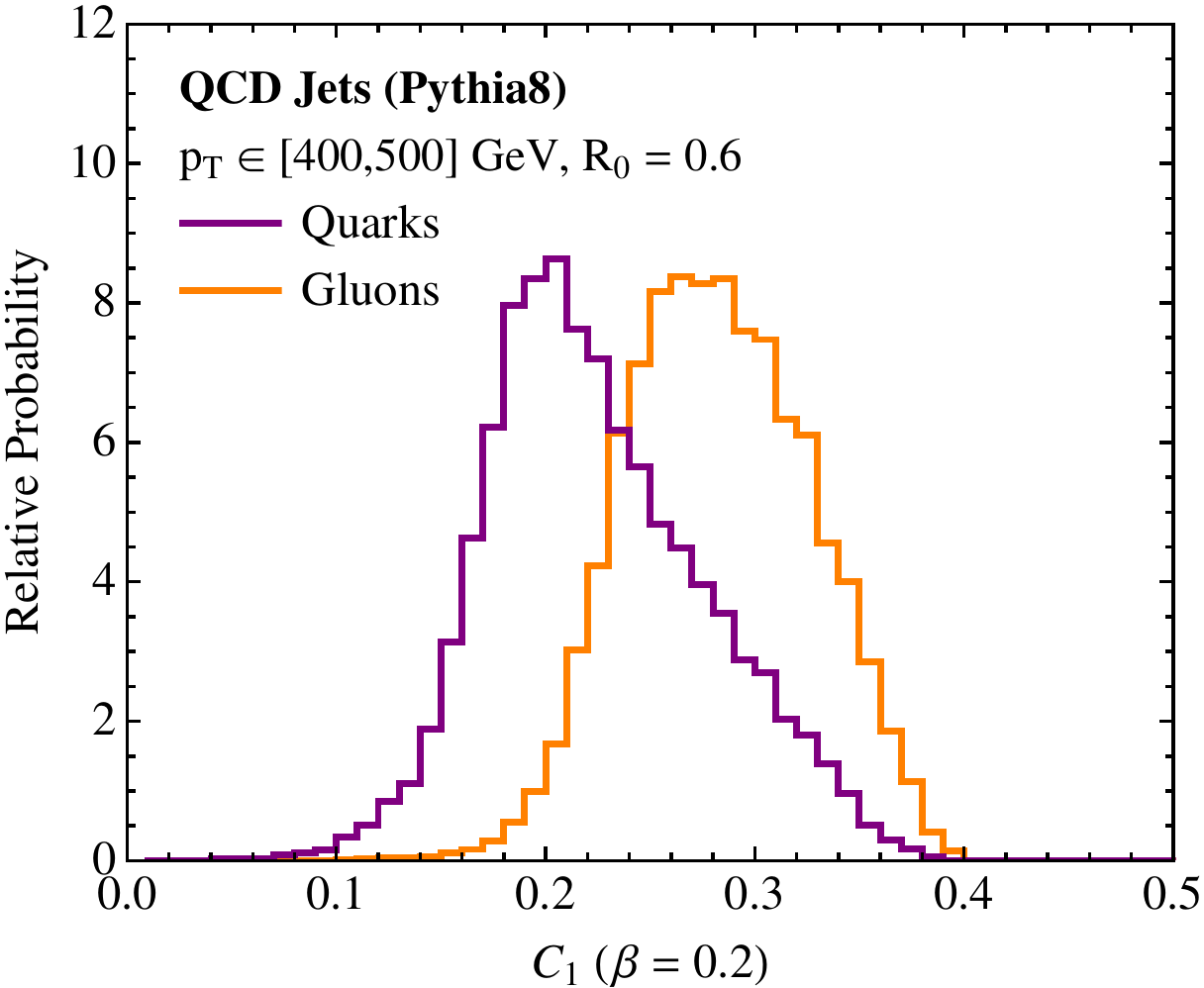}
}
$\qquad$
\subfloat[]{\label{fig:qvg_roc}\includegraphics[width=7.0cm]{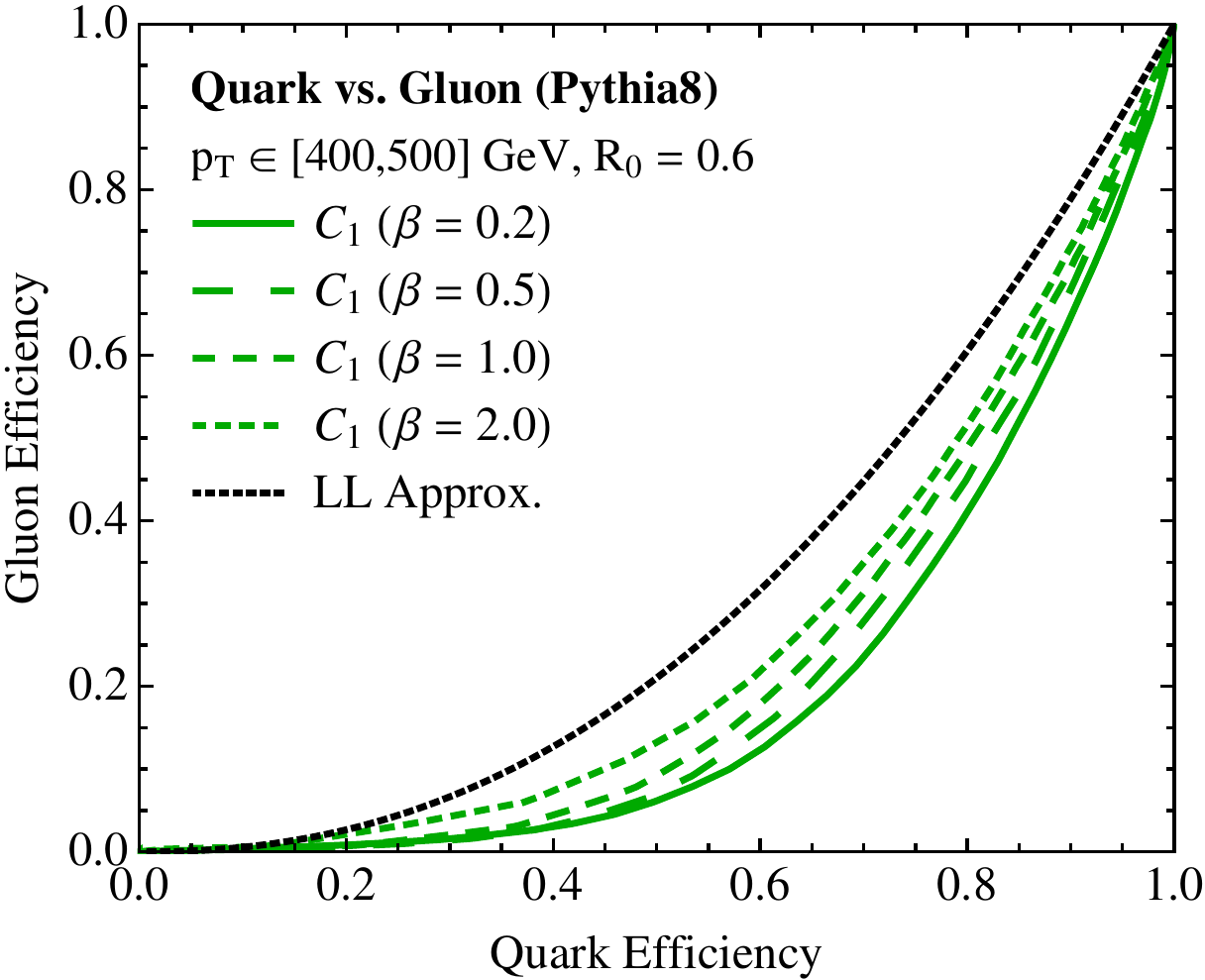}
}
\caption{Left:  Distribution of $\C{1}{0.2}$ for quark jets (purple) and gluon jets (orange) using \pythia{} dijet samples.  The sample consists of anti-$k_T$ jets with radius $R=0.6$ and transverse momentum in the range $[400,500]~\GeV$. Right:  Quark versus gluon discrimination curves using $\C{1}{\beta}$ for several values of $\beta$ in \pythia{}.  Also plotted is the leading log approximation for the discrimination curve, \Eq{eq:rocqvg}.}
\end{figure}

\begin{figure}
\centering
\subfloat[]{ \includegraphics[width=7.0cm]{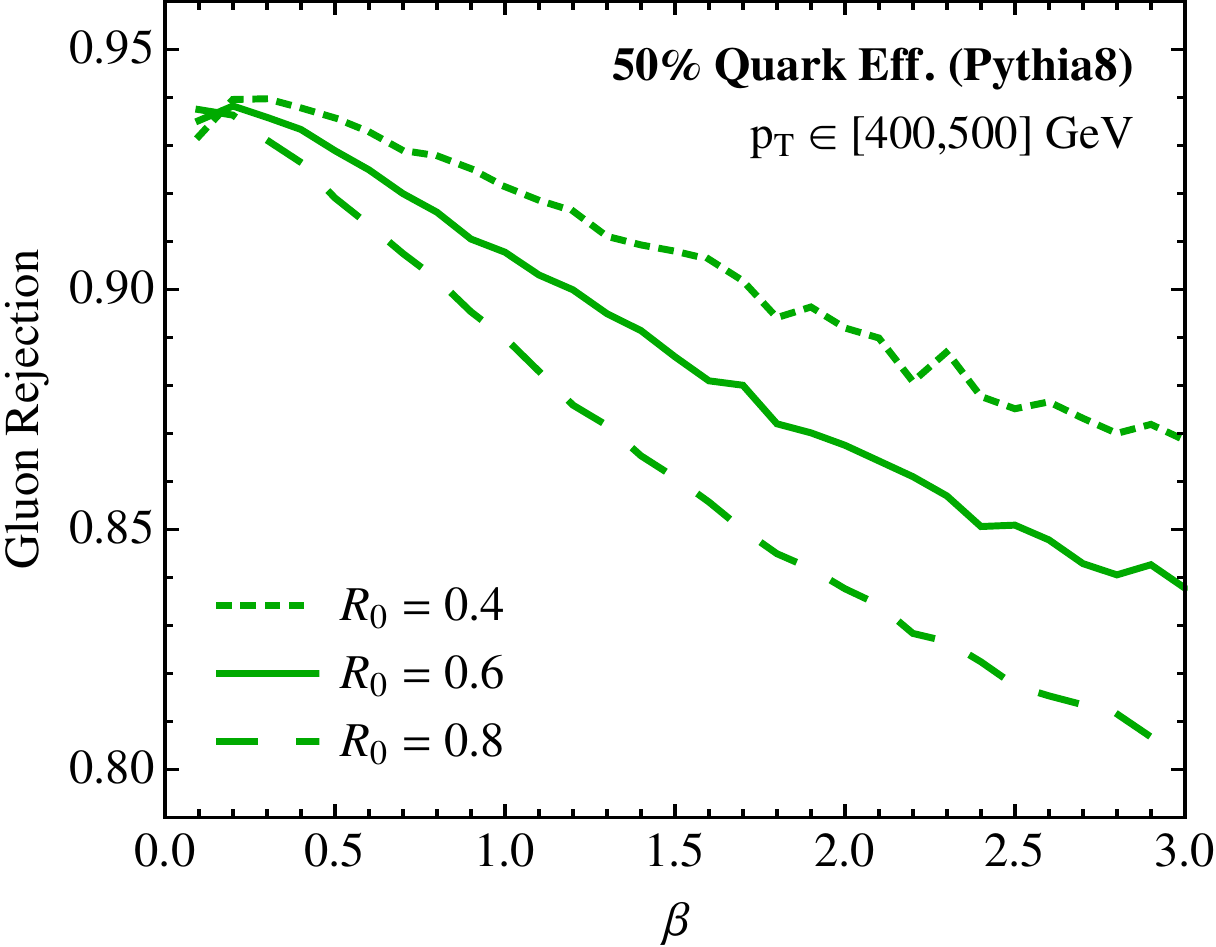}\label{fig:qvg_SB_R}
}
$\qquad$
\subfloat[]{\includegraphics[width=7.0cm]{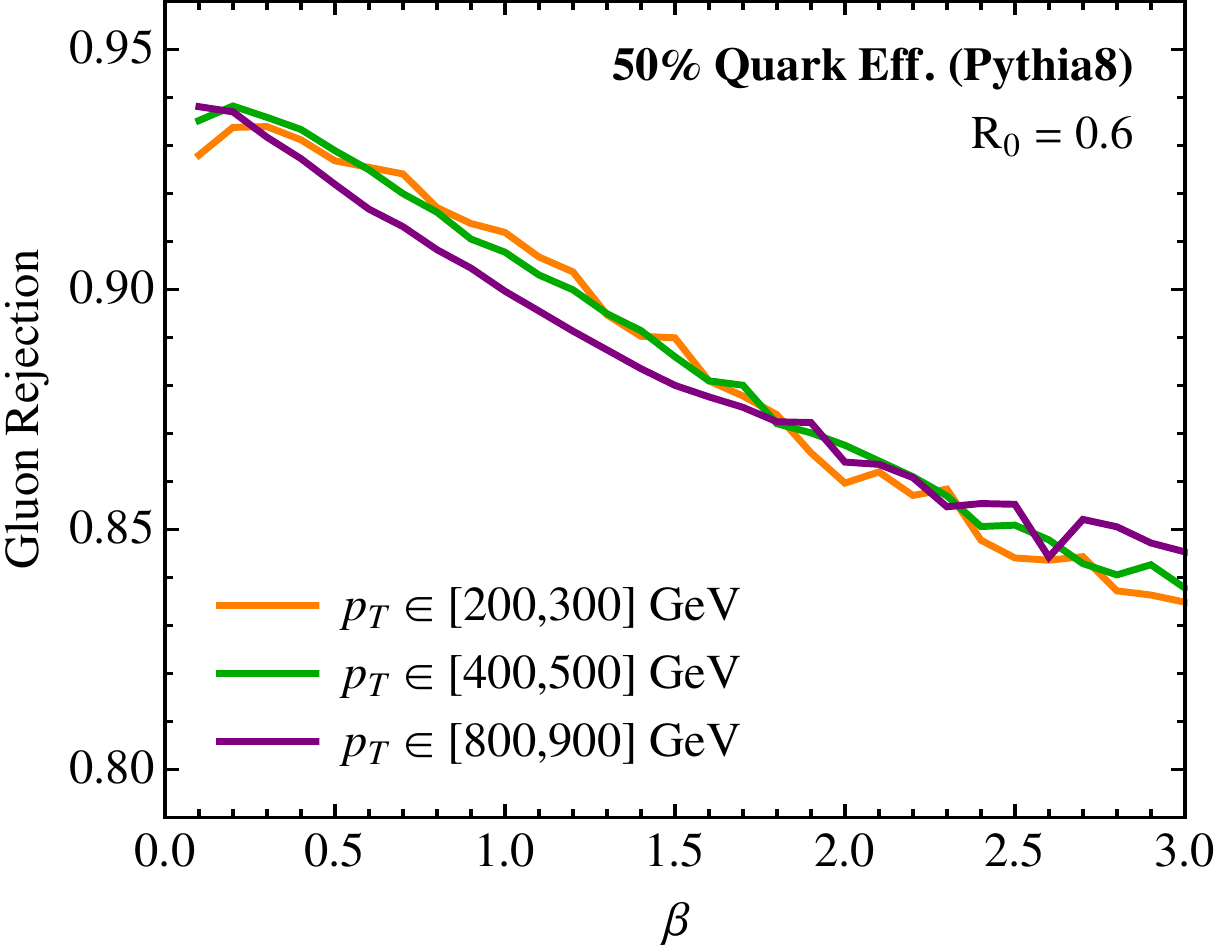}
}
\caption{Gluon rejection rates at 50\% quark efficiency in \pythia{}, as a function of $\beta$.  Left:  fixing the $p_T$ range to be $[400,500]~\GeV$ and sweeping the value of $R_0$.  Right:  fixing $R_0 = 0.6$ and sweeping the $p_T$ range.  For all of these cases, small values of $\beta$ yield the best discrimination.}
\label{fig:qvg_SB}
\end{figure}

In \Fig{fig:qvgC}, we plot the distribution of $\C{1}{0.2}$ for jets
initiated by quarks and gluons with transverse momentum in the range
$[400,500]~\GeV$ and jet radius $R=0.6$ in \pythia{}.  As expected, the gluon curve lies at larger values than the quark curve because of the greater Sudakov suppression in gluon jets.  The quark/gluon discrimination curves for different values of $\beta$ are shown in \Fig{fig:qvg_roc}, which are directly comparable to the NLL results in \Fig{fig:NLLdiscrimination}, up to jet contamination effects included in \pythia{} such as underlying event and initial-state radiation.  Again, we see that $\beta \simeq 0.2$ is the optimal value.  In \Fig{fig:qvg_SB}, we show the gluon rejection rate for 50\% quark efficiency as a function of $\beta$, comparing different $p_T$ ranges and $R_0$ values, all of which favor small values of $\beta$.
Note that the gluon rejection power degrades as the jet radius is
increased, exhibited in \Fig{fig:qvg_SB_R}.  This may be associated
with the increase in the amount of underlying event and initial-state
radiation captured in the jet as the jet radius increases. This
radiation is uncorrelated with the dynamics of the quark or gluon
which initiates the jet.  The degradation is most prominent at large
values of $\beta$, where wide angles in the jet are emphasized (which
is where most of the uncorrelated radiation resides).

\begin{figure}
\centering
\subfloat[]{\label{fig:qvg_tauROC} \includegraphics[width=7.0cm]{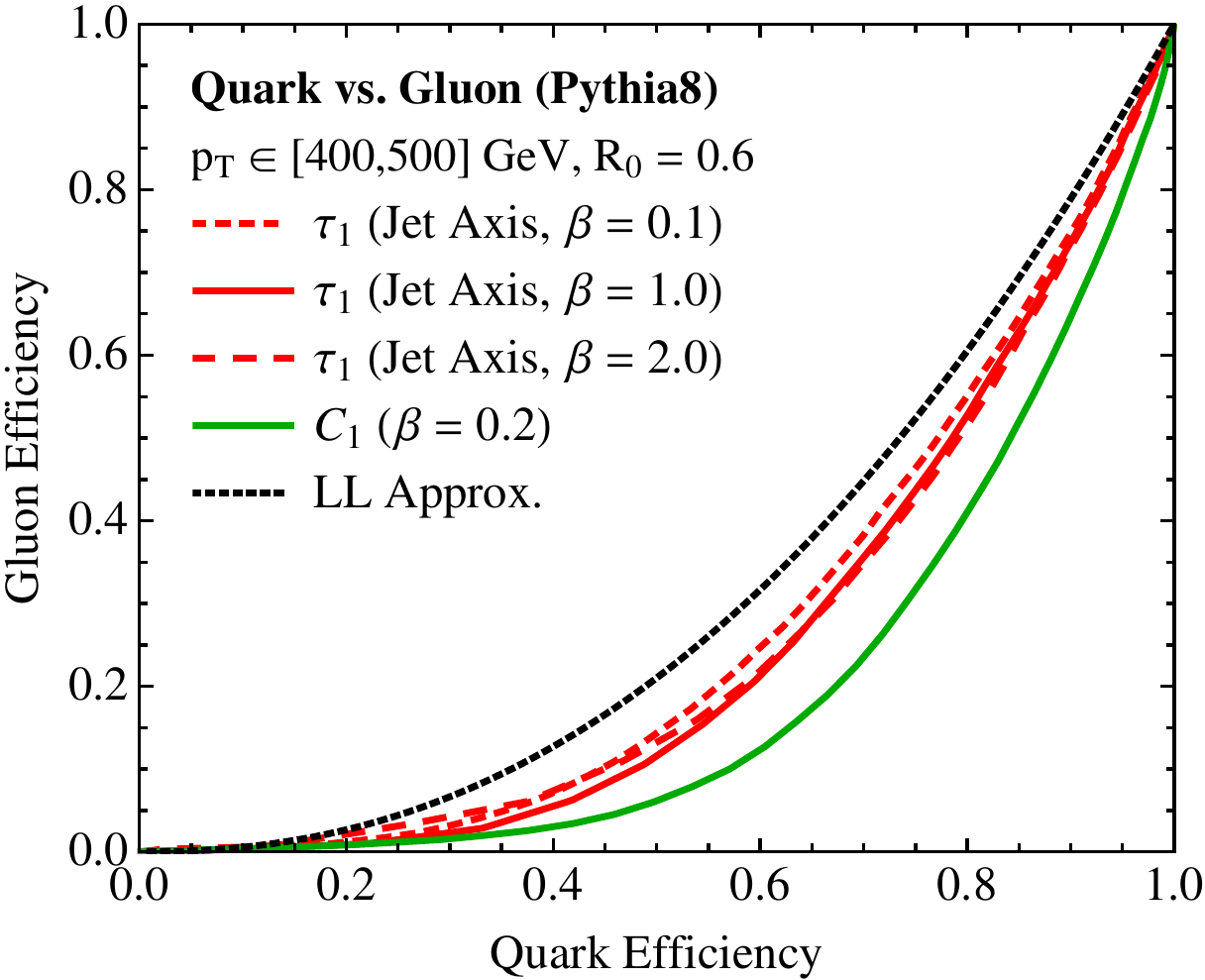}
}
$\qquad$
\subfloat[]{\label{fig:qvg_tauSB}  \includegraphics[width=7.0cm]{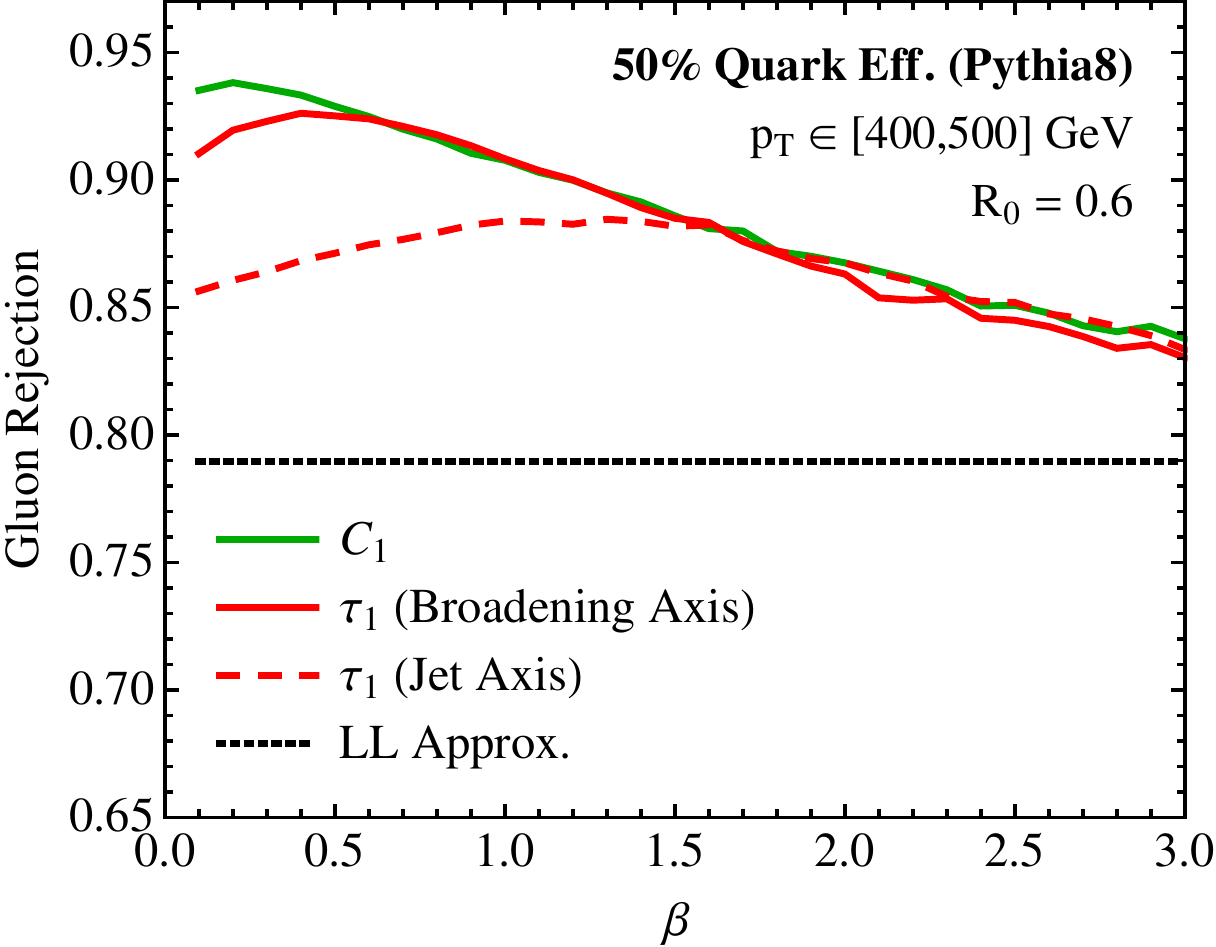}
}
\caption{Left:  Quark/gluon discrimination curves using jet angularities $\tau_1^{(\beta)}$ (i.e.~1-subjettiness measured with respect to the jet axis), for several values of $\beta$ in \pythia{}.  Also plotted is the leading log approximation for the discrimination curve from \Eq{eq:rocqvg} and the discrimination curve for $\C{1}{0.2}$.  The jet sample is the same as used in \Fig{fig:qvg_roc}.  Right:  Gluon rejection rate for 50\% quark efficiency as a function of $\beta$, for angularities, 1-subjettiness measured with respect to the broadening axis, and $\C{1}{\beta}$.  The broadening axis is defined as the axis which minimizes the $\beta = 1$ measure in $N$-subjettiness.  The latter two observables are recoil-free, and therefore give better discrimination power for small values of $\beta$.   
}
\end{figure}

\begin{figure}
\centering
\subfloat[]{\label{fig:qvg_ROC_her} \includegraphics[width=7.0cm]{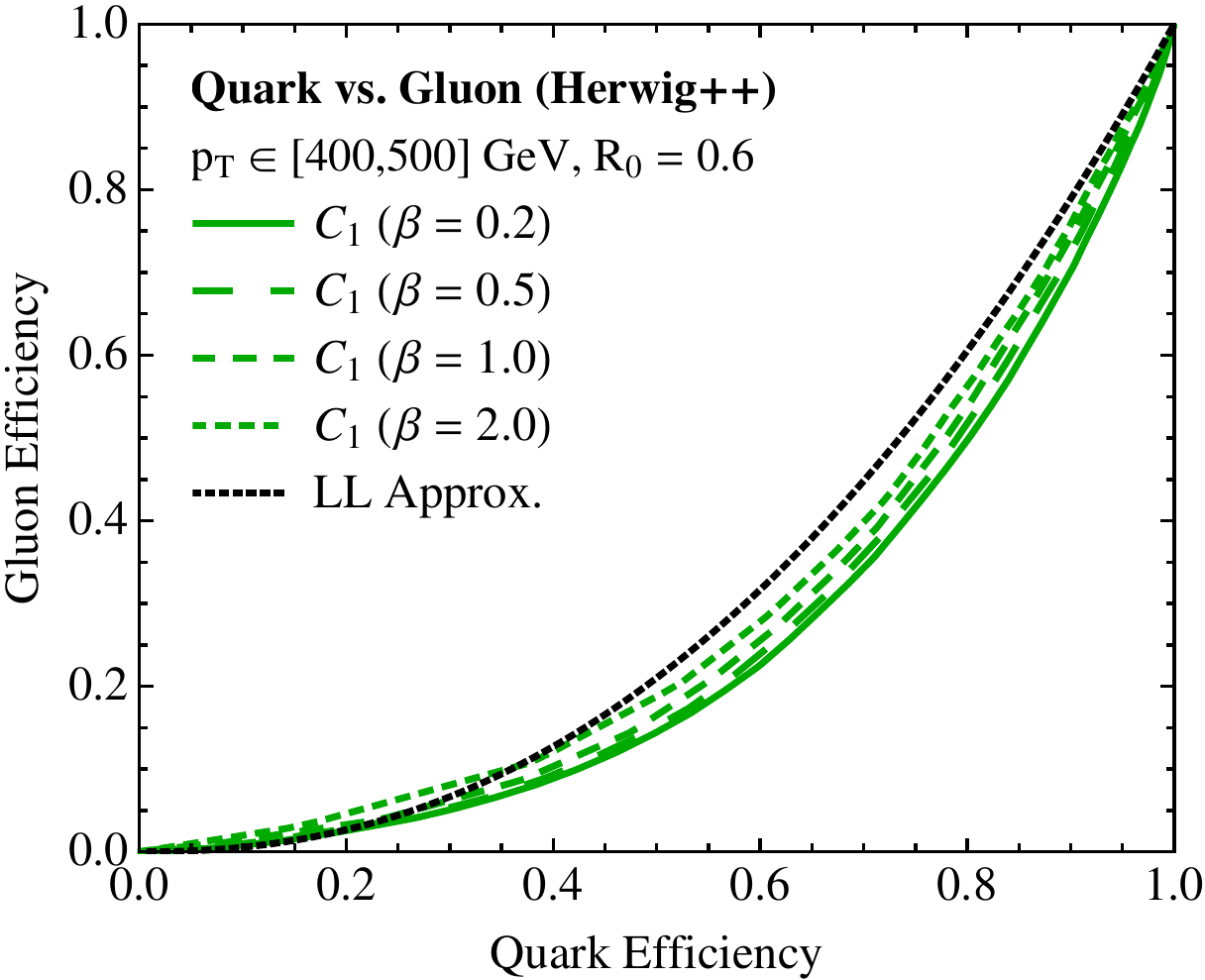}
}
$\qquad$
\subfloat[]{\label{fig:qvg_tauSB_her}  \includegraphics[width=7.0cm]{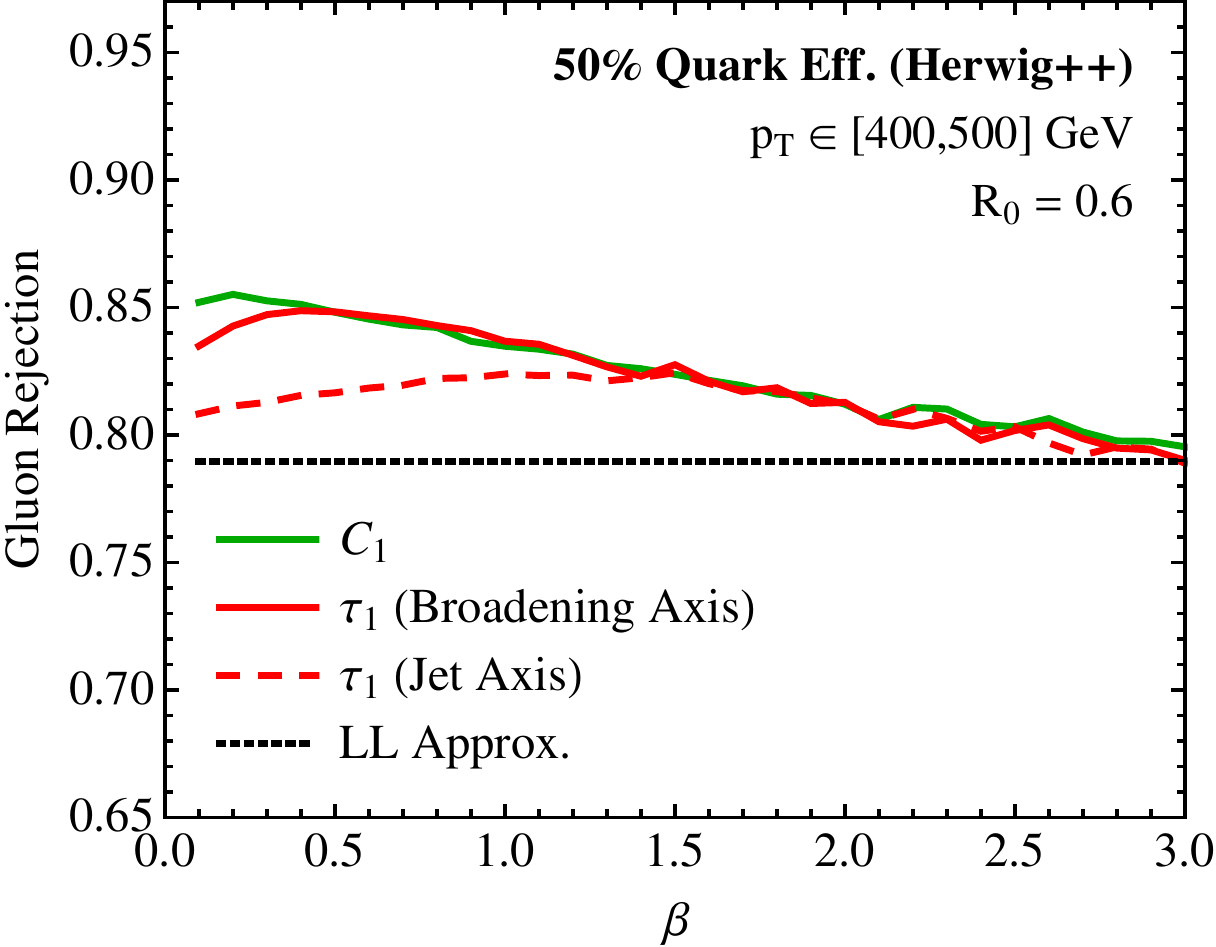}
}
\caption{Left:  Quark versus gluon discrimination curves using $\C{1}{\beta}$ for several values of $\beta$ in \herwigpp (directly comparable to \Fig{fig:qvg_roc}).  Also plotted is the leading log approximation for the discrimination curve, \Eq{eq:rocqvg}.  Right:  Gluon rejection rate for 50\% quark efficiency as a function of $\beta$, for angularities, 1-subjettiness measured with respect to the broadening axis, and $\C{1}{\beta}$ in \herwigpp (directly comparable to \Fig{fig:qvg_tauSB}).   We also tested \pythia~6.425 and \herwig~6.520, whose results lie in between \pythia~8 and \herwigpp. 
}
\end{figure}

To compare the discrimination power of $\C{1}{\beta}$ to other IRC safe observables, we consider 1-subjettiness $\Nsub{1}{\beta}$ defined in \Eq{eq:Nsubdef}.  We allow for two different axis choices:  the jet axis and the broadening axis (i.e.~the axis that minimizes the $\beta = 1$ measure).  When measured with respect to the jet axis, $\Nsub{1}{\beta}$ is essentially the same as the jet angularities $\tau_a$ with $a=2-\beta$.  Angularities coincides with familiar observables for particular values of $\beta$:  $\beta=2$ is jet thrust and $\beta=1$ is jet broadening or girth.  Among the angularities, \Ref{Gallicchio:2011xq} found that jet broadening ($\beta = 1$) was the most powerful angularity for quark/gluon discrimination, and so is a natural benchmark to compare to $\C{1}{\beta}$.  When measured with respect to the broadening axis, $\Nsub{1}{\beta}$ is a recoil-free observable and is therefore expected to behave similarly to $\C{1}{\beta}$.

In \Fig{fig:qvg_tauROC} we plot the discrimination curves for
angularities (i.e.~1-subjettiness measured with respect to the jet
axis) for several values of $\beta$, as well as the discrimination
curve for $\C{1}{0.2}$ in \pythia{}.  Indeed, for most of the range,
the most discriminating angularity is $\beta=1$, but the performance
of all angularities is roughly comparable to and only somewhat better
than the LL expectation.  By contrast, $\C{1}{0.2}$ yields a quark to
gluon efficiency ratio that is about twice as large as any of the
angularities over much of the range.  In \Fig{fig:qvg_tauSB}, we
highlight the importance of working with recoil-free variables, by
plotting the gluon rejection rate at a fixed 50\% quark efficiency.
For $\beta  \ge 1$, $\C{1}{\beta}$ and
1-subjettiness have essentially the same performance.  As $\beta$
approaches 0, however, the discrimination power for the angularities
degrades, while the two recoil-free observables ($\C{1}{\beta}$ and
1-subjettiness with respect to the broadening axis) have improved
performance, as expected from the NLL analysis.\footnote{The reason for the mismatch
  between $\Cnobeta{1}$ and $\tau_1$ with respect to the broadening
  axis at very small values of $\beta$ has not yet been determined.}

To verify the claims made about the performance of $\C{1}{\beta}$ as a quark/gluon discriminator, we also simulate quark and gluon dijet samples in \herwigpp~2.6.3 \cite{Bahr:2008pv,Gieseke:2011na}.  We use the same kinematic cuts and jet algorithm parameters as in the \pythia{} samples.  As the same qualitative conclusions hold in the \herwigpp samples as in \pythia, we only reproduce \Figs{fig:qvg_roc}{fig:qvg_tauSB} for the \herwigpp sample.  In \Fig{fig:qvg_ROC_her}, we plot the quark versus gluon discrimination curve with $\C{1}{\beta}$.  While the discrimination power of $\C{1}{\beta}$ in the \herwigpp sample is not as great as in the \pythia{} sample, 
the behavior that the discrimination increases as $\beta$ decreases is
robust.  In \Fig{fig:qvg_tauSB_her}, we plot the gluon rejection rate
at a fixed 50\% quark efficiency for the three observables considered
earlier.   
As in \Fig{fig:qvg_tauSB}, the discrimination power of the recoil-free
observables increases as $\beta$ decreases and degrades for the
recoil-sensitive angularities (though the $\beta$-dependence is once again
weaker than with \pythia~8).  We also tested  $\C{1}{\beta}$ using
\pythia 6.425~\cite{Sjostrand:2006za} with tunes DW and Perugia 2011
\cite{Skands:2010ak} and \herwig~6.420 \cite{Corcella:2000bw} with
JIMMY \cite{Butterworth:1996zw}, which exhibit discrimination power
that is intermediate between \pythia~8 and \herwigpp.
We also checked that the behavior is robust as underlying event,
initial-state radiation, and hadronization are sequentially turned
off. 

Of course, there is a substantial numerical difference between \pythia~8 and  \herwigpp for quark versus gluon discrimination.   Some of distinction between \pythia~8 and \herwigpp could be due to the fact that different evolution variables are used: \pythia~8 is a $p_T$-ordered shower whereas \herwigpp is angular ordered.  This could in turn affect the flavor content of the quark and gluon jets, thus leading to variations in their ability to discriminate quark and gluon jets.  The energy correlation function observables seem to be particularly sensitive to these differences, especially at relatively small values of the angular exponent $\beta$.  This suggests that any detailed study of the properties of quark and gluon jets should measure $C_1$ at multiple values of $\beta$.  Beyond discriminating quark and gluon jets, measurements of energy correlation functions at both $e^+e^-$ and hadron colliders could be useful for Monte Carlo tuning, especially given the current differences between generators.

In these studies, $\Cnobeta{1}$ has been measured on jet samples which include both charged and neutral hadrons (and we have not applied realistic smearing of energies and angles).  In order to exploit the discrimination power of $C_1$ with $\beta \simeq 0.2$, one needs excellent angular resolution, so in an experimental context, it may be advantageous to measure $C_1$ using only charged hadrons.  A track-only calculation of $C_1$, using e.g.~the methods of \Ref{Chang:2013rca}, is beyond the scope of this work, but we did verify in Monte Carlo that the quark/gluon discrimination power only degrades by a few percent when using a track-only version of $C_1$.
We also note that CMS has successfully made use of $p_{t}D$, related
  to the $\beta \to 0$ limit of $C_1$, using both charged and neutral
  hadrons \cite{Chatrchyan:2012sn,Pandolfi:1480598}.

  Because we observe significant differences in the absolute scale of
  the quark versus gluon discrimination between different Monte Carlo
  generators, the performance of $C_1$ in an experiment may not be as
  optimistic as computed to NLL. However, the increase in the
  discrimination power as $\beta \to 0$ seems robust and would be
  important to verify in data. As discussed in  \App{app:nonpert}, perturbative
  control over $C_1$ ceases to exist for $\beta \lesssim 0.2-0.4$. While
  hadronization will then become significant, separately on the
  distributions of $C_1$ for both quark and gluon jets, it is not
  entirely trivial to relate this to the question of its expected
  impact on the quark/gluon discrimination performance. In any case,
  this kind of questions deserves further investigation, both
  theoretically and experimentally.
    
\section{Boosted Electroweak Bosons with ${\Cnobeta{2}}$}
\label{sec:twoprong}

Our next case study is using $\Cnobeta{2}$ to discriminate between QCD
jets and jets with two intrinsic subjets, such as boosted $W$, $Z$, or
Higgs bosons.\footnote{For related studies, see
  \Refs{Seymour:1993mx,Butterworth:2002tt,Butterworth:2008iy,Almeida:2008yp,Kribs:2009yh,Kribs:2010hp,Chen:2010wk,Hackstein:2010wk,Falkowski:2010hi,Katz:2010mr,Cui:2010km,Kim:2010uj,Thaler:2010tr,Gallicchio:2010dq,Gallicchio:2010sw,Hook:2011cq,Soper:2011cr,Almeida:2011aa,Ellis:2012sn}.
  }  Recall that $\Cnobeta{2}$ involves the
3-point correlator.  To identify a boosted resonance, one first looks
for jets whose mass is compatible with the resonance of interest.
Then $\Cnobeta{2}$ can be used to determine if the jet has two hard
subjets, in which case the jet is tagged as coming from that boosted
resonance.  While we have not carried out analytic calculations to guide us to understand the performance of $\Cnobeta{2}$ (and higher point correlation functions), it is still instructive to study its discrimination power in Monte Carlo.    We use $\pythia~8$ to demonstrate the qualitative behavior and performance of $C_2$, though one must of course be mindful of the quantitative differences in Monte Carlo programs seen already in \Sec{subsec:quarkgluonMCstudy}.  A calculation of $\Cnobeta{2}$ will be left to future work.  

The key finding from this section is that this tagging procedure is
very sensitive to the ratio of the jet mass to the the jet transverse
momentum.  This arises because the structure of the QCD background
depends strongly on the jet mass requirement, and the behavior of
$\C{2}{\beta}$ differs depending on what type of radiation contributes
dominantly to the jet mass.  For a fixed jet $p_T$, we will find that
small values of $\beta \simeq 0.5$ are better for high mass resonance
discrimination, whereas large values of $\beta \simeq 2$ give the
optimal separation at lower masses.  In both regimes, $\Cnobeta{2}$
offers better discrimination power than $\tau_{2,1}$, with the
difference being more pronounced for small $m/p_T$.
After describing this physics for a generic boosted 2-prong resonance, we will specialize to the case of the Higgs boson, where additional $b$-tagging information is available.

\subsection{Dependence on the Mass Criterion}

Consider a quark or gluon jet with invariant mass close to the boosted
resonance of interest (which we will call a $Z$ boson for
concreteness).  For jets with mass comparable to their transverse
momentum, the mass is dominated by a single, relatively hard,
perturbative splitting. Thus, one expects that the QCD jets that can
fake a $Z$ boson are those with two relatively hard cores of energy
surrounded by soft radiation. These jets are straightforward to
analyze in fixed-order perturbation theory (to generate the jet mass) matched to resummed
perturbation theory (to generate the radiation pattern for $\Cnobeta{2}$), since there is a clear ordering of emissions in
the jet.  In particular, QCD jets with large mass should appear
similar to jets initiated from heavy resonance decay, with differences
controlled mainly by the color of the decay products
and the phase space of the hard splitting. 

For many systems of interest, however, the above analysis is not
appropriate. Once the jet mass is less than around a fifth of the jet
transverse momentum times $R$, the mass no longer arises dominantly from a hard perturbative splitting.  For jets in the low to intermediate mass ranges, a significant mass can be generated by a single soft emission from a single hard core. At lower masses, the mass of a jet is generated by multiple soft emissions.  Jets in the low and intermediate mass regimes require resummation of these soft emissions to accurately model their dynamics as fixed-order perturbation theory is no longer accurate.  For this reason, we expect QCD jets in this mass regime to look very different from boosted resonances with two hard cores, and the discrimination power of $C_2$ should improve as the ratio of the jet mass to the transverse momentum decreases.  In addition, as discussed in \Sec{subsubsec:softwide}, $C_2$ is better able to exploit the color singlet nature of the $Z$ boson when $m/p_T$ is small.

\begin{figure}
\centering
    \includegraphics[width=7cm]{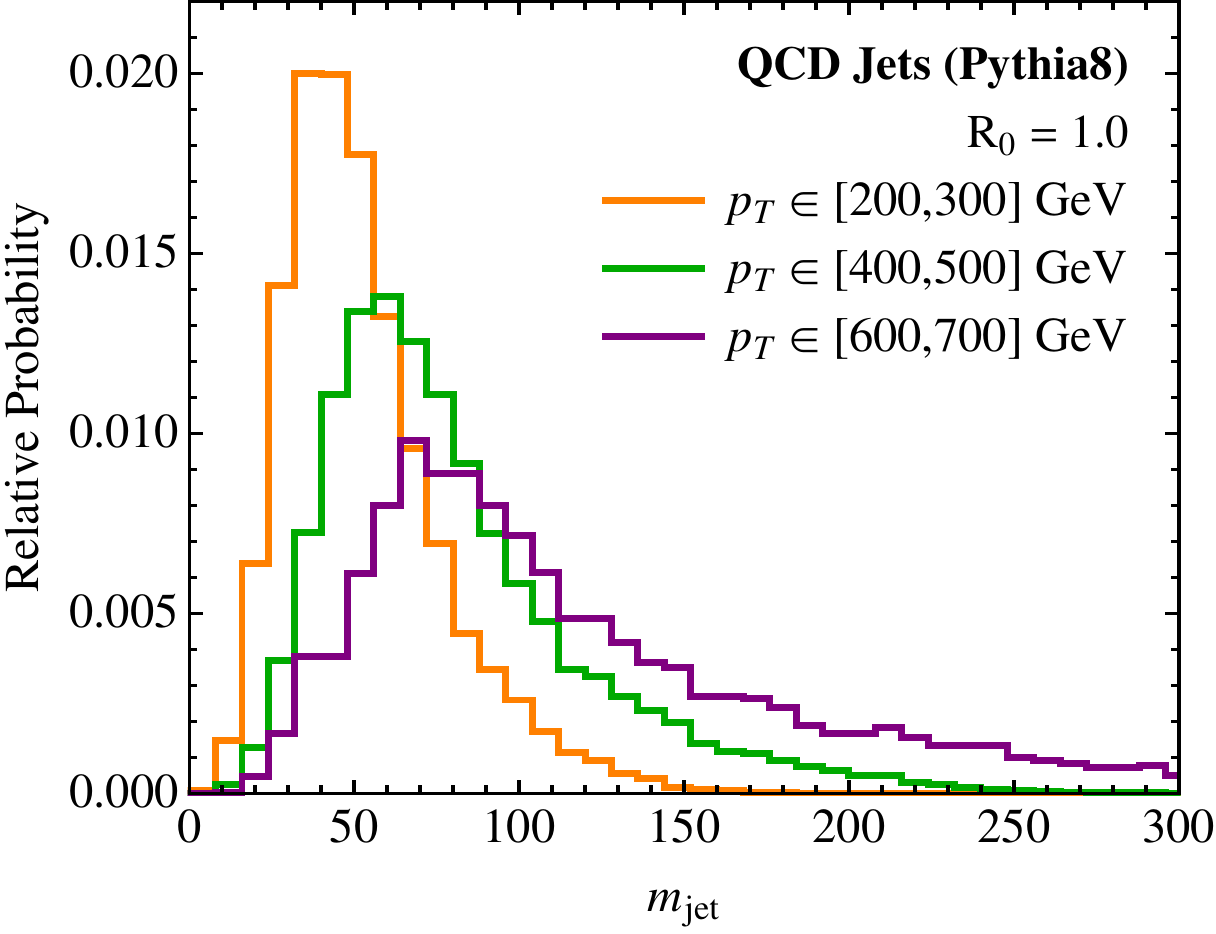}
\caption{Distribution of mass of QCD jets in the process $pp\to Z j$ as simulated in \madgraph~1.5.0 and showered in \pythia~8.165.  The transverse momentum
of the jets lie in one of the ranges of $[200,300]$, $[400,500]$ or $[600,700]$ GeV, as labeled on the plot.}\label{mass_distro}
\end{figure}

To illustrate this, we generate a mixed sample of quark and gluon jets
from $pp\to Z j$ with the $Z$ decaying to leptons.  These are
simulated at the 8 TeV LHC in \madgraph~1.5.0 \cite{Alwall:2011uj},
showered in \pythia~8.165 \cite{Sjostrand:2006za,Sjostrand:2007gs},
and we identify the hardest anti-$k_T$ $R_0=1.0$ jet.  In
\Fig{mass_distro} we plot the invariant mass spectrum of QCD jets in
three different transverse momentum bins, $p_T \in [200,300]~\GeV$,
$[400,500]~\GeV$, and $[600,700]~\GeV$.  We see that the mass
distributions in each $p_T$ bin have steeply falling tails extending
to masses of about $p_T/2$.  In the tail region, we expect fixed-order
perturbation theory to accurately describe the origin of mass of the
jet.  At lower masses, below about $p_T/5$, Sudakov suppression
becomes important as the distributions peak and decrease toward zero
mass. In this mass regime, fixed-order
perturbation theory is no longer adequate to describe the
distribution. 

\begin{figure}
\centering
\subfloat[]{\label{C2_distro_qcd} \includegraphics[width=7.0cm]{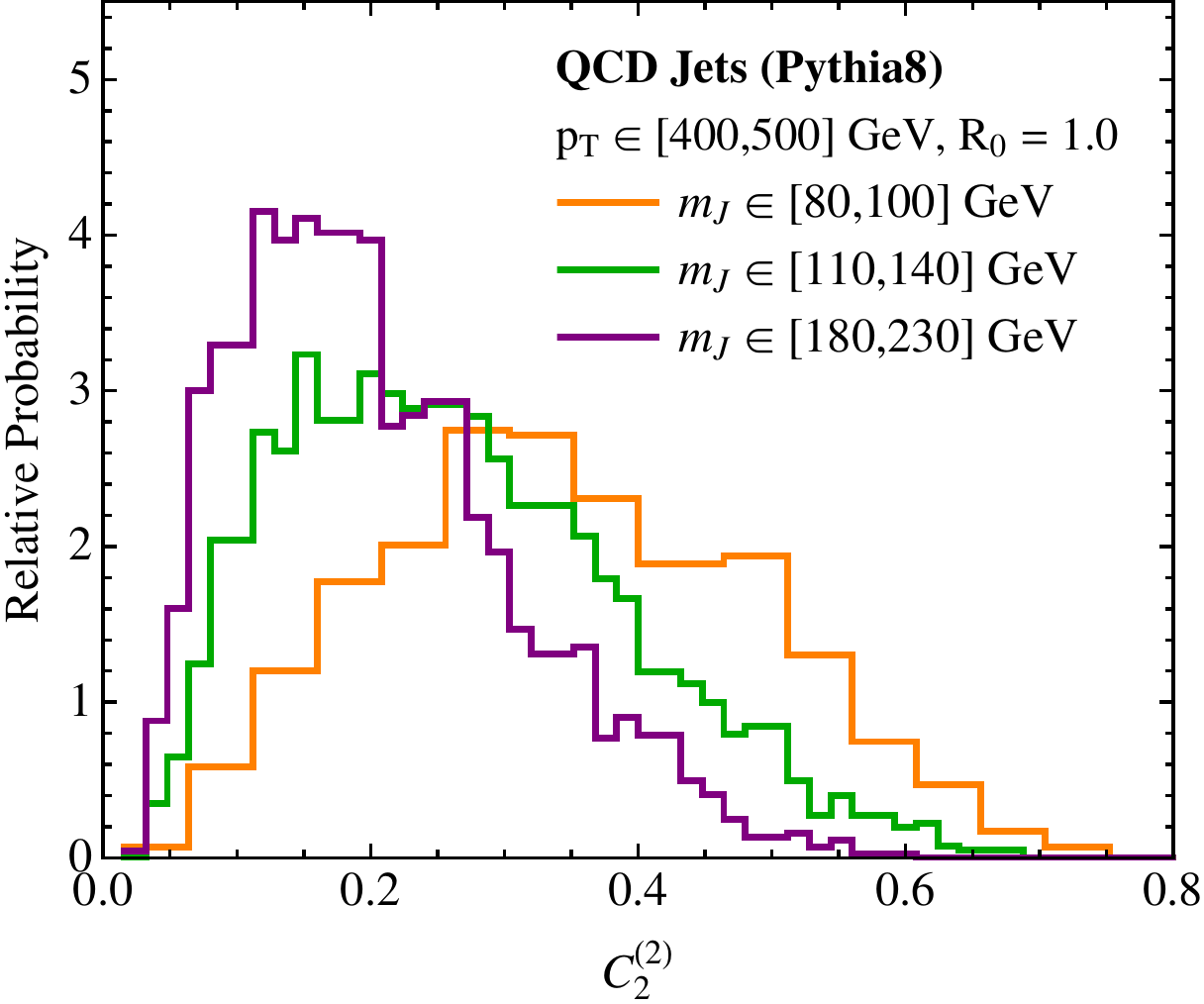}
}
$\qquad$
\subfloat[]{\label{C2_distro_Z}  \includegraphics[width=7.0cm]{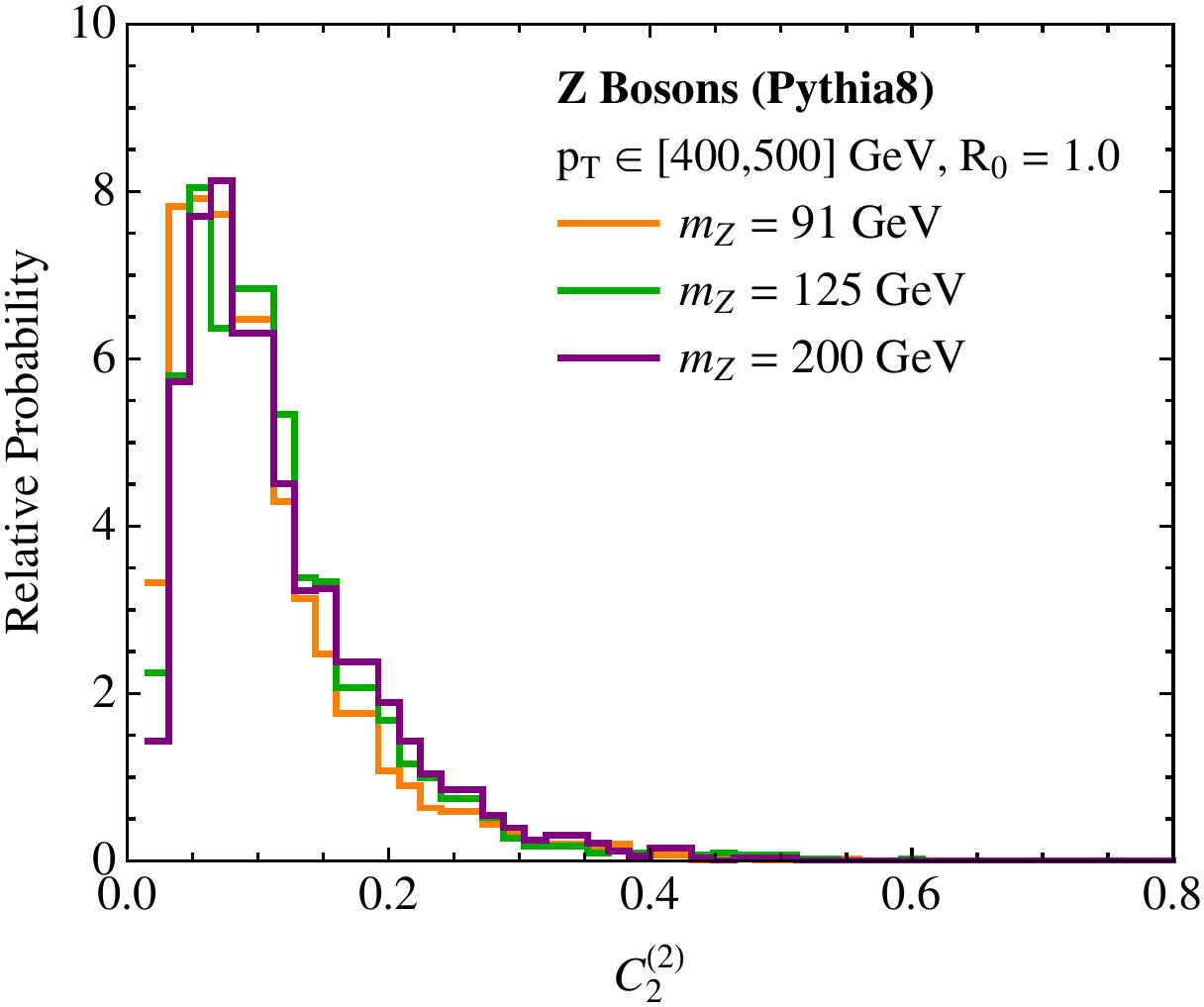}
}
\caption{Distributions of $\C{2}{2}$ for QCD jets (left) and $Z$ bosons decaying to jets (right) with different masses of the $Z$.  The transverse
momentum of the jets for all masses lies in the range of $[400,500]$ GeV.  The different curves correspond to different event samples according
to the mass of the resonance. 
}
\end{figure}

This differing origin of the jet mass is reflected in the
$\C{2}{\beta}$ distributions.  Because small values of $\C{2}{\beta}$
correspond to 2-subjet-like jets, the $\C{2}{\beta}$ distribution
moves to lower values as the mass of a QCD jet increases, as shown in
\Fig{C2_distro_qcd} for $\beta = 2$ in the $p_T$ range $[400,500]$
GeV.\footnote{The labelled jet masses of $m_Z = \{80, 91, 110, 125,
  150, 200 \}~\GeV$ correspond to the jet mass ranges $[70,90]~\GeV$,
  $[80,100]~\GeV$, $[100,120]~\GeV$, $[110,140]~\GeV$,
  $[140,170]~\GeV$, and $[180,230]~\GeV$.}  In contrast, for a boosted
heavy particle that decays to two partons, the $\C{2}{\beta}$
distribution is relatively insensitive to the resonance mass, since
the mass of such a jet comes mostly from two partons from the decay
regardless of the boost factor.  Shown in \Fig{C2_distro_Z} is the
signal $\C{2}{2}$ distribution for $p p \to Z Z$, where one of the $Z$
bosons decays to leptons and the other decays to jets.   We can
manually adjust the mass of the $Z$ in \madgraph to study several
different mass to transverse momentum ratios. For $m_Z = \{91, 125, 200 \}~\GeV$,  the $\C{2}{2}$ distributions are
remarkably similar.\footnote{$\Cnobeta{2}$ is not invariant to transverse boosts, so for more extreme values of $m/p_T$, the distribution will move to smaller values.  However, because of underlying event and initial state radiation, $\Cnobeta{2}$ does not change as much as one would na\"ively expect under boosts.}

\begin{figure}
\centering
\subfloat[]{\label{C2_roc_91} \includegraphics[width=7.0cm]{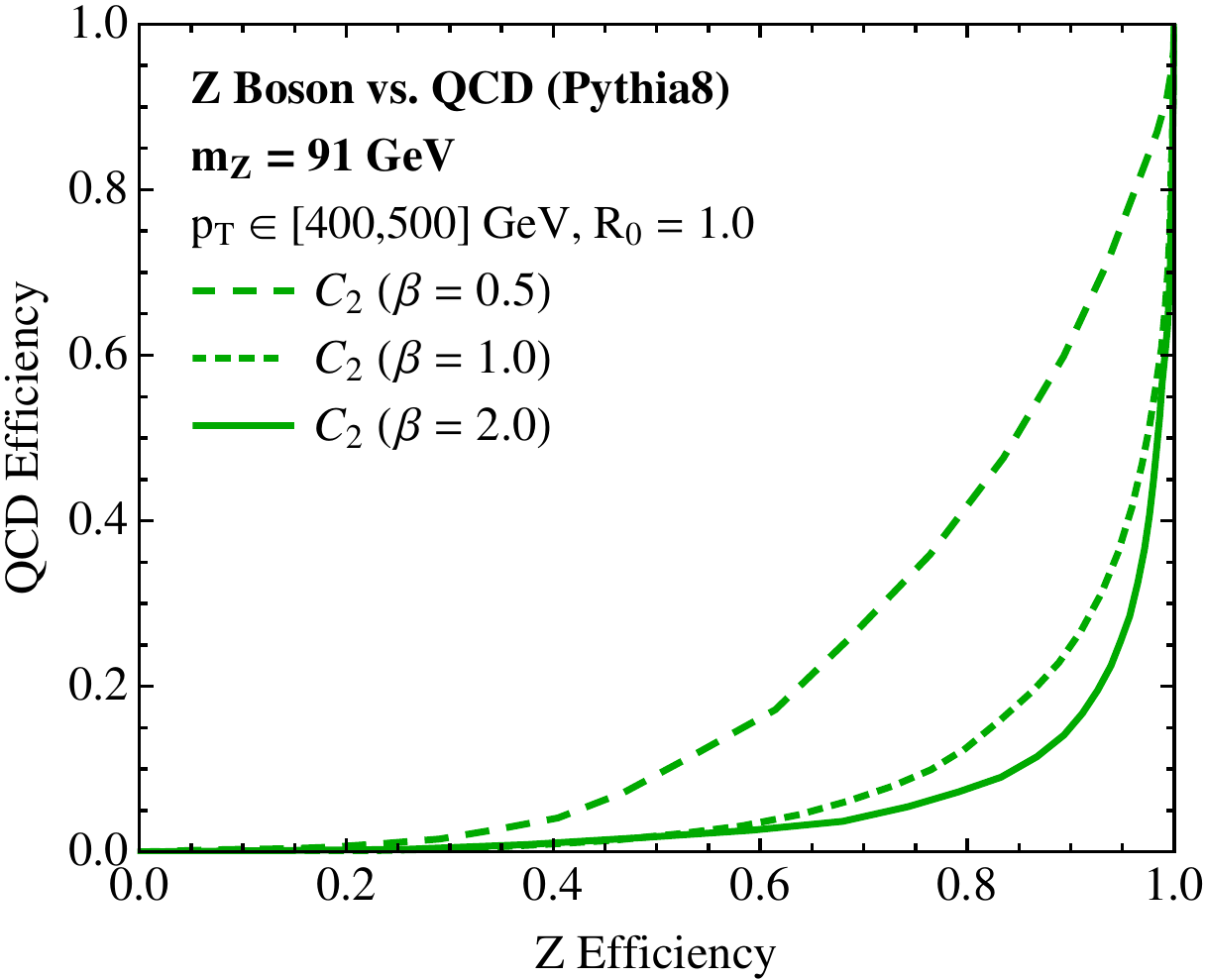}
}
$\qquad$
\subfloat[]{\label{C2_SB} \includegraphics[width=7.0cm]{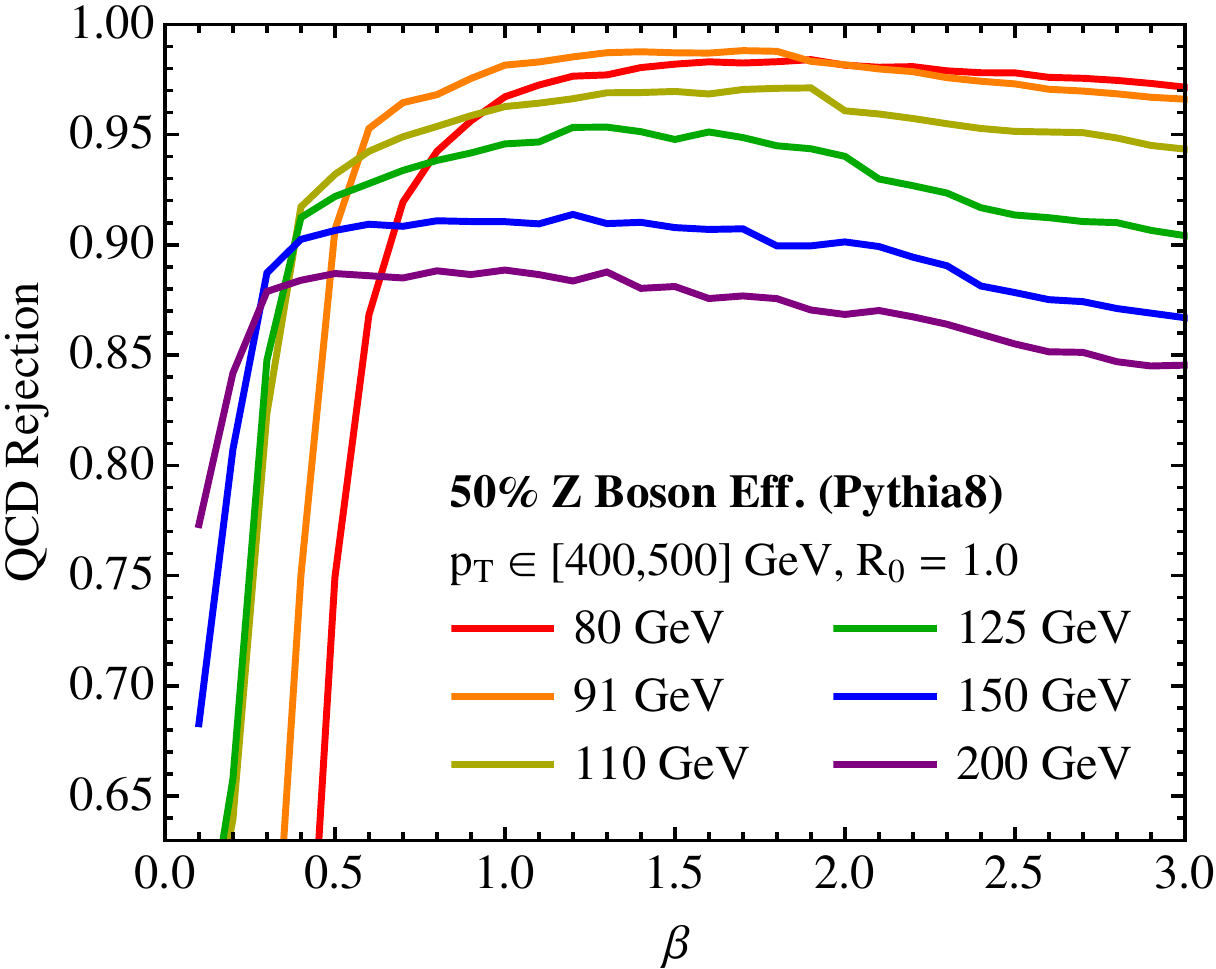}
}
\caption{Left:  the discrimination curves for boosted hadronic $Z$ bosons ($m_Z = 91~\GeV$) compared to QCD jets with $\C{2}{\beta}$ for various values of $\beta$.  The transverse momentum of all jets was required to lie in the range of $[400,500]$ GeV.  Right:  QCD rejection rate for 50\% boosted $Z$ efficiency as a function of $\beta$, sweeping the value of the $Z$ boson mass to $m_Z = \{80, 91, 110, 125, 150, 200 \}~\GeV$.  The optimal value of $\beta$ depends strongly on the resonance mass.
}\label{C2_ROC}
\end{figure}

In \Fig{C2_roc_91}, we show the QCD jet versus $Z$ boson discrimination curve for $m_Z = 91~\GeV$ with $p_T \in [400,500]~\GeV$ for several values of $\beta$.  To see how the physics changes as the resonance mass changes, we plot the QCD rejection rate for 50\% boosted $Z$ efficiency in \Fig{C2_SB} as a function of $\beta$, for $m_Z = \{80, 91, 110, 125, 150, 200 \}~\GeV$.   
At low masses, the most powerful discriminant is $\beta \simeq 1.5-2$.  This is expected, since large values of $\beta$ emphasize soft wide-angle emissions where there is more of a penalty for QCD jets in the Sudakov peak.  However, we do not have a quantitative way to understand why the discrimination power saturates at $\beta \simeq 2$, as opposed to even higher values.  At intermediate masses, a wide range of $\beta$ values yield very similar results.  At the high masses where QCD jets are in the tail region, the discrimination dependence on $\beta$ inverts, with the most powerful discrimination for $\beta \simeq 0.5$.  This is likely  to do with the same quark/gluon color factor discrimination as in \Sec{sec:oneprong}.  In particular, high mass QCD jets are formed by a hard perturbative splitting, which is most likely to be a gluon, whereas the $Z$ jet has two hard quark subjets.  That said, we have not yet performed a NLL calculation to understand why $\beta \simeq 0.5$ is preferred, as opposed to even smaller values. 

\begin{figure}
\centering
\subfloat[]{\label{C2_comp_91} \includegraphics[width=7.0cm]{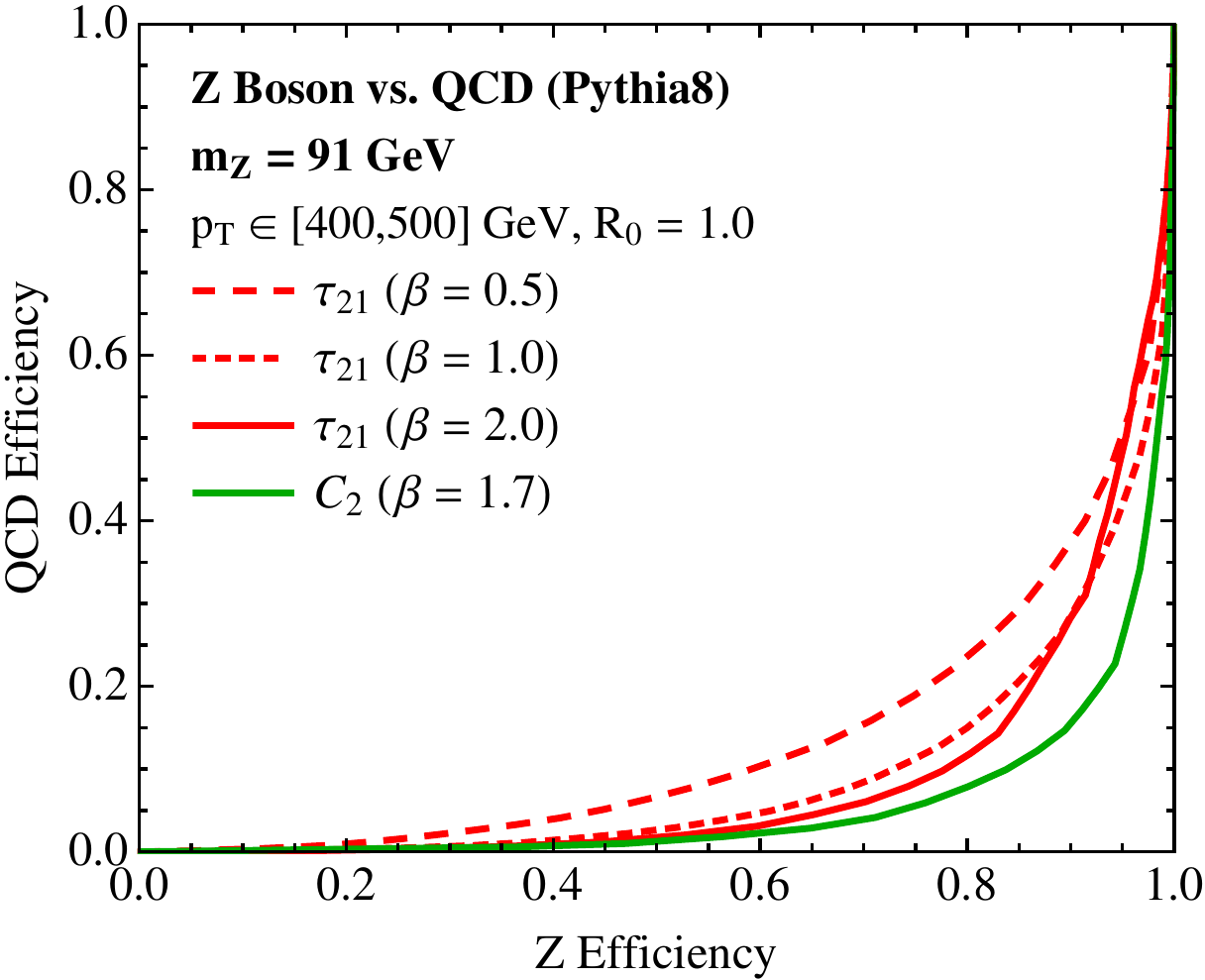}
}
$\qquad$
\subfloat[]{\label{C2_compSB}  \includegraphics[width=7.0cm]{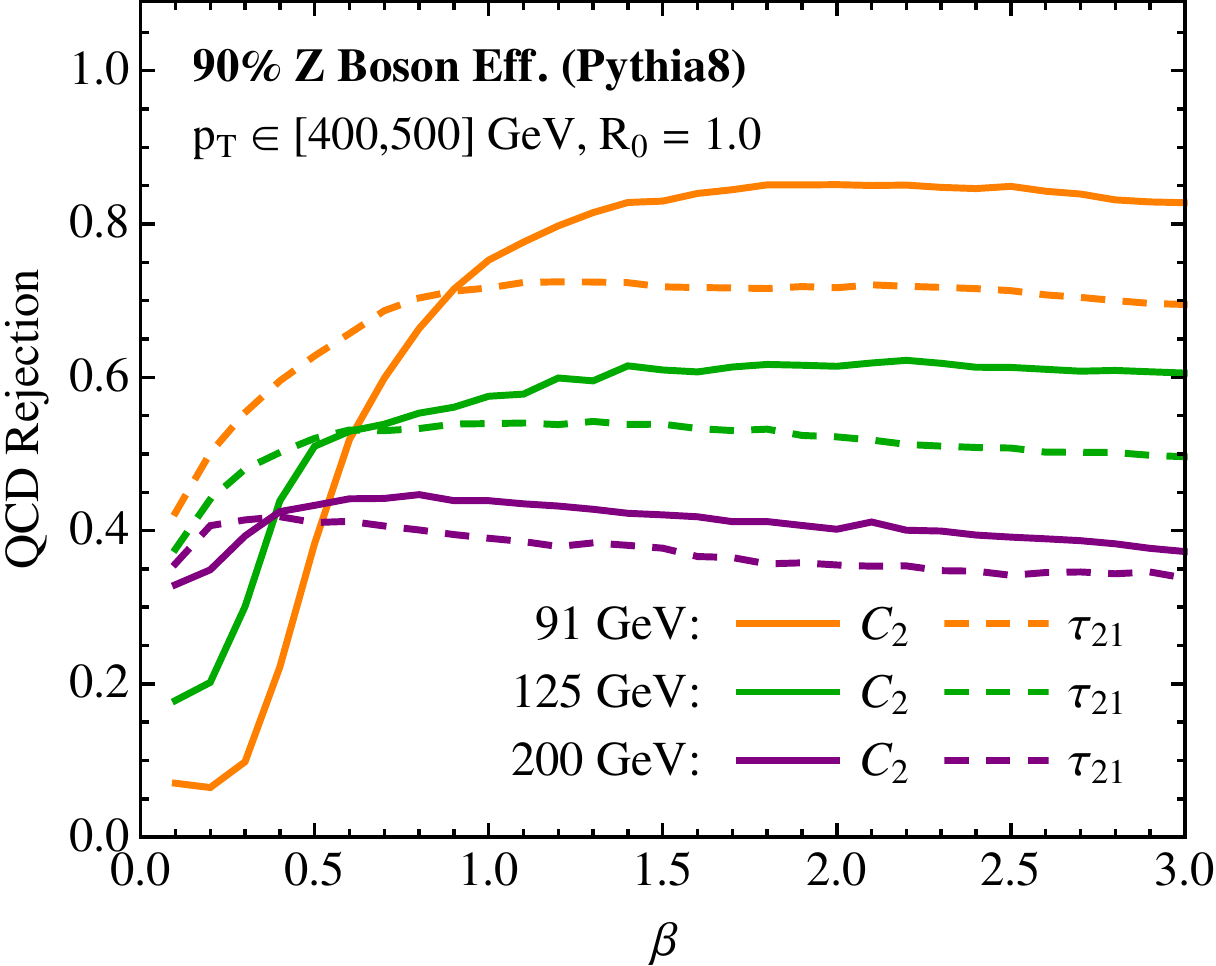}
}
\caption{Left:  the discrimination curves for boosted hadronic $Z$ bosons ($m_Z = 91~\GeV$) compared to QCD jets with $\tau_{2,1}^{\beta}$ for various values of $\beta$.  For comparison is shown the $\C{2}{\beta}$ curve with the best discrimination ($\beta = 1.7$).  The subjet axes for $N$-subjettiness are defined as those that minimize the $\beta = 1$ measure (broadening axes).  Right:  QCD rejection rate for 90\% boosted $Z$ efficiency as a function of $\beta$, sweeping the value of the $Z$ boson mass to $m_Z = \{91, 125, 200\}~\GeV$.  Because these curves are with 90\% $Z$ efficiency, they are not directly comparable to \Fig{C2_SB}.   Note that as $m/p_T$ decreases, the performance of $\Cnobeta{2}$ improves relative to $\Nsubnobeta{2,1}$. 
}
\end{figure}

Finally, it is instructive to compare the discrimination power of
$\Cnobeta{2}$ to $N$-subjettiness.  The ratio of 2-subjettiness to
1-subjettiness $\tau_{2,1}^{(\beta)}$ is defined in \Eq{eq:NsubRatio}
and can be used to identify $Z$ bosons decaying to two jets.  To
eliminate ambiguities in minimum axes finding at small values of
$\beta$, we choose to define the subjet axes by those that minimize
the $\beta=1$ measure (i.e.~the broadening axes).  The discrimination
curves of $\tau_{2,1}^{(\beta)}$ for $m_Z = 91~\GeV$ is plotted in
\Fig{C2_comp_91}, with the $\C{2}{\beta}$ curve with the most
discriminating value from \Fig{C2_roc_91} shown for comparison. We
also show the QCD rejection rate for 90\% boosted $Z$ efficiency in
\Fig{C2_compSB}.  At low masses, $\C{2}{2}$ performs as well as or better than $\tau_{2,1}^{(\beta)}$ over the entire range of $\beta$, except at very small values of $\beta$.  
At high masses, the discrimination power of $\tau_{2,1}^{(\beta)}$ becomes comparable to $\C{2}{\beta}$, since both observables lock onto the hard subjets in the $Z$ decay of the massive QCD jet.  The increase in the relative discrimination power of $\Cnobeta{2}$ with respect to $\tau_{2,1}$ as the ratio $m/p_T$ decreases is expected from the discussion of \Sec{subsubsec:softwide}.  As $m/p_T$ decreases, soft wide-angle subjets become more important for determining the structure of the jet and $\Cnobeta{2}$ emphasizes these emissions more than $\tau_{2,1}$.

\subsection{Boosted Higgs Identification}
  
One key application for 2-prong jet substructure observables is for identifying boosted Higgs bosons in the decay $H \to b \bar{b}$.  Compared to the case of $Z$ bosons, there is additional information from the presence of $b$ quarks (and the resulting $B$ hadrons) in the final state, which can be used to mitigate QCD backgrounds.  Thus, to identify boosted Higgs bosons decaying to bottom quarks, we employ three criteria.  First, we require the jet to have a mass comparable to the Higgs boson.  Second, we demand that two $B$ hadrons are tagged in the jet.  Third, we use a sliding cut on $\C{2}{\beta}$ to test for two hard subjets in the jet.

Because we demand that the jet have two $B$ hadrons, the largest QCD background to Higgs decays to bottoms is gluon splitting to bottoms.  The splitting function $g \to b \bar{b}$ does not have a soft singularity, so the bottoms from this splitting will have comparable energies.  This is also the case for Higgs decay, so we do not expect the same discrimination power for Higgs bosons compared to $Z$ bosons studied above.  That said, because of the difference in the total color of the jets, there is an additional handle on Higgs versus gluon discrimination; the bottom quarks from the gluon splitting will be in a color octet state, so there will be significantly more radiation at wide angles compared to Higgs jets.  

This color octet versus color singlet distinction can be exploited in two ways.  First, more wide-angle radiation can be included in the jet if the jet radius is increased.  Larger jet radii improve the contrast for $\C{2}{\beta}$, since more wide-angle radiation is included in the (octet) gluon jets compared to the (singlet) Higgs jets.  Second, the value of $\beta$ can be set to accentuate the importance of wide-angle emissions in the jet.  As $\beta$ increases, more weight is given to wide-angle emissions, further penalizing gluon jets compared to Higgs jets when using $\C{2}{\beta}$.

A full study of boosted Higgs identification using $\C{2}{\beta}$ is
beyond the scope of this work, but we can get a sense for the
discrimination power of $\C{2}{\beta}$ by comparing the boosted Higgs
signal $pp\to Z H$ to the leading QCD background of $pp\to Z b\bar{b}$
where both bottom quarks happen to be clustered into the same jet.  We
generate both the signal and background distributions for the 8 TeV
LHC in \madgraph~1.5.0 \cite{Alwall:2011uj} plus  \pythia~8.165
\cite{Sjostrand:2006za,Sjostrand:2007gs}, with all ground state $B$
hadrons stable to allow for na\"ive $b$-tagging of the jets (with
100\% efficiency and no mistags).  The mass of the Higgs is set to
$125~\GeV$, and the $Z$ is decayed to leptons and the Higgs is decayed
to $b\bar{b}$.  We consider anti-$k_T$ jets with various values of the
jet radius $R_0 = \{0.6, 0.8, 1.0, 1.2 \}$.  The leading jet is
required to have  transverse momentum in the range $[400,500]~\GeV$
with exactly two $B$-hadrons as constituents.  To approximate
realistic $b$-tagging within jets, we recluster the jet with the $k_T$
algorithm to find two exclusive subjets, and we require that each
subjet contain exactly one identified $B$-hadron.  Finally, the mass
of the jet is required to be in the window of $m_J \in [110,140]~\GeV$
(i.e.~within 15 GeV of the Higgs mass).  From the leading jet, we
compute $\C{2}{\beta}$ for various values of $\beta$ and determine the
discrimination power. 

\begin{figure}
\centering
\subfloat[]{\label{C2_distro_bb} \includegraphics[width=7.0cm]{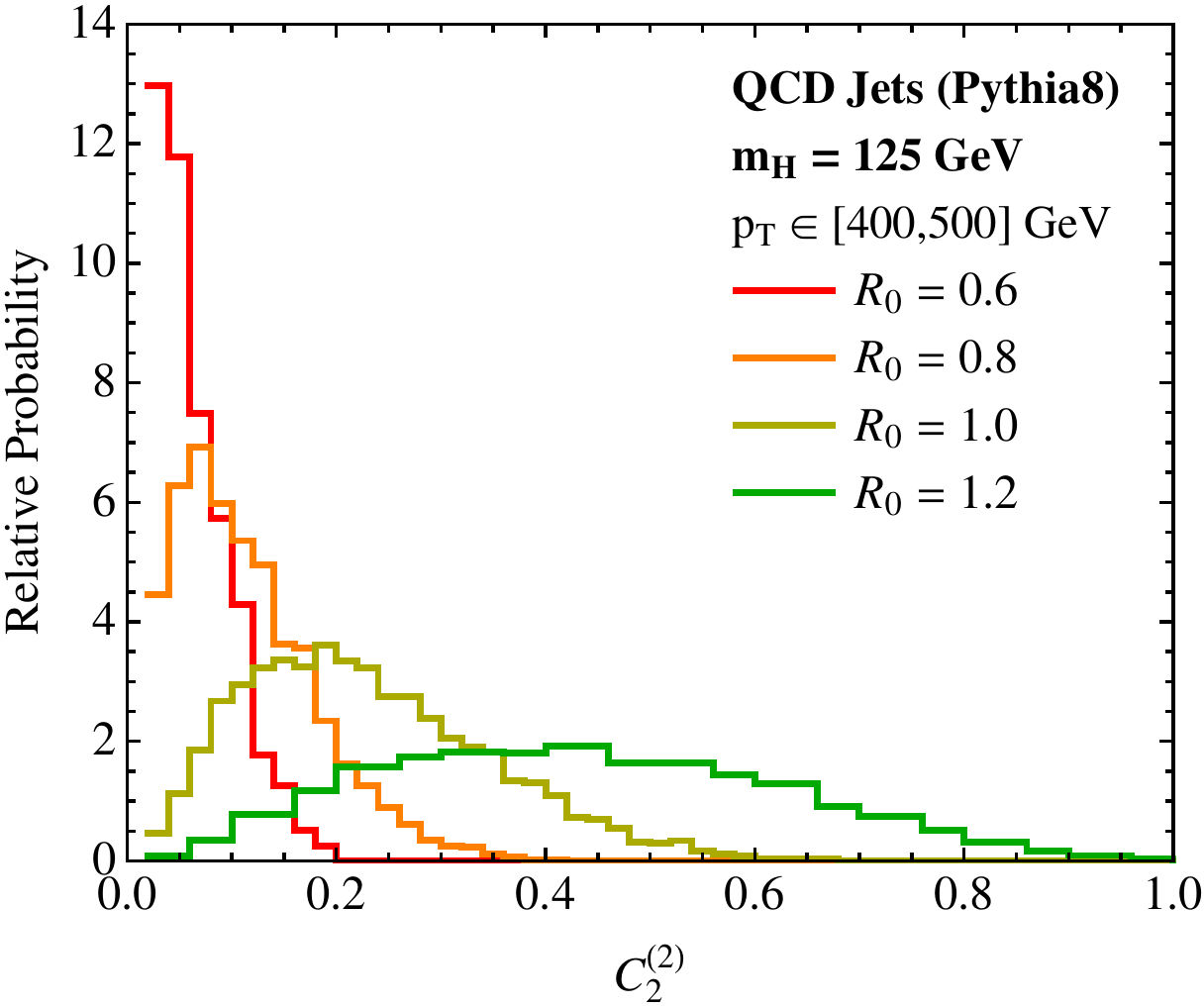}
}
\subfloat[]{\label{C2_distro_h}  \includegraphics[width=7.0cm]{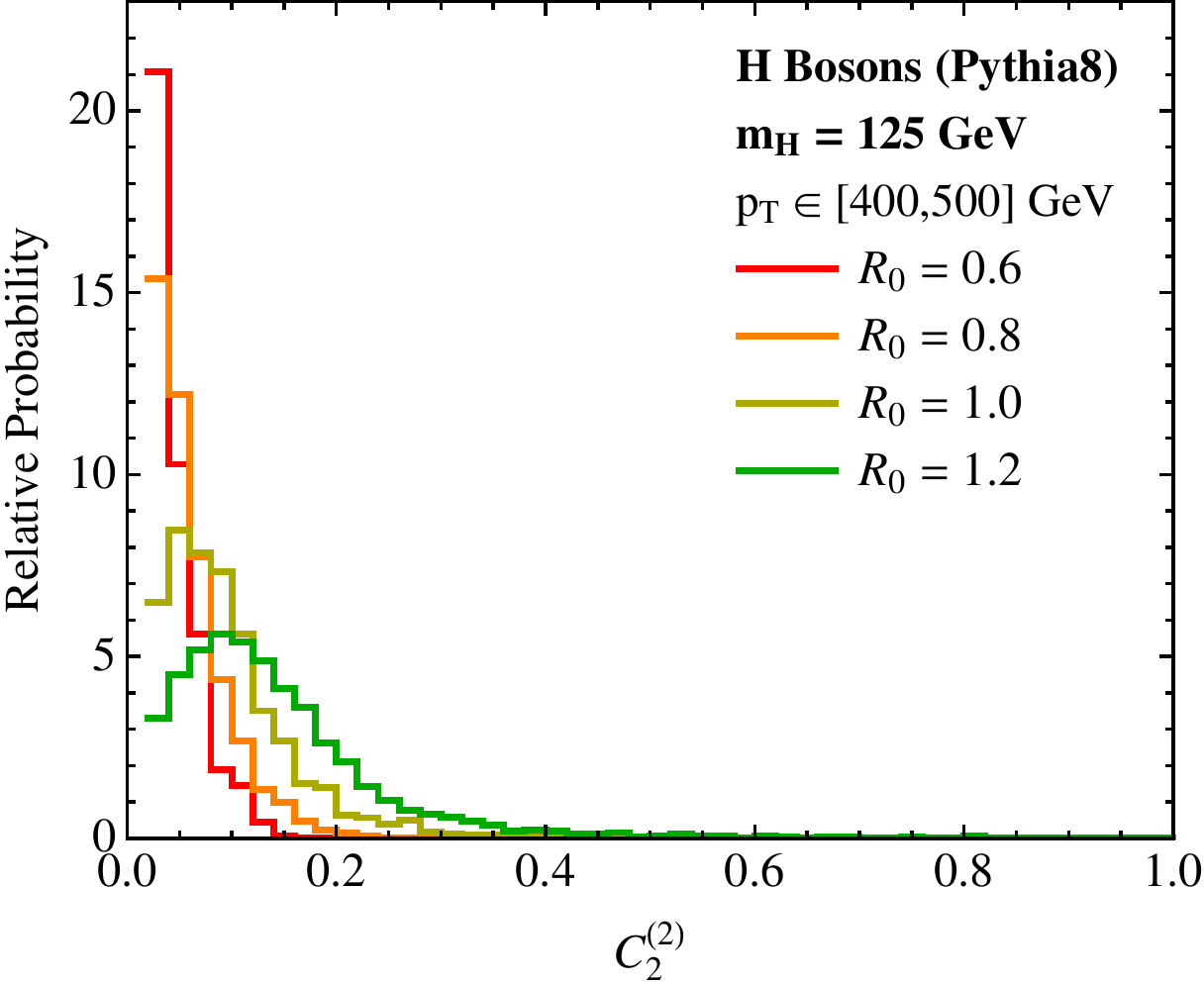}
}
\caption{Distributions of $\C{2}{2}$ for $b\bar{b}$ jets from QCD (left) and Higgs bosons decaying to $b\bar{b}$ (right) with different jet radii.  The plotted events are in the mass window $m_J \in [110, 140]~\GeV$ and the transverse momentum window $p_T \in [400, 500]~\GeV$.
}\label{C2_distroHiggs}
\end{figure}

\begin{figure}
\centering
\subfloat[]{\label{hvg_roc} \includegraphics[width=7.0cm]{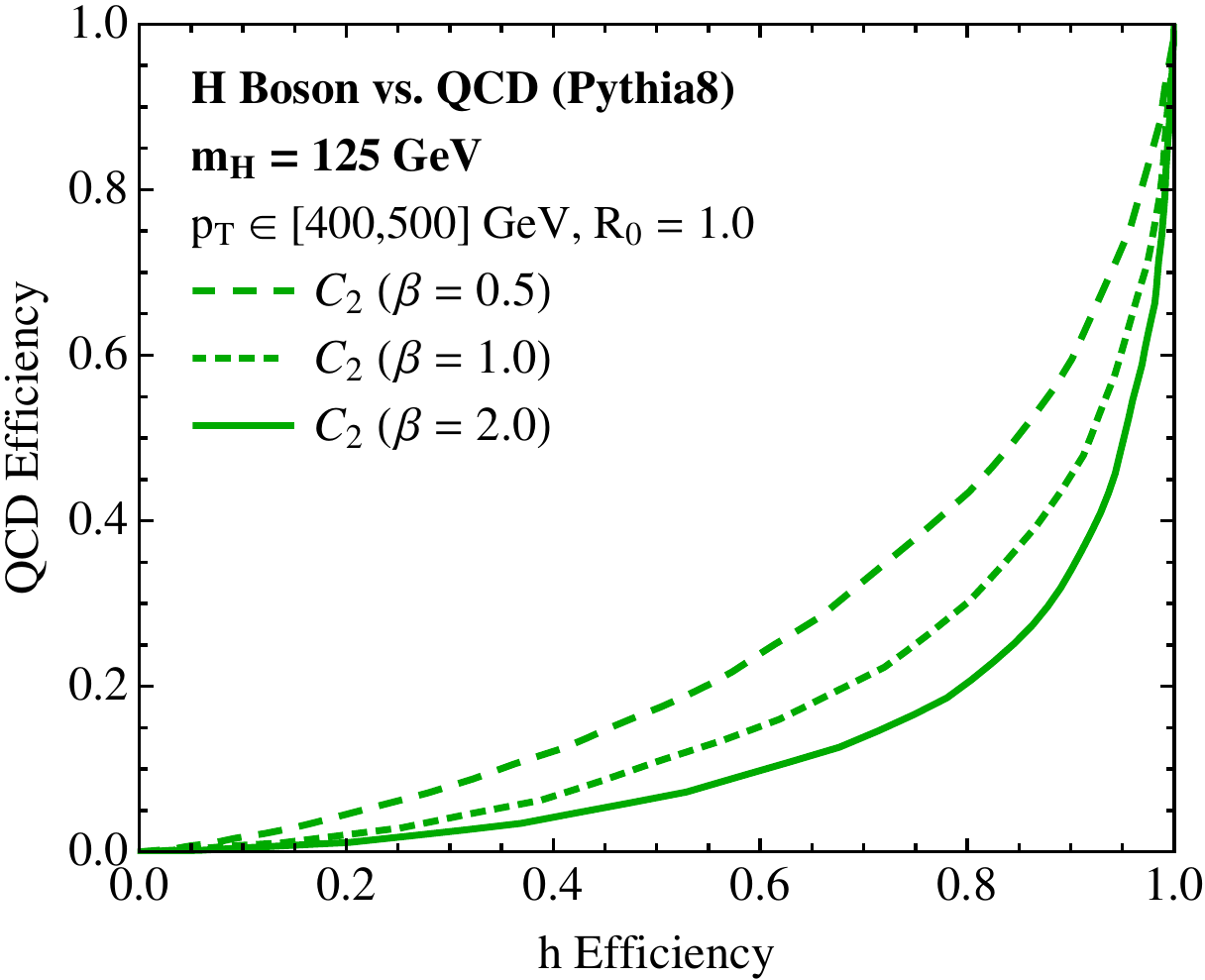}
}
\subfloat[]{\label{hvg_sb}  \includegraphics[width=7.0cm]{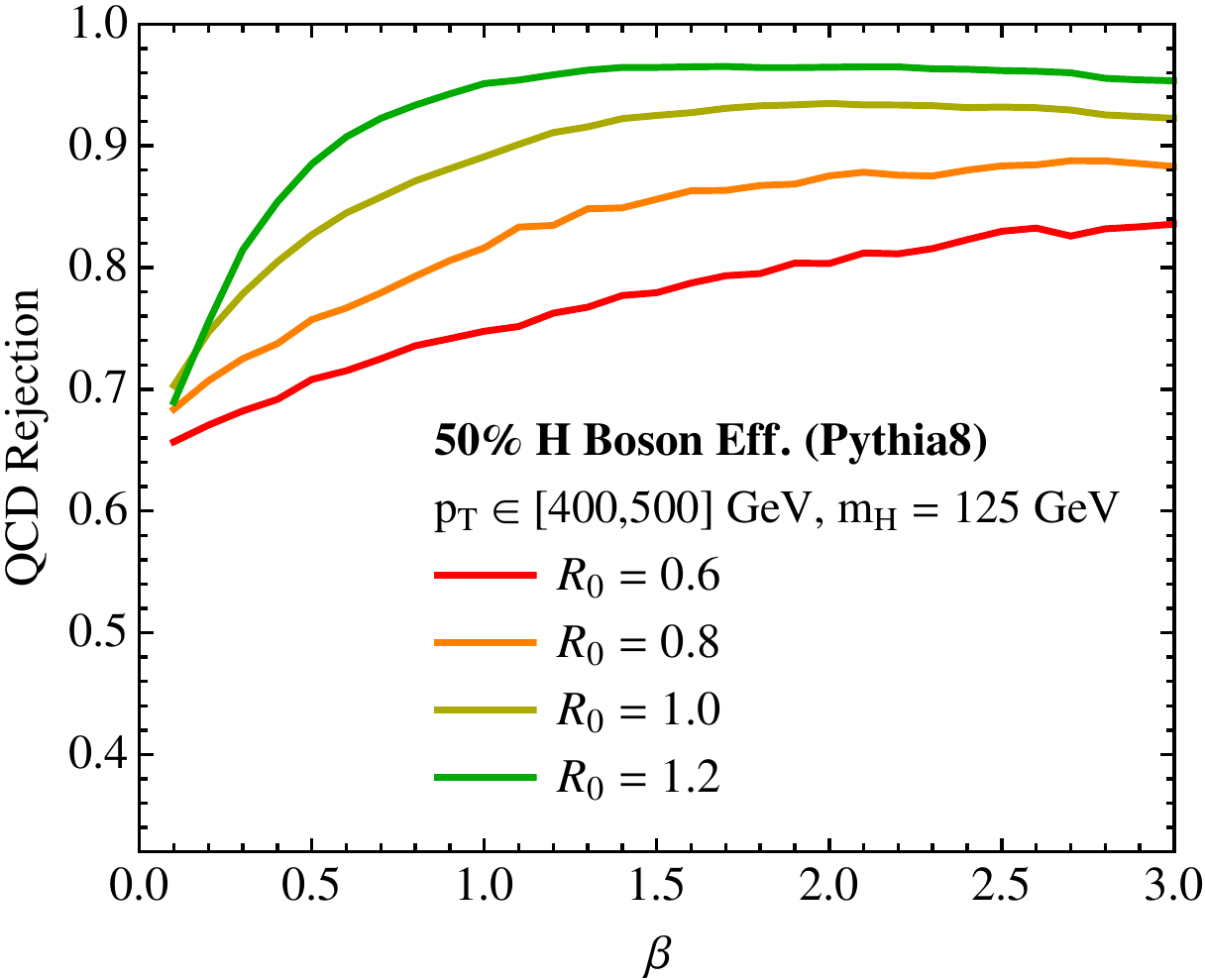}
}
\caption{Left:  Discrimination curves of $H\to b\bar{b}$ jets versus $b\bar{b}$ jets from QCD with $\C{2}{\beta}$ for several values of $\beta$ with jet radius $R_0=1.0$.
Right:  QCD $b \bar{b}$ rejection rate for 50\% boosted $H\to b\bar{b}$ efficiency as a function of $\beta$, sweeping the value of the jet radius $R_0= \{0.6,0.8,1.0,1.2\}$.
}\label{hvg_ROC}
\end{figure}

\begin{figure}
\centering
\subfloat[]{\label{hvg_comp_10} \includegraphics[width=7.0cm]{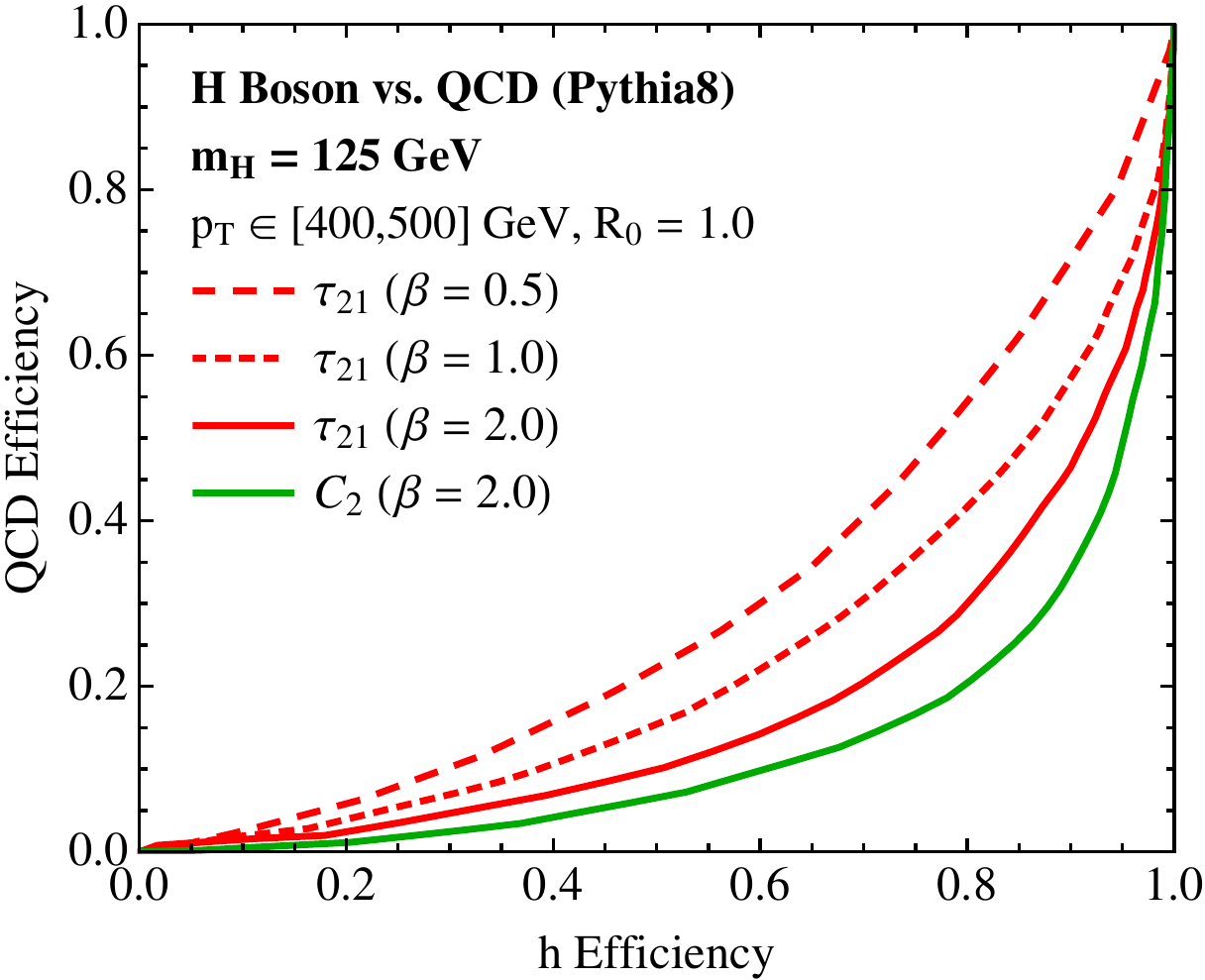}
}
$\qquad$
\subfloat[]{\label{hvg_comp_sb}  \includegraphics[width=7.0cm]{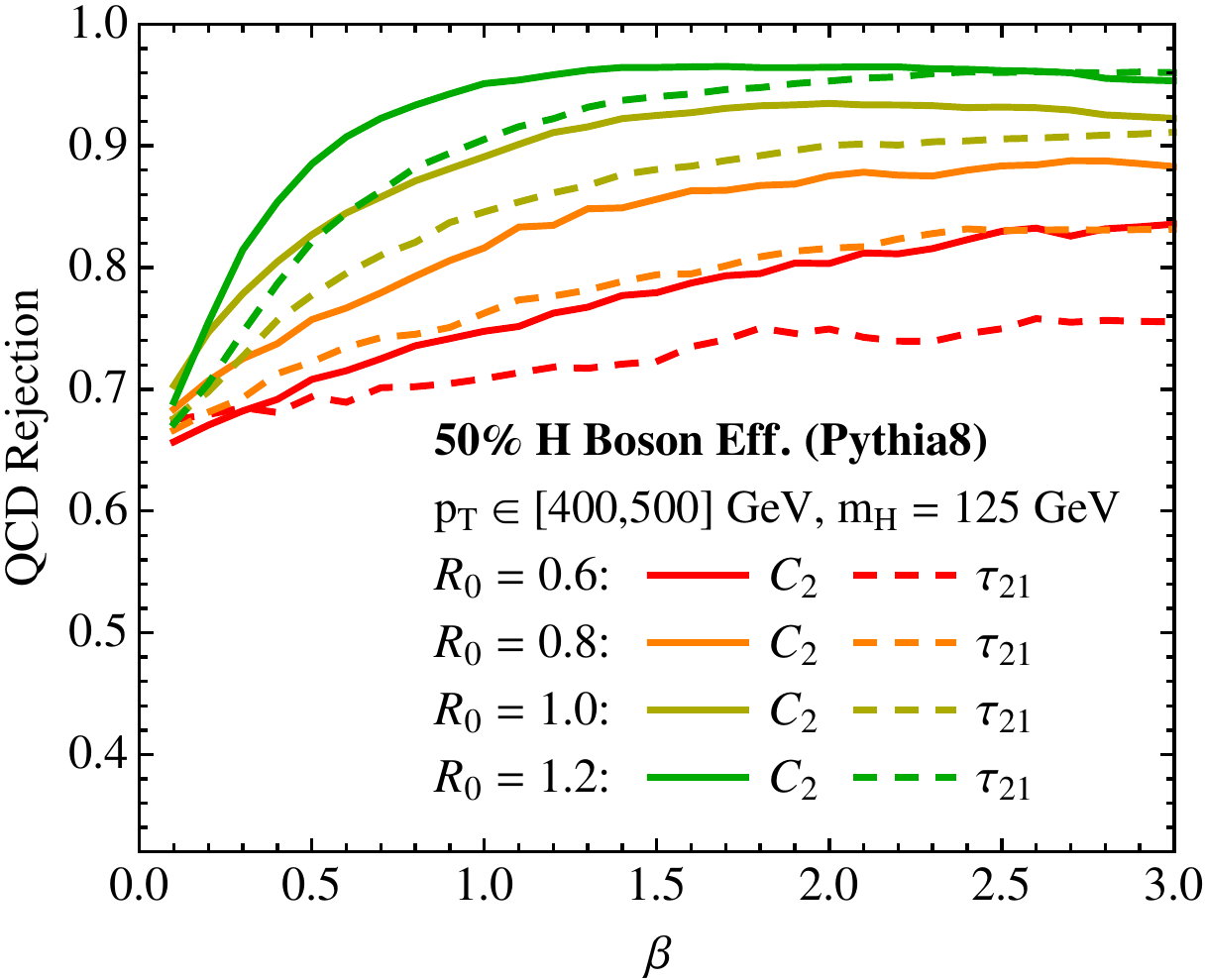}
}
\caption{Left: the discrimination curves for boosted $H\to b\bar{b}$ compared to QCD $b \bar{b}$ jets with $\tau_{2,1}^\beta$ for various values of $\beta$. For comparison is shown the $\C{2}{\beta}$ curve with the best discrimination ($\beta = 2.0$).
Right:  QCD $b \bar{b}$ rejection rate for 50\% boosted $H\to b\bar{b}$ efficiency as a function of $\beta$, sweeping the value of the jet radius $R_0= \{0.6,0.8,1.0,1.2\}$.
}\label{hvg_ROC_comp}
\end{figure}

In \Fig{C2_distroHiggs}, we plot the distributions of $\C{2}{2}$ for
Higgs jets and the QCD background for various jet radii.  As expected,
the $\C{2}{2}$ distributions dramatically increase as the jet radius
increases for QCD jets, while they only increase slightly for Higgs jets.  In \Fig{hvg_roc}, we plot the discrimination curves of Higgs jets versus QCD using $\C{2}{\beta}$ for several values of $\beta$ for the jet radius $R_0 = 1.0$.  As expected, the discrimination power increases both as the angular exponent increases, again, a consequence of the greater amount of wide-angle radiation in the QCD jets.  
\Fig{hvg_sb} shows the $\beta$ dependence on the QCD rejection rate for 50\% boosted Higgs efficiency for jet radii of $R_0=\{0.6,0.8,1.0,1.2\}$.  The rejection rate increases dramatically as the jet radius increases.
At small jet radii, large values of $\beta$ lead to the best
discrimination, as large $\beta$ emphasizes wide-angle emissions which
differ for QCD and boosted Higgs jets.  As the jet radius increases,
the largest QCD rejection rate moves to intermediate values of
$\beta$.  This may be because a large jet radius will tend to include
more initial state radiation or underlying event, which is independent
of the dynamics of the jet.  

In \Fig{hvg_comp_10}, we compare the discrimination performance of the $N$-subjettiness ratio $\Nsub{2,1}{\beta}$ to the most discriminating $\C{2}{\beta}$ ($\beta=2.0$) with jet radius equal to $R_0 = 1.0$.
Over the entire range of signal efficiencies, $\C{2}{2}$ performs better than $\Nsub{2,1}{\beta}$ for any value of $\beta$.  In \Fig{hvg_comp_sb}, we plot the QCD rejection rate for 50\% boosted Higgs efficiency at several
jet radii.  For a jet with a small jet radius, $\C{2}{\beta}$ performs significantly better than any $N$-subjettiness, with the distinction decreasing as the jet radius increases.  

\section{Boosted Top Quarks with ${\Cnobeta{3}}$}
\label{sec:threeprong}

Our final case study tests the discrimination power of even
higher-point correlation functions, namely using $\Cnobeta{3}$ to
distinguish boosted top quarks from QCD jets.\footnote{For related
  studies, see
  \Refs{Seymour:1993mx,Kaplan:2008ie,Brooijmans:1077731,Almeida:2008tp,Almeida:2008yp,Thaler:2008ju,CMS:2009lxa,CMS:2009fxa,Rappoccio:1358770,ATL-PHYS-PUB-2009-081,ATL-PHYS-PUB-2010-008,Plehn:2009rk,Plehn:2010st,Almeida:2010pa,Thaler:2010tr,Thaler:2011gf,Jankowiak:2011qa,Soper:2012pb}.}  Unlike the previous two case studies,
this observable is significantly more challenging than lower point
correlation functions.  From a computational point of view,
$\Cnobeta{3}$ involves a 4-point correlator, so its computational cost
is expensive since it scales like $k^4$, where $k$ is
the number of particles in the system.  That said, our \fastjet add-on only requires a few milliseconds to analyze a boosted top event at one value of $\beta$.  From an analytical point of
view, each term in $\EC{4}{\beta}$ involves a product of 4 energies
and $\binom{4}{2}=6$ angles, complicating an understanding of how $\Cnobeta{3}$ behaves in various limits.

We will find that $\Cnobeta{3}$ performs significantly worse than one might expect from the strong performance seen in $\Cnobeta{1}$ and 
$\Cnobeta{2}$. While it is possible that this is an artifact of choosing the particular double ratio combination in the definition of $\Cnobeta{3}$, we suspect that the proliferation of energy and angular factors in $\EC{4}{\beta}$ is reducing the sensitivity of $\Cnobeta{3}$ to any individual soft emission.  
In particular, for a soft-collinear emission, $\Cnobeta{1}$ and $\Cnobeta{2}$ are independent of the kinematics of the hard structure of the jet.
By contrast, even for a soft-collinear emission, $\Cnobeta{3}$ retains dependence on the hard kinematics of the jet.  This is because the correlation functions in the ratio defining $\Cnobeta{3}$ are dominated by possibly different subsets of the hard emissions in the jet.
Nevertheless, it is illustrative to see that even with these limitations, there is still discrimination power in $\Cnobeta{3}$.

To study the performance of $\Cnobeta{3}$ as a top tagger, we use the boosted top and QCD background event samples created for the BOOST 2010 workshop \cite{Abdesselam:2010pt}.\footnote{The events can be found at \url{http://www.lpthe.jussieu.fr/~salam/projects/boost2010-events/} and \url{http://tev4.phys.washington.edu/TeraScale}.
While updated event samples are available from the BOOST 2011 report \cite{Altheimer:2012mn}, the comparison includes a $W$ subjet tagger which would artificially improve the performance of $\Cnobeta{3}$.
}
These events come from 7 TeV LHC collisions simulated with \herwig~{6.510} \cite{Corcella:2002jc} with underlying event simulated with JIMMY \cite{Butterworth:1996zw} with an ATLAS tune \cite{atlasmc}.
The event samples consist of $2\to 2$ QCD processes, either all hadronic $t\bar{t}$ production or dijet production.  For direct comparison to other top tagging procedures, we follow the analysis procedures used in \Ref{Abdesselam:2010pt}.   We identify anti-$k_T$ jets with radius $R_0=1.0$ and demand that the jets have $p_T \in [500,600]~\GeV$.  No detector simulation is performed at this stage, other than removing muons and neutrinos before jet finding.  We impose three cuts to discriminate top jets from QCD.  First, we demand that the jets have mass in the fixed window of $160 < m_J < 240$ GeV, and second, we apply a sliding cut on $\C{3}{\beta}$.  In addition, it was noted in \Ref{Soyez:2012hv} that ratio observables such as $\Cnobeta{3}$ can be IR-unsafe without an additional cut.  We therefore apply a third cut that $\C{2}{\beta} > 0.1$, which makes $\Cnobeta{3}$ explicitly IR-safe.

\begin{figure}
\centering
\subfloat[]{\label{topvqcd}  \includegraphics[width=7.0cm]{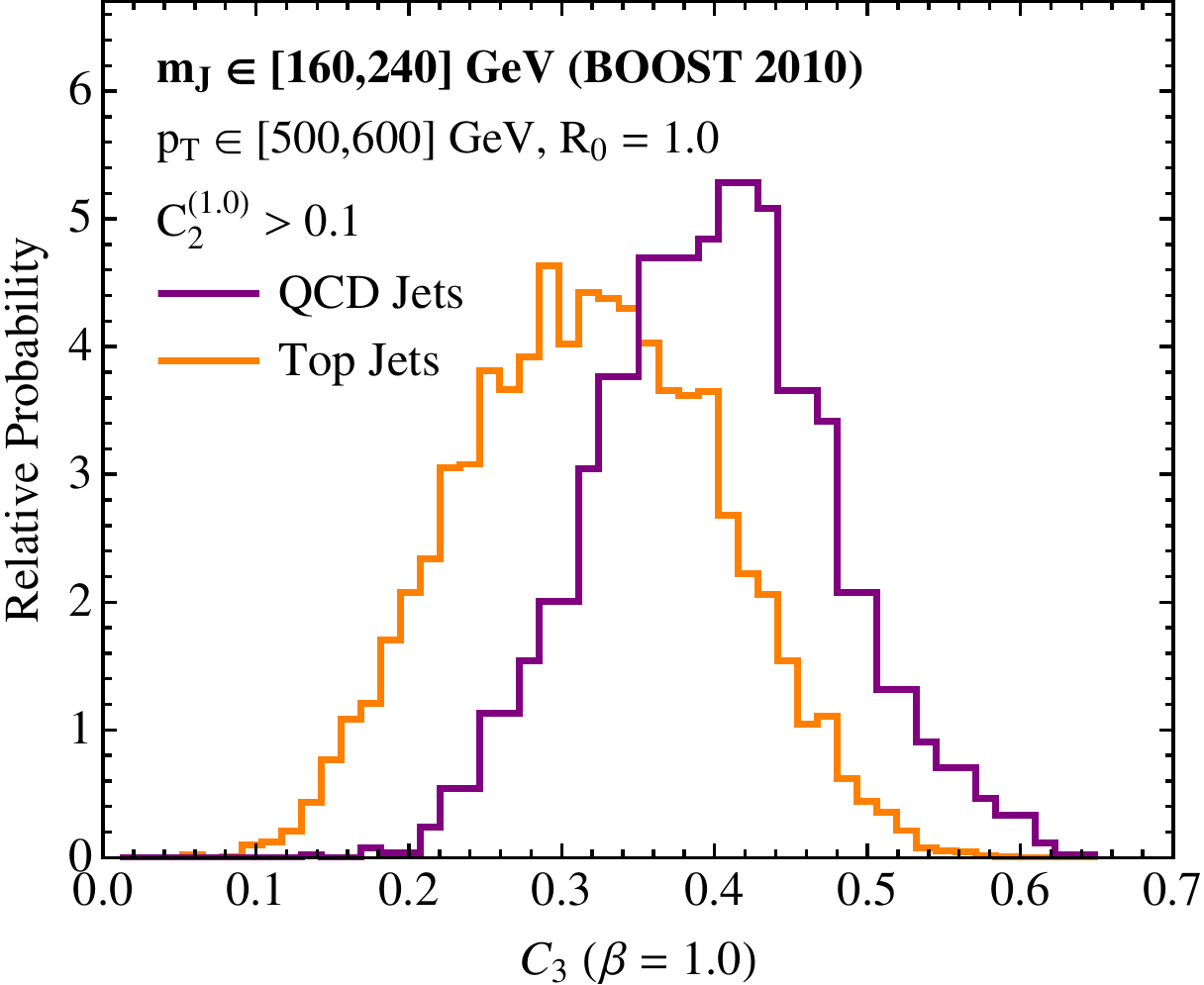}
}
\subfloat[]{\label{top_tag_roc_beta} \includegraphics[width=6.5cm]{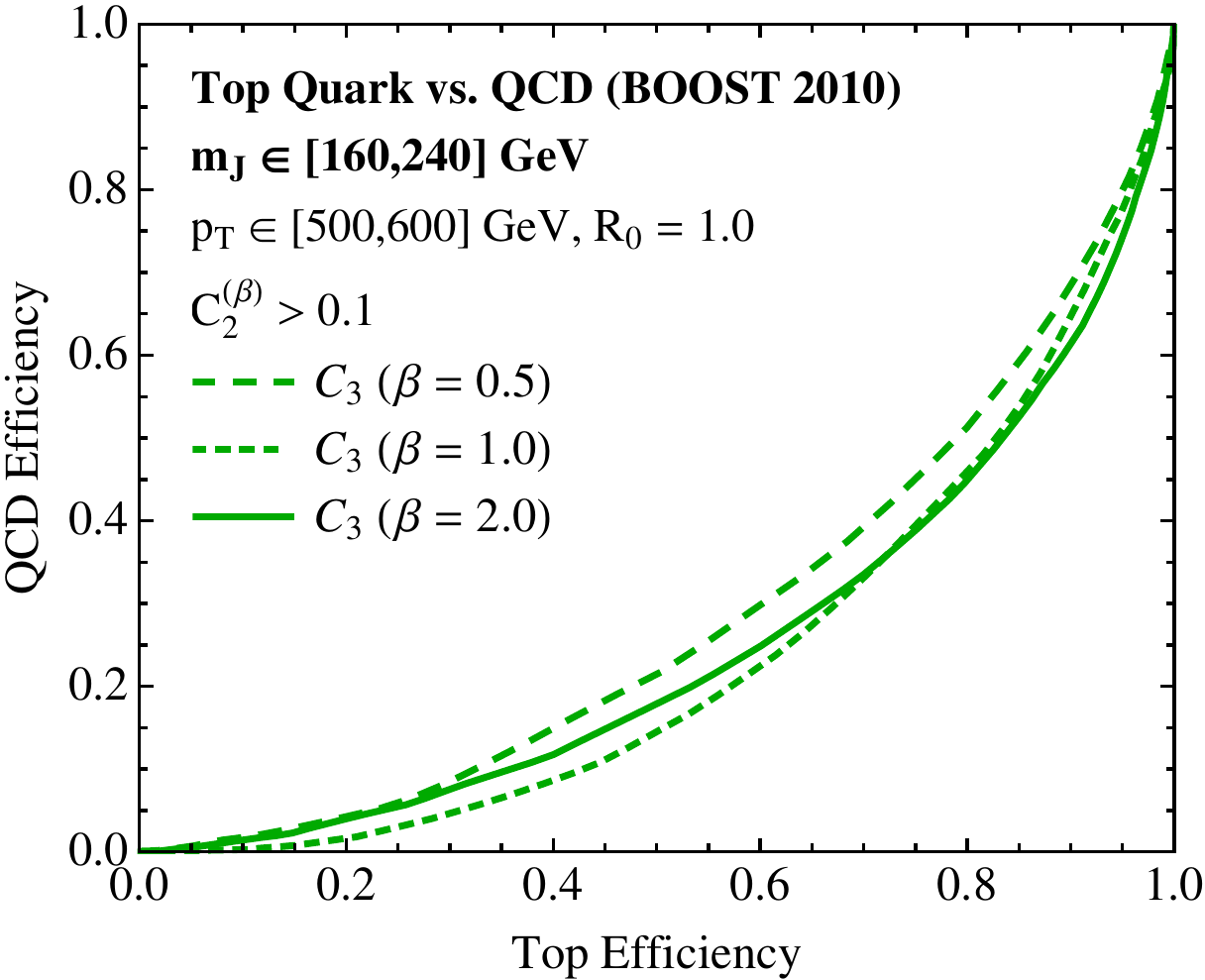}
}

\caption{Left:  Distribution of $\C{3}{1}$ comparing top jets and QCD jets.  The plotted events are in the mass window $m_J \in [160, 240]~\GeV$ and the transverse momentum window $p_T \in [500, 600]~\GeV$.  Right:  Discrimination curves for top jets versus QCD jets, using $\C{3}{\beta}$ for several values of $\beta$.  These efficiencies 
only include the effect of the cut on $\Cnobeta{3}$.
}
\end{figure}

\begin{figure}
\centering
\includegraphics[width=8.5cm]{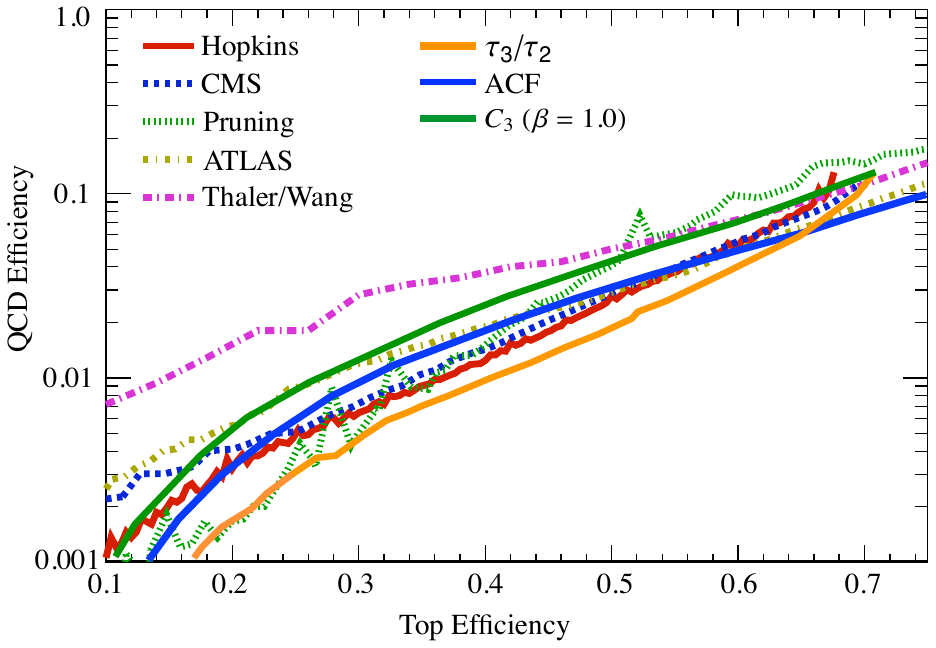}
\caption{Comparing the performance of $\C{3}{1}$ to other methods studied in the BOOST 2010 report \cite{Abdesselam:2010pt}.  The efficiency curves for $N$-subjettiness ($\tau_3/\tau_2$) \cite{Thaler:2011gf} and the angular correlation function (ACF) \cite{Jankowiak:2011qa} were added later.  Here, the efficiencies 
include both the effect of a mass cut as well as a cut on $\C{3}{1}$.}
\label{top_tag_roc} 
\end{figure}

In scanning over the range of $0.5 < \beta < 2.5$, we found that the best discrimination over a wide range of signal efficiencies using $\C{3}{\beta}$ is obtained for $\beta=1.0$.  This is the same $\beta$ value that is optimal for $N$-subjettiness $\tau^{(\beta)}_{3,2}$ \cite{Thaler:2011gf}. A plot of the distribution of $\C{3}{1}$ for top jets and QCD jets in the kinematic and mass windows from above is shown in \Fig{topvqcd}.  The discrimination curves for the different values of $\beta$ are shown in \Fig{top_tag_roc_beta}, where the quoted efficiencies only include the effect of a cut on the observable $\Cnobeta{3}$ for jets in the mass window of $160$ to $240$ GeV.

Finally, we compare the performance of $\C{3}{1}$ against several other top tagging procedures in \Fig{top_tag_roc}.  For this plot, the quoted efficiencies include both the effect of the mass cut as well as the effect from a cut on $\C{3}{1}$. While not as powerful as methods like $N$-subjettiness, the energy correlation function yields comparable discrimination power to other methods.  Of course, the performance may be improved by combining information from different values of $\beta$, as well as including additional $\Cnobeta{2}$ and $\Cnobeta{1}$ information.

\section{Conclusions}
\label{sec:conc}

In this paper, we have introduced arbitrary-point energy correlators that are sensitive to $N$-prong substructure.  These correlators are effective when used as part of an energy correlation double ratio $\C{N}{\beta}$, though more general combinations deserve further exploration.  Through an NLL calculation, we have seen how $\Cnobeta{1}$ yields excellent quark/gluon discrimination, with $\beta \simeq 0.2$ being most effective at capturing the differences in color charges.  We have also shown the power of $\Cnobeta{2}$ for boosted 2-prong objects like Higgs bosons, and the potential power of $\Cnobeta{3}$ for boosted 3-prong objects like top quarks.

Given the explosion of jet substructure methods over the past few
years, it is worth asking whether $\Cnobeta{N}$ is sufficiently novel
to merit further experimental and theoretical studies.  Indeed, it is
a quite unique variable that combines a number of desirable features.
Like $N$-subjettiness, $\Cnobeta{N}$ is a variable which tests for
$N$-prong substructure, but can behave more continuously in situations
with soft subsets. Like planar flow and related jet shapes,
$\Cnobeta{N}$ can be calculated directly from the energies and angles
of the jet constituents without a separate axes finding step, but it
is designed for identifying $N$-prong substructure instead of just
exotic kinematic configurations.  Finally, like jet angularities,
$\Cnobeta{N}$ is sensitive to higher-order radiation about LO substructure,
but because it is a recoil-free observable, it can better probe
the collinear physics that distinguishes a jet's color 
with $0.2 \lesssim \beta \lesssim 1.0$.
  Because $\Cnobeta{N}$ has a high computational cost
for $N > 3$, we expect $\Cnobeta{N}$ will be most useful in practice
for 1-, 2-, and 3-prong jet studies.

To gain further confidence in the behavior and performance of these observables, further analytic studies are needed.  Of particular need is to calculate $C_2$ for QCD backgrounds.  We already saw that the behavior of $C_2$ for QCD jets depends strongly on the jet's mass over $p_T$ ratio, and it is likely that different theoretical descriptions will be needed for $C_2$ as a function of $m/p_T$.  While $C_2$ is built as a ratio of IRC safe observables, $C_2$ itself is only IRC safe with a cut on the jet mass (which acts like a cut on the denominator), and it is an interesting question how to best perform NLL resummation for generic ratio observables.   Like all jet shape observables, $C_2$ is sensitive to underlying event, initial state radiation, and pileup, which must be accounted for in determining the optimal $\beta$ value.  Ideally, theoretical progress on understanding $C_2$ and other jet shapes will match the rapid experimental progress in implementing them, such that jet substructure observables can truly be a robust tool for LHC physics.

\begin{acknowledgments}
  We thank Duff Neill, Gregory Soyez, and Iain Stewart for helpful conversations.  We thank
  David Lopez for performing preliminary studies of $\Cnobeta{1}$ and Andreas Hinzmann
  for performing preliminary studies of $\Cnobeta{2}$.
  A.L.\ and J.T.\ are supported by the U.S. Department of Energy (DOE)
  under cooperative research agreement DE-FG02-05ER-41360 and by
  the DOE Early Career research program DE-FG02-11ER-41741.
  G.P.S.\ is supported in part by the French Agence Nationale de la
  Recherche, under grant ANR-09-BLAN-0060 and
  by the EU ITN grant LHCPhenoNet, PITN-GA-2010-264564.
  A.L., G.P.S., and J.T.\ are also
  supported in part by MISTI global seed funds.
  A.L.\ and G.P.S.\ are grateful to Perimeter Institute for Theoretical
  Physics for hospitality and support; research at Perimeter Institute
  is supported by the Government of Canada through Industry Canada and
  by the Province of Ontario through the Ministry of Economic
  Development \& Innovation.
  G.P.S.\ wishes to thank MIT for hospitality while this work was
  being initiated.
  A.L.~thanks the CERN TH division for support and hospitality while this work was being completed.
\end{acknowledgments}

\appendix

\section{Fixed-Order Calculation}\label{app:match}

In this appendix, we present the details of the fixed-order
calculation of $\Cnobeta{1}$ and matching to the NLL result from
\Sec{subsec:NLLanalysis}.  
The calculation is valid for any jet algorithm that, for
configurations involving exactly two partons in some neighborhood,
combines two partons into a single jet if they are separated by an
angle less than $R_0$ and otherwise places them in separate jets.
At this order we define a quark jet to be a jet that contains a
single quark, or a quark and a gluon.
A gluon jet is a jet that contains a single gluon, a gluon pair, or
a quark-antiquark pair (of identical flavor).
\footnote{Algorithms that satisfy the condition for when they pair
  partons into a single jet include all members of the
  generalized-$k_T$ family, notably the anti-$k_T$
  algorithm~\cite{Cacciari:2008gp}.
  One subtlety is that the flavor of jets from such algorithms is not
  infrared safe for configurations with three or more particles in a
  common neighborhood. 
  There exist algorithms designed specifically to guarantee a
  safe jet flavor to all orders, the ``flavor-$k_T$''
  algorithms~\cite{Banfi:2006hf}. These, however, have the property
  that a quark-antiquark pair can be combined into a jet even for
  angular separations larger than $R_0$, and so they do not yield the
  same jets at $\order{\alpha_s}$ as the generalized-$k_T$ family and
  as we assume in the calculation here.
  We have reason to believe that it is possible to design an algorithm
  that is both equivalent to generalized-$k_T$ at $\order{\alpha_s}$
  and flavor safe to all orders, but leave an investigation of this
  question to future work.  }

In the limit where the jet radius $R_0$
is small, the ${\cal O}(\alpha_s)$ cumulative distribution of the
observable $\C{1}{\beta}$ can be computed from 
\begin{align}
\Sigma(\Cnobeta{1}) &= 1+\frac{\alpha_s}{\pi}\int_0^{R_0}\frac{d\theta}{\theta} \int_0^1 dz\, P(z) \, \Theta\left( \Cnobeta{1} - z(1-z)\theta^\beta\right) \nonumber\\
&=1 -\frac{\as}{\pi}\frac{1}{\beta} \int\limits_{\frac{1-u}{2}}^{\frac{1+u}{2}} dz \, P(z)\, \ln\frac{z(1-z)R_0^\beta}{\Cnobeta{1}} \ ,
\end{align}
where
\begin{equation}
  u\equiv \sqrt{1-\frac{4\Cnobeta{1}}{R_0^\beta}} \ ,
\end{equation}
and we have approximated the full matrix element by the appropriate
splitting function, $P(z)$, as is legitimate for $R_0 \ll 1$.  
The splitting functions are
\begin{equation}
P_q(z) = C_F\frac{1+z^2}{1-z} \ ,
\end{equation}  
for quarks and
\begin{equation}
P_g(z) = C_A\left( \frac{z}{1-z} +\frac{1-z}{z}+z(1-z)\right)+\frac{n_F}{2} \left( z^2 + (1-z)^2 \right) \ ,
\end{equation}
for gluons, including combinatoric factors.  
For quarks, it follows that
\begin{align}
\Sigma_q(\Cnobeta{1}) = & \ 1-\frac{\alpha_s}{\pi}\frac{C_F}{\beta} \left\{-4\, \text{Li}_2\left(\frac{1+u}{2}\right)+3 u  + \ln ^2\left(1-u\right)  -2 \ln \left(u+1\right) \ln\left(1-u\right)\right.  \nonumber \\
& \left.+\ \left[4 \ln  2 -\ln \left(u+1\right)\right] \ln\left(u+1\right)   - \ 3 \tanh^{-1}\left(u\right)+\frac{\pi ^2}{3}-2 \ln^2 2  \right\} \ .
\end{align}
For gluons, the fixed-order cumulative distribution is
\begin{align}
\Sigma_g(\Cnobeta{1}) =&\ 1 - \frac{\alpha_s}{\pi}\frac{1}{\beta}\left\{C_A\left[-4\, \text{Li}_2\left(\frac{1+u}{2}\right)-\frac{2}{9} \frac{\Cnobeta{1}}{R_0^\beta} u \right. \right. +\frac{67}{18} u+\ln ^2\left(1-u\right)-\ln^2\left(u+1\right)\nonumber \\
&\left.-\ 2 \ln \left(1-u\right) \ln \left(u+1\right)+4 \ln 2 \ln \left(u+1\right) - \frac{11}{3} \tanh ^{-1}\left(u\right)+\frac{\pi ^2}{3}-2 \ln ^2 2 \right] \nonumber \\
&\left. + \ n_F  \left[u \left(\frac{2}{9} \frac{\Cnobeta{1}}{R_0^\beta}-\frac{13}{18}\right)+\frac{2}{3} \tanh ^{-1}\left(u\right)
\right]
\right\} \ .
\end{align}
Here, $\text{Li}_2(x)$ is dilogarithm function
\begin{equation}
\text{Li}_2(x) = -\int_0^x dt \, \frac{\ln(1-t)}{t} \ .
\end{equation}

To match the fixed-order cumulative distribution to the NLL cumulative distribution, we use the ``Log-R'' matching scheme \cite{Catani:1992ua}.  In this matching scheme, the fixed-order corrections are exponentiated with the NLL distribution.  The logarithms that appear in the fixed-order expression must be properly subtracted so as to eliminate a double counting with the logarithms that were resummed.  Also, at large values of $\Cnobeta{1}$, we want the distribution to agree with the fixed-order result.  This requires ``turning off'' the logarithms of the resummation properly.

Matching ${\cal O}(\alpha_s)$ fixed-order to NLL is straightforward.  The matching scheme can be written as
\begin{equation}
\Sigma_{\text{match}} = \Sigma\left( L \right)_{\text{resum}}\, e^{-\frac{\alpha_s}{\pi} \left( R_1 - G_0-G_1L - G_2L^2 \right)} \ .
\end{equation}
Here, $R_1$ is defined from the fixed-order cumulative distribution as
\begin{equation}
\Sigma = 1 - \frac{\alpha_s}{\pi} R_1 + \order{\alpha_s^2},
\end{equation}
and $G_0$, $G_1L$ and $G_2L^2$ are placeholders representing the constant terms, single logarithms and double logarithms that have been resummed, respectively.  For quarks, the logarithms are
\begin{equation}
\left( G_1L +G_2L^2 \right)_q = \frac{C_F}{\beta} \ln ^2\frac{R_0^{\beta
   }}{\Cnobeta{1}}-\frac{3}{2}\frac{C_F}{\beta} \ln \frac{R_0^{\beta }}{\Cnobeta{1}},
\end{equation}
and for gluons the logarithms are
\begin{equation}
\left(G_1L +G_2L^2  \right)_g =\frac{C_A}{\beta} \ln ^2 \frac{R_0^{\beta }}{\Cnobeta{1}}
  -\frac{11}{6}\frac{C_A}{\beta}\ln \frac{R_0^{\beta }}{\Cnobeta{1}}+\frac{1}{3}\frac{n_F}{\beta} \ln \frac{R_0^{\beta
   }}{\Cnobeta{1}} \ .
\end{equation}
The choice of $G_0$ in $\Sigma(L)_\text{resum}$ is arbitrary because
these terms are subleading to the NLL resummation.
Subtracting these logarithms from $R_1$, in addition to the constant terms $G_0$, eliminates double counting.  Also, to verify that the distribution agrees with the fixed-order result at large values of $\Cnobeta{1}$, we can shift the argument of the logarithms appropriately to vanish when $\Cnobeta{1} = \frac{R_0^\beta}{4}$, which is the maximum value of $\Cnobeta{1}$.  That is, we replace the logarithms in the resummation and subtraction to be
\begin{equation}
L\to \tilde{L} = \ln\left( \frac{R_0^\beta}{\Cnobeta{1}} - \frac{R_0^\beta}{C_1^{\max}} + 1   \right)  = \ln\left(  \frac{R_0^\beta}{\Cnobeta{1}} - 3 \right) \ .
\end{equation}
$\tilde{L}$ vanishes when $\Cnobeta{1} = \frac{R_0^\beta}{4}$ and smoothly interpolates to $L$ in the small $\Cnobeta{1}$ region.  The final NLL resummed cumulative distribution matched to fixed-order is
\begin{equation}
\Sigma_{\text{match}} = \Sigma\bigl( \tilde{L} \bigr)_{\text{resum}}\, e^{-\frac{\alpha_s}{\pi} \left( R_1 - G_0-G_1\tilde{L}-G_2\tilde{L}^2 \right)} \ .
\end{equation}
We use this expression to determine the quark versus gluon discrimination in \Sec{subsec:NLLanalysis}.

\section{Breakdown of Perturbative Calculation}\label{app:nonpert}

In this appendix, we provide some simple quantitative arguments for the breakdown of the perturbative calculation from \Sec{subsec:NLLanalysis} for values of $\beta$ less than about $0.2$.  There are two effects that we will consider:  the QCD Landau pole and the breakdown of the independent emission approximation.  Of course, there may be other effects that become important at small values of $\beta$, but these nevertheless suggest that our perturbative calculation of the quark versus gluon discrimination power ceases to make sense at very small values of $\beta$.

First, the QCD Landau pole.  At small $\beta$, the smallest scale $Q_0$ that the running coupling is sensitive to is
\begin{equation}
Q_0 = p_T R_0 e^{-L/\beta} \ .
\end{equation}
The perturbative calculation can be trusted when $Q_0 \gg
\Lambda_{\text{NP}}$, where $\Lambda_{\text{NP}}$ is a scale at which
$\alpha_s$ becomes non-perturbative. 
We can estimate the value of $\beta$ at which the non-perturbative
effects become important as follows.
The logarithm of the observable $\Cnobeta{1}$ can be roughly estimated
from the LL, fixed-coupling expression for the cumulative distribution
$\Sigma$ for quarks, where
\begin{equation}\label{eq:appcum}
\Sigma \simeq e^{-\frac{\alpha_s}{\pi}\frac{C_F}{\beta}L^2} \ .
\end{equation}
Then, $L$ in terms of $\Sigma$ is
\begin{equation}
L \simeq \left( \frac{\pi}{\alpha_s} \frac{\beta}{C_F} \ln1/\Sigma  \right)^{1/2} \ .
\end{equation}
Demanding that $Q_0 > \Lambda_{\text{NP}}$ and using the expression for $L$ from the above equation we find that
\begin{equation} 
  \label{eq:betamin-fxied-coupling}
\beta_{\min} \simeq \frac{\pi \ln1/\Sigma}{\alpha_s C_F}\frac{1}{\ln^2 \frac{p_T R_0}{\Lambda_{\text{NP}}}} \ .
\end{equation}
Because we have used a fixed-coupling approximation, it is not
immediately clear at what scale $\alpha_s$ should most appropriately
be evaluated.
Taking it at the geometric mean of $p_T R_0$ and $\Lambda_{\text{NP}}
\simeq 0.5$~GeV gives $\alpha_s \simeq 0.17$.
Plugging this into \Eq{eq:betamin-fxied-coupling}, for a quark
efficiency of $50\%$ and a jet selection as in
\Sec{subsec:NLLanalysis}, yields
\begin{equation}
\beta_{\min} \simeq \frac{\pi \ln2}{0.17\, C_F}\frac{1}{\ln^2 \frac{400\times0.6}{0.5}} \simeq 0.25  \ ,
\end{equation} 
suggesting that non-perturbative effects become critical for $\beta
\lesssim 0.2$--$0.3$. 
One can also perform such an analysis numerically using the full NLL
expressions for $\Sigma_q$, including all running-coupling effects,
and one reaches a similar conclusion. 

Second, the NLL calculation assumed that emissions could be treated as independent, but multiple emissions cannot be regarded as independent when each emission can take an ${\cal O}(1)$ fraction of the energy of the jet.  That is, if the logarithm of $\Cnobeta{1}$ is not large then our analysis (appropriate for soft-collinear emissions) breaks down.  Assuming as above that the cumulative distribution can be written as in \Eq{eq:appcum}, the minimal value of $\beta$ is
\begin{equation}
\beta_{\min} \simeq \frac{\alpha_s}{\pi}\frac{C_F}{\ln 1/\Sigma}L_{\min}^2 \ ,
\end{equation}
where $L_{\min}$ is the minimal value for the logarithm at which we trust the soft-collinear analysis.
Assuming that the soft-collinear analysis fails when $L_{\min}\simeq
2$, with the same choice of parameter values as above, including
$\alpha_s \sim 0.17$, $\beta_{\min}$ is
\begin{equation}
\beta_{\min} \simeq \frac{0.17}{\pi}\frac{4/3}{\ln 2} \,2^2 \simeq 0.41 \ .
\end{equation}
The precise value of $L_{\min}$ at which the soft-collinear analysis
is deemed to break down will change this value.  
Nevertheless,
multiple hard, collinear emissions become important and result in a
breakdown of the analysis when $\beta$ is too small.  
To include the leading effect of energy conservation among emissions, one must match the NLL resummation to NLO (${\cal O}(\alpha_s^2)$) splitting functions.

It should be
noted that the fact that the non-perturbative analysis and the
multiple emissions analysis give the same ballpark of $\beta_{\min}$
is a coincidence due to the choice of parameters that were made. 

\bibliography{EnC}

\end{document}